\documentclass[twocolumn]{aastex62}
\usepackage[utf8]{inputenc}

\usepackage{amsmath}
\newcommand{\schw}{Schwarzschild }
\newcommand{\fdm}{$f_{\rm{DM}}$ }
\newcommand{\lam}{$\lambda_{Re, EO}$}
\usepackage{natbib}
\usepackage{graphicx}
\usepackage{multirow}

\begin{document}

    \correspondingauthor{Giulia Santucci}
	\email{g.santucci@student.unsw.edu.au}
	
    \author[0000-0003-3283-4686]{Giulia Santucci}
	\affil{School of Physics, University of New South Wales, NSW 2052, Australia}
	\affil{ARC center of Excellence for All Sky Astrophysics in 3 Dimensions (ASTRO 3D)}
	
	\author[0000-0002-9796-1363]{Sarah Brough}
	\affil{School of Physics, University of New South Wales, NSW 2052, Australia}
	\affil{ARC center of Excellence for All Sky Astrophysics in 3 Dimensions (ASTRO 3D)}
	
	\author[0000-0003-2552-0021]{Jesse van de Sande}
	\affil{Sydney Institute for Astronomy, School of Physics, University of Sydney, NSW 2006, Australia}
	\affil{ARC center of Excellence for All Sky Astrophysics in 3 Dimensions (ASTRO 3D)}
	
	\author[0000-0002-8175-7229]{Richard M. McDermid}
	\affil{School of Mathematical and Physical Science, Macquarie University, NSW 2109, Australia}
	\affil{Astronomy, Astrophysics and Astrophotonics Research center, Macquarie University, Sydney, NSW 2109, Australia}
	\affil{ARC center of Excellence for All Sky Astrophysics in 3 Dimensions (ASTRO 3D)}

    \author[0000-0003-4546-7731]{Glenn van de Ven}
	\affil{Department of Astrophysics, University of Vienna, Türkenschanzstrasse 17, 1180 Vienna, Austria}
	
	\author{Ling Zhu}
	\affil{Shanghai Astronomical Observatory, Chinese Academy of Sciences, 80 Nandan Road, Shanghai 200030, China}
    
    \author[0000-0003-2388-8172]{Francesco D'Eugenio}
    \affil{Cavendish Laboratory and Kavli Institute for Cosmology, University of Cambridge, Madingley Rise, Cambridge, CB3 0HA, United Kingdom}
	\affil{Sterrenkundig Observatorium, Universiteit Gent, Krijgslaan 281 S9, B-9000 Gent, Belgium}
	
	\author{Joss Bland-Hawthorn}
	\affil{Sydney Institute for Astronomy, School of Physics, University of Sydney, NSW 2006, Australia}
	\affil{ARC center of Excellence for All Sky Astrophysics in 3 Dimensions (ASTRO 3D)}
	
	\author{Stefania Barsanti}
	\affil{Research School of Astronomy and Astrophysics, The Australian National University, Canberra, ACT 2611, Australia}
	\affil{ARC center of Excellence for All Sky Astrophysics in 3 Dimensions (ASTRO 3D)}
	
	\author{Julia J. Bryant}
	\affil{Sydney Institute for Astronomy, School of Physics, University of Sydney, NSW 2006, Australia}
	\affil{ARC center of Excellence for All Sky Astrophysics in 3 Dimensions (ASTRO 3D)}
	\affil{Australian Astronomical Optics, AAO-USydney, School of Physics, University of Sydney, NSW 2006, Australia}
	
	\author{Scott  M. Croom}
	\affil{Sydney Institute for Astronomy, School of Physics, University of Sydney, NSW 2006, Australia}
	\affil{ARC center of Excellence for All Sky Astrophysics in 3 Dimensions (ASTRO 3D)}
	
	\author{Roger L. Davies}
	\affil{Astrophysics, Department of Physics, University of Oxford, Denys Wilkinson Building, Keble Rd., Oxford, OX1 3RH, UK}
    
    \author{Andrew W. Green}
	\affil{Atlassian, C/O Scott Croom, University of Sydney, NSW 2006 Australia}
	
	\author{Jon S. Lawrence}
	\affil{Australian Astronomical Optics, Faculty of Science \& Engineering, Macquarie University. 105 Delhi Rd, North Ryde, NSW 2113, Australia}
	
	\author{Nuria P.F. Lorente}
	\affil{Australian Astronomical Optics, Faculty of Science \& Engineering, Macquarie University. 105 Delhi Rd, North Ryde, NSW 2113, Australia}
	
	\author{Matt S. Owers}
	\affil{Department of Physics and Astronomy, Macquarie University, NSW 2109, Australia}
	\affil{Astronomy, Astrophysics and Astrophotonics Research center, Macquarie University, Sydney, NSW 2109, Australia}
	\affil{ARC center of Excellence for All Sky Astrophysics in 3 Dimensions (ASTRO 3D)}
	
	\author{Adriano Poci}
	\affil{center for Extragalactic Astronomy, University of Durham, Stockton Road, Durham DH1 3LE, United Kingdom}
	\affil{Astronomy, Astrophysics and Astrophotonics Research center, Macquarie University, Sydney, NSW 2109, Australia}
	
	\author{Samuel N. Richards}
	\affil{Sydney Institute for Astronomy, School of Physics, University of Sydney, NSW 2006, Australia}
	
	\author[0000-0003-1820-2041]{Sabine Thater}
	\affil{Department of Astrophysics, University of Vienna, Türkenschanzstrasse 17, 1180 Vienna, Austria}
	
	\author{Sukyoung Yi}
	\affil{Department of Astronomy and Yonsei University Observatory, Yonsei University, Seoul 03722, Republic of Korea}

\title{The SAMI Galaxy Survey: The internal orbital structure and mass distribution of passive galaxies from triaxial orbit-superposition Schwarzschild models}

\begin{abstract}
Dynamical models are crucial for uncovering the internal dynamics of galaxies, however, most of the results to date assume axisymmetry, which is not representative for a significant fraction of massive galaxies. Here, we build triaxial Schwarschild orbit-superposition models of galaxies taken from the SAMI Galaxy Survey, in order to reconstruct their inner orbital structure and mass distribution. The sample consists of 161 passive galaxies with total stellar  masses in the range $10^{9.5}$ to $10^{12} M_{\odot}$. We find that the changes in internal structures within 1$R_{\rm e}$ are correlated with the total stellar mass of the individual galaxies. The majority of the galaxies in the sample ($73\% \pm 3\%$) are oblate, while $19\% \pm 3\%$ are mildly triaxial and $8\% \pm 2\%$ have triaxial/prolate shape. Galaxies with $\log M_{\star}/M_{\odot} > 10.50$ are more likely to be non-oblate. We find a mean dark matter fraction of $f_{\rm{DM}} = 0.28 \pm 0.20$, within 1$R_{\rm e}$. Galaxies with higher intrinsic ellipticity (flatter) are found to have more negative velocity anisotropy $\beta_r$ (tangential anisotropy). $\beta_r$ also shows an anti-correlation with the edge-on spin parameter \lam, so that $\beta_r$ decreases with increasing \lam, reflecting the contribution from disk-like orbits in flat, fast-rotating galaxies. We see evidence of an increasing fraction of hot orbits with increasing stellar mass, while warm and cold orbits show a decreasing trend. We also find that galaxies with different ($V/\sigma$ - $h_3$) kinematic signatures have distinct combinations of orbits. These results are in agreement with a formation scenario in which slow- and fast-rotating galaxies form through two main channels.\\
\end{abstract}

\section{Introduction}
The assembly history of a galaxy is thought to be one of the major factors that determines its internal kinematic structure \citep[e.g.,][]{White1979, Fall1980, Park2019} and so observations of the internal kinematic structure should give an indication of a galaxy's past. 

Our current understanding of galaxy formation suggests that massive galaxies form in a two-phase process \citep[e.g.,][]{Naab2009, Oser2010}. During the first phase, at high redshift, they grow by a rapid episode of in-situ star formation, resulting in compact massive systems.
After $z \approx 2$, these massive, $\log_{10} (M_{\star}/ M_{\odot}) > 10.5$, compact galaxies are predicted to be quiescent and grow mostly by accreting mass through gas-poor galaxy mergers that add stars mainly to their outskirts.

Early-type galaxies (ETGs) have been separated into two classes, based on their stellar kinematics: fast rotators and slow rotators \citep[e.g.][]{Emsellem2004,Emsellem2007,Cappellari2007,Emsellem2011}. \cite{Cappellari2016} suggested that these two classes also indicate two major channels
of galaxy formation where fast-rotating ETGs start their life as star-forming disks and evolve through a set of processes dominated by gas accretion,
bulge growth and quenching. In contrast, slow-rotating ETGs
assemble near the centers of massive halos, via intense star formation at high redshift, and evolve from a set of processes dominated
by gas-poor mergers. However, \cite{Naab2014} showed that the detailed
formation history of a galaxy cannot be constrained from the slow-fast rotator classification alone, but when combined with the higher-order kinematic signatures, different merger scenarios can be distinguished.

In order to understand the evolutionary history of galaxies, we need a detailed analysis of its intrinsic structure. The Schwarzschild orbit-superposition method \citep{Schwarzschild1979} is a powerful dynamical modelling technique that allows dynamical substructures in galaxies to be revealed. Several different implementations of the Schwarzschild method, with varying degrees of symmetry, have been described \citep[e.g.][]{Cretton1999,Gebhardt2003, Valluri2004,vandenBosch2008, Vasiliev2015, Vasiliev2020, Neureiter2021}. The Schwarschild method has been used to model supermassive black holes \citep{vanderMarel1998, Verolme2002,Gebhardt2003,Valluri2004,Krajnovic2009, Rusli2013,Seth2014,Thater2017,Thater2019,Liepold2020, Quenneville2021}, the internal orbital structures of globular clusters \citep{vandeVen2006,Feldmeier2017}, early-type galaxies \citep{Cappellari2006,Thomas2007,vandeVen2008, Thomas2014, Fahrion2019, Poci2019, Jin2020, denBrok2021, Thater2021} and recently expanded to galaxies of all morphologies \citep{Vasiliev2015,Zhu2018a,Zhu2018b, Vasiliev2020, Lipka2021}. The orbit distributions obtained by these models have also been used to identify different dynamical components in these stellar systems \citep[e.g.][]{vandeVen2006, Cappellari2007, vandenBosch2008, Lyubenova2013, Breddels2014,Krajnovic2015}. \cite{Zhu2018a} separated orbits into four different components: a cold component with near circular orbits (with strong rotation), a hot component with near radial orbits (characterized by random motions), a warm component in-between (characterized by weak rotation) and a counter-rotating component (similar to the warm and cold components, but with reversed angular momentum). The inferred internal orbital distributions were then used to reconstruct the observed photometry and stellar kinematics of each component. However, the majority of these studies only had a few objects available (less than 30 galaxies). A large sample of galaxies, observed with good radial coverage and spatial resolution, is required in order to understand the average evolution history of the general galaxy population.

In the last two decades, Integral Field Spectroscopy (IFS) surveys such as SAURON (Spectroscopic Areal Unit for Research on Optical Nebulae; \citealt{deZeeuw2002}), ATLAS\(^{3D}\) \citep{Cappellari2011}, CALIFA (Calar Alto Legacy Integral Field Array survey; \citealt{Sanchez2012}), SAMI (Sydney-Australian-Astronomical-Observatory Multi-object Integral-Field Spectrograph) Galaxy Survey \citep{Croom2012, Bryant2015,Croom2021}, MASSIVE \citep{Ma2014}, MaNGA (Mapping Nearby Galaxies at Apache Point Observatory; \citealt{Bundy2015}) and the  Fornax 3D survey \citep{Sarzi2018} have provided us with rich observational datasets of galaxies, allowing their structure and evolution to be investigated in detail through the mapping of stellar kinematics across individual galaxies. These IFS surveys have made possible the use of techniques such as Schwarschild orbit-superposition method to dynamically decompose IFS observations to estimate the internal mass distribution, intrinsic stellar shapes and orbit distributions of galaxies across the Hubble sequence \citep[e.g.,][] {Zhu2018c, Zhu2018a, Zhu2018b, Zhuang2019,Jin2020, Aquino-Ortiz2020}.

\cite{Zhu2018b} studied a sample of 250 galaxies in the CALIFA survey, with total stellar masses between $10^{8.5}$ and $10^{12}  M_{\odot}$, spanning all morphological types. About 95\% of the galaxies in their sample had stellar kinematic maps with $R_{\rm max} > 1 R_{\rm e}$, and $\sim$ 8\% with $R_{\rm max} > 3 R_{\rm e}$. They found that, within 1 $R_{\rm e}$, galaxies have more stars in warm orbits than in either cold or hot orbits. 
Similar results were also found in a sample of 149 early-type galaxies in the MaNGA survey \citep{Jin2020}, with stellar masses ranging between $10^{9.9}$ and $10^{11.8}  M_{\odot}$ and observations up to 1.5 - 2.5 $R_{\rm e}$ per galaxy. These studies also found that the changes of internal structures within 1$R_{\rm e}$ are correlated with the stellar mass of the galaxies.

The number of galaxies considered for \schw model studies to date has been limited and they have often not incorporated higher-order kinematic moments to further constrain the orbital models. Higher-order kinematic signatures are defined as the deviations from a Gaussian line-of-sight velocity distribution (LOSVD).
When the LOSVD is parametrized as a Gauss--Hermite series \citep{vanderMarel1993, Gerhard1993}, its skewness and excess kurtosis are parametrized by the coefficients of the 3\textsuperscript{rd}- and 4\textsuperscript{th}-order Hermite polynomials ($h_3$ and $h_4$, respectively). Given the connection between the higher-order stellar kinematic moments and a galaxy's assembly history \citep{Naab2014}, their inclusion in dynamical modelling can help distinguish between different formation scenarios.

In this paper we will apply \schw modelling to the SAMI Galaxy Survey \citep{Croom2012, Bryant2015, Owers2017} to investigate the evolutionary histories of passive galaxies by studying their internal structures. The SAMI Galaxy Survey data allows us to study the internal orbits of a significant number of galaxies for the first time and allows us to further constrain the \schw models by adding information on the higher-order kinematic moments.

Throughout the paper, we adopt a $\Lambda CDM$ cosmology with $\Omega_m=0.3$, $\Omega_{\Lambda} = 0.7$, and $H_0 = 70$ km s$^{-1}$ Mpc$^{-1}$.

\section{Observations} \label{sec:style}
The Sydney-AAO Multi-object Integral field spectrograph (SAMI) Galaxy Survey is a large, optical 
Integral Field Spectroscopic \citep{Croom2012,Bryant2015,Owers2017} survey of low-redshift 
($0.04 < z < 0.095$) galaxies covering a broad range in stellar mass,  $7 < \log_{10} (M_{\star}/ M_{\odot}) < 12$, 
morphology and environment. The sample, with $\approx$ 3000 galaxies, is selected from the Galaxy and Mass Assembly survey (GAMA; \citealt{Driver2011}) regions (field and group galaxies), as well as eight additional clusters to probe higher-density environments \citep{Owers2017}.

The SAMI instrument \citep{Croom2012}, on the 3.9m Anglo-Australian telescope, consists of 13 ``hexabundles" 
\citep{Bland-Hawthorn2011, Bryant2014}, across a 1-degree field of view. Each hexabundle consists of 61 individual 1.\textsuperscript{$\prime \prime$}6 fibres, and covers a $\sim$ 15\textsuperscript{$\prime \prime$} diameter region on the sky. In the typical configuration, 12 hexabundles are used to observe 12 science targets, with the 13\textsuperscript{th} one allocated to a secondary standard star used for calibration. Moreover, SAMI also has 26 individual sky fibres, to enable accurate sky subtraction for all observations without the need to observe separate blank sky frames. The SAMI fibres are fed to the dual-beam AAOmega spectrograph \citep{Sharp2006}.

\subsection{IFS Spectra and kinematic maps}
SAMI data consist of 3D data cubes: two spatial dimensions and a third spectral dimension.

The wavelength coverage is from 3750 to 5750 \AA\ in the blue arm, and from 6300 to 7400 \AA\ 
in the red arm, with a spectral resolution of R = 1812 (2.65 \AA\ full-width half maximum; FWHM) and R = 4263 (1.61 \AA ~FWHM), respectively \citep{vandeSande2017a}, so that two data cubes are produced for each galaxy target. 

Each galaxy field was observed in a set of approximately seven 30 minute exposures, that are aligned together by fitting the galaxy position within each hexabundle with a two-dimensional Gaussian and by fitting a simple empirical model describing the telescope offset and atmospheric
refraction to the centroids. The exposures are then combined to produce a spectral cube with regular $0.5^{\prime\prime}$ spaxels, with a median seeing of $2.1^{\prime\prime}$.
More details of the Data Release 3 reduction can be found in \cite{Croom2021}\footnote{Reduced data-cubes and stellar kinematic data products for
all galaxies are available on: \href{https://datacentral.org.au}{https://datacentral.org.au}.}. 

Stellar kinematic measurements were derived using the penalized pixel fitting code (pPXF; \citealt{Cappellari2004, Cappellari2017}), after combining the blue- and red-arm spectra by matching their spectral resolution. A detailed description of the method used to derive the stellar kinematic measurement can be found in \cite{vandeSande2017a, vandeSande2017b}. In particular, for our analysis, we use the Voronoi-binned kinematic measurements. Bins are adaptively generated to contain a target S/N of 10 \AA$^{-1}$, using the Voronoi binning code of \cite{Cappellari2003}. 

The available stellar kinematic measurements consist of 2D maps of stellar rotational velocity $V$, velocity dispersion $\sigma$, and the high kinematic orders ($h_3$ and $h_4$). In addition, each kinematic map has kinematic position angle and FWHM of the Point Spread Function (PSF - taken from a star observed at the same time as the galaxies) provided. 

\subsection{Multi Gaussian Expansion profiles and effective radius}
Multi-Gaussian Expansion (MGE; \citealt{Emsellem1994, Cappellari2002}) profile fits for the SAMI Galaxy Survey are produced from the $r-$band photometry by \cite{Deugenio2021}.
The MGE method consists of a series expansion of galaxy images using 2D Gaussian functions. This method enables us to take the PSF into account; given a value of the inclination and assuming an intrinsic shape, the MGE model can be deprojected \emph{analytically,} which is orders of magnitude faster than the general, integral-based method.

The fits are applied to re-analysed Sloan Digital Sky Survey (SDSS; \citealt{York2000}) images for GAMA galaxies, reprocessed as described in \cite{Hill2011}, as well as VST/ATLAS (VLT Survey Telescope - ATLAS; \citealt{Shanks2015}) and SDSS DR9 \citep{Ahn2012} observations for cluster galaxies, with VST/ATLAS data reprocessed as described in \cite{Owers2017}. The images are square cutouts with $400^{\prime\prime}$ side, centerd on the center of the galaxy, and the MGE fits are calculated using \texttt{MgeFit} \citep{Cappellari2002} and the regularisation feature described in \cite{Scott2009}. For a given galaxy, each Gaussian component has its PA fixed to that of the host's major axis. As such, the stellar mass distribution is assumed to be axisymmetric in projection, but can be intrinsically triaxial. A more detailed description of the fitting process can be found in \cite{Deugenio2021}. From the MGE best fit, we use the projected luminosity, size, and flattening of each Gaussian component to model the surface density of each galaxy and to deproject the stellar component into a 3D density. The effective radius, $R_{\rm e}$, used here is that of the major axis in the $r$-band. The semi-major axis values were taken from MGE fits.

\subsection{Stellar Mass} \label{sec:stellarmass}
Stellar masses are estimated assuming a \cite{Chabrier2003} initial mass function (IMF), from the K-corrected $g-$ and the $i-$ magnitudes using an empirical proxy developed from GAMA photometry \citep{Taylor2011, Bryant2015}. For cluster galaxies, stellar masses are derived using the same approach \citep{Owers2017}. We use the photometric stellar masses for our analysis in order to be consistent with previous SAMI studies and to have consistent comparisons with previous results in the literature (e.g. from CALIFA and MaNGA).

\subsection{Sample Selection}
We use data from the final SAMI data release (described in the Data Release 3 publication \citealt{Croom2021}). This data release consists of 3068 unique galaxies. Of these, we have MGE profiles from \cite{Deugenio2021} for 2957 galaxies ($r-$band images are not available for some galaxies or they have been affected by a bright star in the field of view). Following \cite{vandeSande2017a}, we exclude all galaxies whose kinematics are influenced by mergers, that have strong bars or that have a bright secondary object within one effective radius in their stellar velocity field. This leaves us with 2834 galaxies with stellar kinematic and MGE measurements.

We exclude all galaxies with masses below $\log_{10} (M_{\star}/ M_{\odot}) = 9.5$, because the incompleteness of the stellar kinematic sample is larger than 50\% of the SAMI galaxy survey sample observed in this mass range. We further exclude 433 galaxies where $R_{\rm e} < 2^{\prime\prime}$ (due to their spatial size being smaller than the instrumental spatial resolution). This leaves us with 1649 galaxies.

\begin{figure}[!ht]
\centering
\includegraphics[scale=0.43]{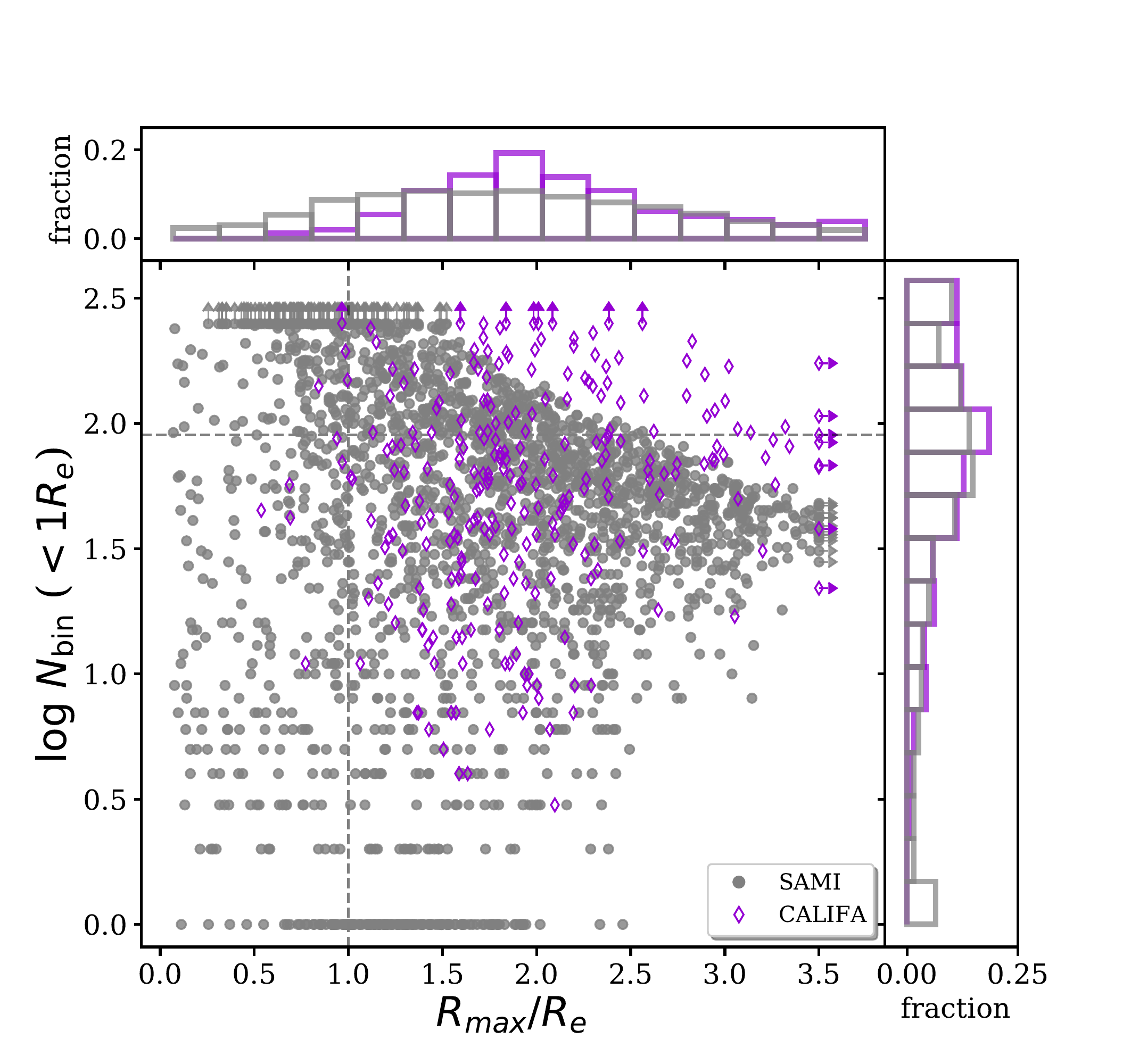}
\caption{Number of Voronoi bins within 1$R_{\rm e}$ that meet our quality criteria versus the maximum radius available for stellar kinematics (in units of $R_{\rm e}$) for the galaxies in the SAMI Galaxy Survey (1589 galaxies; grey circles) and in the CALIFA survey (259 galaxies; violet diamonds). Black dashed lines indicate $R_{\rm max}/R_{\rm e} = 1$ and Voronoi bins $= 85$.  We calculate the marginalized fractions of galaxies to the total number in each sample, by mass and size, and show them in the top and left panels of the figure. Grey lines are for SAMI galaxies, while the violet lines are for CALIFA galaxies. The CALIFA and the SAMI samples have similar distributions in Voronoi bins and radial coverage, although there are more CALIFA galaxies with measurements up to 2$R_{\rm e}$.For this analysis we select galaxies in the top right corner ($R_{\rm max}> 1R_{\rm e}$ and Voronoi bins $> 85$).}
\label{fig:califa_sami}
\end{figure}

Following the recommendations of \cite{vandeSande2017a}, for each galaxy we select spaxels that meet the following quality criteria:
\begin{itemize}
\item[$Q_1$)] $S/N > 3$ \AA$^{-1}$ \& $\sigma_{obs}> 35 $ km/s;
\item[$Q_2$)] $V_{ERR} < 30$ km/s \& $\sigma _{ERR} < \sigma_{obs} \times 0.1 + 25$ km/s.

\end{itemize}
$Q_3$ in \cite{vandeSande2017a} is for measurements with $S/N < 20$ \AA$^{-1}$ and $\sigma_{obs}<70 $ km/s. We cautiously include these in this analysis and increase the errors on the measurements that do not meet this criterion to down-weight their contributions. The 1589 galaxies that meet these criteria are shown in Fig. \ref{fig:califa_sami}. 

In this paper we focus on passive galaxies, because the long-term goal of this project is to study the effects of galaxy environment on passive galaxies (Santucci et al. in prep). We use the SAMI spectroscopic classification presented in \cite{Owers2019} to select a homogeneous sample. The SAMI spectroscopic classification labelled galaxies as star-forming, passive, or H$\delta$-strong, using the absorption- and emission-line properties of each SAMI spectrum. We select 738 passive galaxies.

\subsubsection{Radial coverage and spatial sampling selection}
We compare the spatial resolution and radial extent of our sample to the sample from \citep{Zhu2018a} who used CALIFA data to derive orbital parameters using the Schwarzschild method. SAMI Voronoi bins are generated to contain a target $S/N$ of 10 \AA$^{-1}$. Since the target $S/N$ is the only requirement for the bins, individual spaxels of 0.5$^{\prime\prime}$ are left unbinned when they meet this requirement. For these single-spaxel bins, the covariance is larger (since they are smaller than the SAMI spatial resolution). In Fig. \ref{fig:califa_sami} we show the number of Voronoi bins within 1$R_{\rm e}$ versus the radial coverage avalable (in units of $R_{\rm e}$) for the 1589 SAMI galaxies (in grey) that meet our quality criteria. CALIFA galaxies (in purple; from \citealt{Zhu2018a}) have a similar distribution in number of bins to SAMI, however their bins were generated with different criteria (their minimum $S/N=20$ and their spaxel size is consistent with their spatial resolution), therefore a direct comparison is not possible. CALIFA and SAMI also show a similar distribution in radial coverage, although there are more CALIFA galaxies with measurements up to 2$R_{\rm e}$.

In this analysis, the first in a series, we select a high-quality subsample of SAMI galaxies, identified by good spatial resolution and good radial coverage (top right corner of Fig. \ref{fig:califa_sami}). This region is selected as the optimal compromise between best quality data and reasonable sample size, and corresponds to galaxies with 85 Voronoi bins within 1$R_{\rm e}$ and $R_{\rm max}> R_{\rm e}$. More details about the radial coverage tests we performed can be found in Appendix \ref{sec:radial_test}. 

\begin{figure}[!ht]
\centering
\includegraphics[scale=0.45,trim=0.5cm 0cm 0cm 1cm,clip=true]{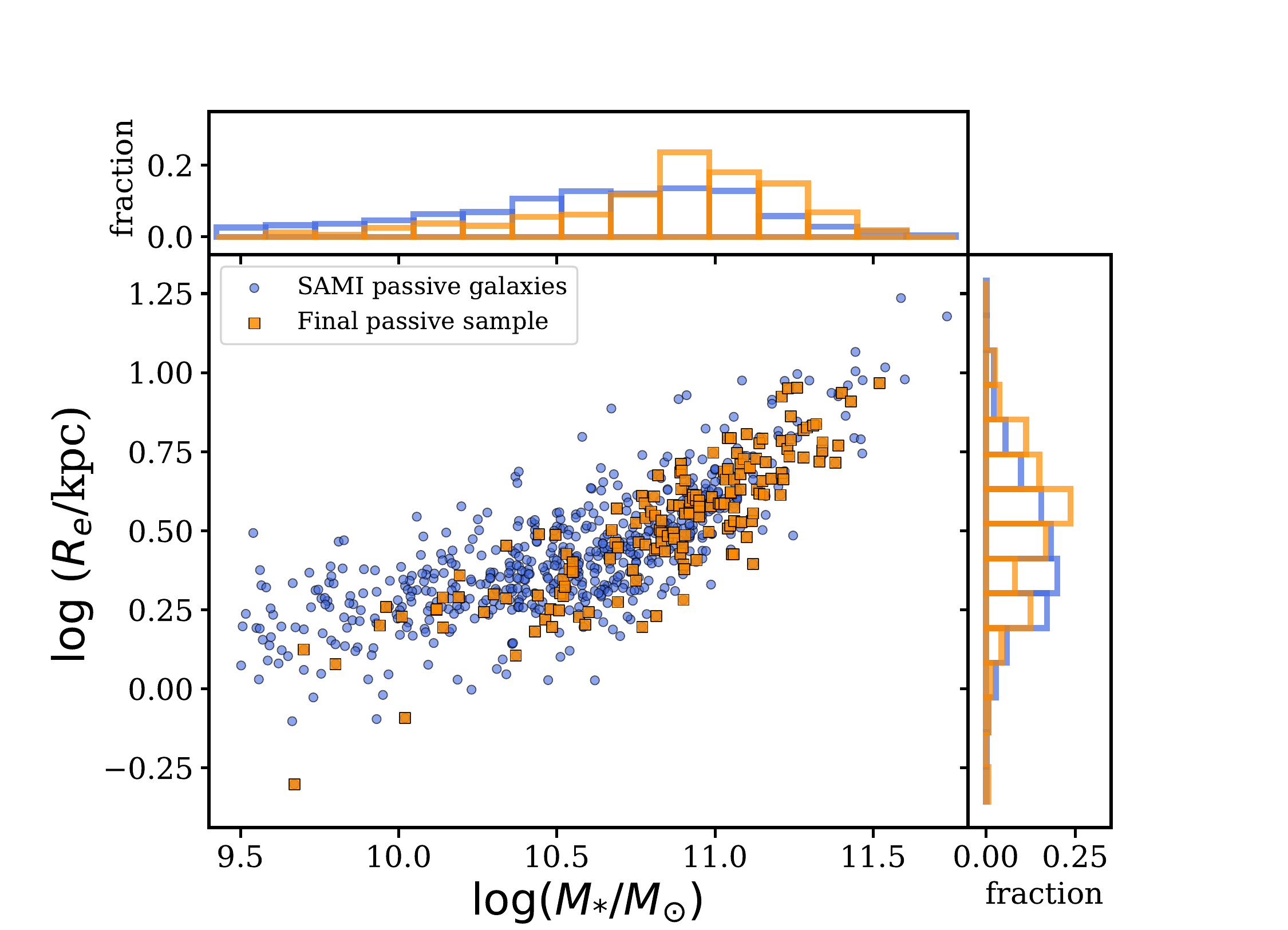}
\caption{Effective radius, $R_{\rm e}$, versus stellar mass. Blue circles are the passive galaxies in the SAMI sample with $\log_{10} (M_{\star}/ M_{\odot}) > 9.5$ and $R_{\rm e} > 2^{\prime\prime}$  (738), orange squares are the galaxies included in the final sample (161). We calculate the marginalized fractions of galaxies with respect to the total number in each sample, by mass and size, and show them in the top and left panels of the figure. Blue lines are for the passive galaxies in the SAMI Galaxy Survey, while the orange lines are for our final sample. The two samples are slightly different in the marginalized mass and size distributions, so that we have higher fractions of massive and large galaxies in the final sample compared to the initial sample. This is due to selecting galaxies with more than 85 Voronoi bins. }
\label{fig:sample_sel}
\end{figure}

This quality cut gives us a sample of 179 passive galaxies. We visually inspect the galaxies in this sample using HSC images and exclude the face-on strongly barred galaxies that were not identified as barred from the square cutouts used for the MGE modelling. This cut gives us a final sample of 161 galaxies. These are shown in Fig. \ref{fig:sample_sel} and used hereafter in this analysis. The majority of the galaxies in our sample are early-type galaxies ($\sim 85\%$), $\sim 11\%$ are S0/Early-spirals and $\sim 4\%$ are late-type galaxies (visual morphological classification from \citealt{Cortese2016}). We note that our final sample is biased toward galaxies that are more massive and larger than the general SAMI passive population. This bias is caused by selecting galaxies with at least 85 Voronoi bins within 1 $R_e$.

\section{Schwarzschild orbit-superposition technique}
\subsection{Schwarzschild's models and free parameters}
We use the Schwarzschild orbit-superposition technique \citep{Schwarzschild1979} to model our individual galaxies, using the implementation from \cite{vandenBosch2008} with correct orbital mirroring from \cite{Quenneville2021b}. This code allows us to model triaxial stellar systems \footnote{A new implementation of this code, DYNAMITE (DYnamics, Age and Metallicity Indicators Tracing Evolution), has recently been released \citep{DYN2020}. This was not available at the beginning of this analysis. Internal tests have been carried out which have verified the consistency between the two implementations.}, while many previous applications of this technique assume axisymmetry.
There are three main steps required to create a \schw model:
\begin{enumerate}
\item Construct a model for the underlying gravitational potential;
\item Calculate a representative library of orbits using the gravitational potential previously modelled;
\item Find a combination of orbits that can reproduce the observed kinematic maps and luminosity distribution.
\end{enumerate}
These steps are fully described in \cite{vandenBosch2008} and \cite{Zhu2018c} and are summarized in the following subsections.

\subsection{Gravitational Potential} \label{sec:param}

The model gravitational potential of each galaxy is generated using the combination of three components: a stellar and a dark matter distribution and a central super-massive black hole. The triaxial stellar component mass is calculated from the best-fit two-dimensional MGE luminosity density \citep[from][]{Deugenio2021} which is de-projected assuming the orientation in space of the galaxy, described by three viewing angles ($\theta$, $\phi$, $\psi$), to obtain a three-dimensional luminosity density. The space orientation  ($\theta$, $\phi$, $\psi$) can be converted directly to the intrinsic shape ($p_i$, $q_i$, $u_i$), where $p_i = B_i/A_i$, $q_i = C_i/A_i$ and $u_i= \sigma_{{\rm Gauss},i}^{\rm obs}/\sigma_{{\rm Gauss},i}$. $A_i$, $B_i$, $C_i$ represent the major, medium and minor axes of the 3D triaxial Gaussian component and $\sigma_{{\rm Gauss},i}$ represents the size of each Gaussian component. Moreover, the flattest Gaussian component, having the minimum {\em observed} flattening \(q_{\rm min}^\prime\), dictates the allowed space orientation for the de-projection, so that we can take ($p_{\rm min}$, $q_{\rm min}$, $u_{\rm min}$) as our free parameters.
The 3D density defined by these intrinsic shapes is then converted into a stellar mass distribution using a radially constant stellar mass-to-light ratio $M_{\star}/L$ (note that $M_{\star}/L$ is a free parameter in our modelling). The corresponding stellar gravitational potential $\Phi_{\star}$ is calculated using the classical formula from \cite{Chandrasekhar1969}.

The dark matter halo distribution is assumed to follow a spherical Navarro-Frenk-White profile (NFW; \citealt{Navarro1996}). The mass, $M_{200}$ (mass enclosed within a radius, $R_{200}$, where the average density is 200 times the critical density), in a NFW dark matter halo is determined by two parameters. These are the concentration parameter, $c$, and the fraction of dark matter within $R_{200}$, $f = M_{200}/ M_{\star}$ (where $M_{200}$ is as defined above and $M_{\star}$ is the total stellar mass). 

The spatial resolution of SAMI data is poorer than the influence radius of the black hole, so its mass leaves no imprint on the stellar kinematic maps and therefore does not affect our results. We therefore fix the black hole mass to the value derived from the stellar velocity dispersion, measured within an aperture of 1$R_{\rm e}$, assuming the relation between black hole mass and the stellar velocity dispersion of a galaxy from \cite{McConnell2011}.

Combining the components used to describe the gravitational potential, we have six free parameters (stellar mass-to-light ratio, $M_{\star}/L$, the intrinsic shape of the flattest Gaussian component ($p_{\rm min}$, $q_{\rm min}$, $u_{\rm min}$), the dark matter halo concentration, $c$, and dark matter fraction, $f$) that must be determined. To determine these best-fit parameters for each galaxy, we run an optimized grid-based parameter
search as described in \cite{Zhu2018c} and summarized in Sec. \ref{sec:bfmodel}.

\subsection{Orbit library}
To fit a model to our observed data we need an orbit library. To create the orbit library we use a separable triaxial potential, where all orbits are regular and conserve three integrals of motion (energy $E$, second integral $I_2$ and third integral $I_3$) which can be calculated analytically. Our Schwarzschild implementation considers four different families of orbits: three types of tube orbits (short axis tubes, outer and inner long axis tubes) and box orbits. We create initial conditions for our orbits by sampling from the three integrals of motion. We refer to \cite{vandenBosch2008} for the details of the orbit sampling. 

The number of points we sample across the three integrals is $n_E \times n_{\theta} \times n_R = 21 \times 10 \times 7$, where $n_E$, $n_{\theta}$, $n_R$ are the number of intervals taken across the energy $E$, the azimuthal angle $\theta$ and radius $R$ on the $(x, z)$ plane. However, this orbit library includes mostly short axis tubes, long axis tubes and a relatively low fraction of box orbits in the inner region. Since box orbits are essential for creating triaxial shapes, we construct an additional set of box orbits.
Box orbits always touch equipotentials \citep{Schwarzschild1979}, so they can be described by combining the energy $E$ with two spherical angles ($\theta$ and $\phi$). The number of points included in the box orbit set are $n_E \times n_{\theta} \times n_{\phi} = 21 \times 10 \times 7$.

We add an additional set of orbits to account for retrograde stars commonly found in early-type galaxies \citep{Bender1988, Kuijken1996}. This set contains $21 \times 10 \times 7$ orbits to describe the initial conditions for counter-rotating orbits.
To summarize, we use three sets of $21 \times 10 \times 7$ orbits: a typical set of ($E$, $I_2$, $I_3$), a box orbits set of ($E$, $\theta$, $\phi$) and a counter-rotating set of also ($E$, $- I_2$, $I_3$).

As in \cite{vandenBosch2008} and \cite{Zhu2018a}, we dither every orbit to give $5^3$ orbits by perturbing the initial conditions slightly, in order to smooth the model. The orbit trajectories created by the dithering will be co-added to form a single orbit bundle in our orbit library.

We then use Schwarzschild’s method to weight the various orbit contributions to the LOSVD in each bin to construct a model with observational parameters that can be fit to the data \citep[the description of how kinematic maps are fitted can be found in][]{Zhu2018c}. The quantities that will be compared to observations are spatially convolved with the same PSF as the observations. The model and the observed values are then divided by the observational error so that a $\chi^2$
comparison is achieved. The weights are determined by the \cite{vandenBosch2008} implementation, using the \cite{Lawson1974} non-negative least squares (NNLS) implementation. 

\subsection{Best-fit model}\label{sec:bfmodel}
In order to find the best-fit model, which contains six free parameters, we run a grid based parameter search. We use a parameter grid with intervals of 0.5, 0.1, 0.2, 0.05, 0.05 and 0.01 in $M_{\star}/L$, $\log(c)$, $\log(f)$, $q_{\rm min}$, $p_{\rm min}$ and $u_{\rm min}$, respectively, and perform an iterative search for the best-fitting models. After each iteration, the best-fit model is selected by using a $\chi^2$ comparison.
The best-fit model is defined as the model with minimum kinematic $\chi^2$:

\begin{multline}\label{eq:chi2}
\chi^2 = \sum\limits_{n=1}^{N_{kin}} \Bigg[ \Bigg.   \left(\frac{V_{mod}^n-V_{obs}^n}{V_{obserr}^n}\right)^2 + \left(\frac{\sigma_{mod}^n- \sigma_{obs}^n}{\sigma_{obserr}^n}\right)^2 + \\ 
\left(\frac{h_{3,\ mod}^n- h_{3,\ obs}^n}{h_{3,\ obserr}^n}\right)^2 + \left(\frac{h_{4,\ mod}^n- h_{4,\ obs}^n}{h_{4,\ obserr}^n}\right)^2 \Bigg. \Bigg]
\end{multline}

where $V_{mod}^n$, $\sigma_{mod}^n$, $h_{3,\ mod}^n$ and $h_{4,\ mod}^n$ are the model values for each bin $n$, $V_{obs}^n$, $\sigma_{obs}^n$,  $h_{3,\ obs}^n$ and $h_{4,\ obs}^n$ are the observed values in each bin and $V_{obserr}^n$, $\sigma_{obserr}^n$, $h_{3,\ obserr}^n$ and $h_{4,\ obserr}^n$ represent the observational errors. $N_{kin}$ is the number of bins in the kinematic maps.
We define a confidence level around that minimum value and select all the models whose $\chi^2$ is within that confidence level:
$\chi^2 - \chi_{min}^2 < \chi_s^2 \times \sqrt{(N_{obs}-N_{par})}$, with $\chi_s^2 = 2$, $N_{obs} = 4N_{kin}$, as we use $V,\ \sigma, \ h_3$ and $h_4$ as model constraints, and $N_{par}$ is the number of free parameters (6 here). We then create new models around the existing models with lower kinematic $\chi^2$ values by walking two steps in every direction of the parameter grid from each of the selected models. In this way, the searching process goes in the direction of smaller $\chi^2$ on the parameter grid, and it stops when the minimum $\chi^2$ model is found. Next, we continue the iteration by using a larger value of $\chi_s^2$, to ensure all the models within 1$\sigma$ confidence are calculated before the iteration finishes. The values of $\chi_s^2$ are chosen empirically so that it is neither too small (finding only local minimums) nor too large. For the final step, we reduce the parameter intervals by half to get a better estimate of the best-fit parameters. 
The models whose $\chi^2$ are within the confidence level are included for calculating the statistical uncertainties of the model parameters for single data analysis. The maximum and minimum values of the parameters or properties in these models are treated as upper and lower limits in 1$\sigma$ error regions.

\begin{figure*}
\centering
\includegraphics[scale=0.5, trim=2.2cm 1.8cm 2.5cm 2cm,clip=true]{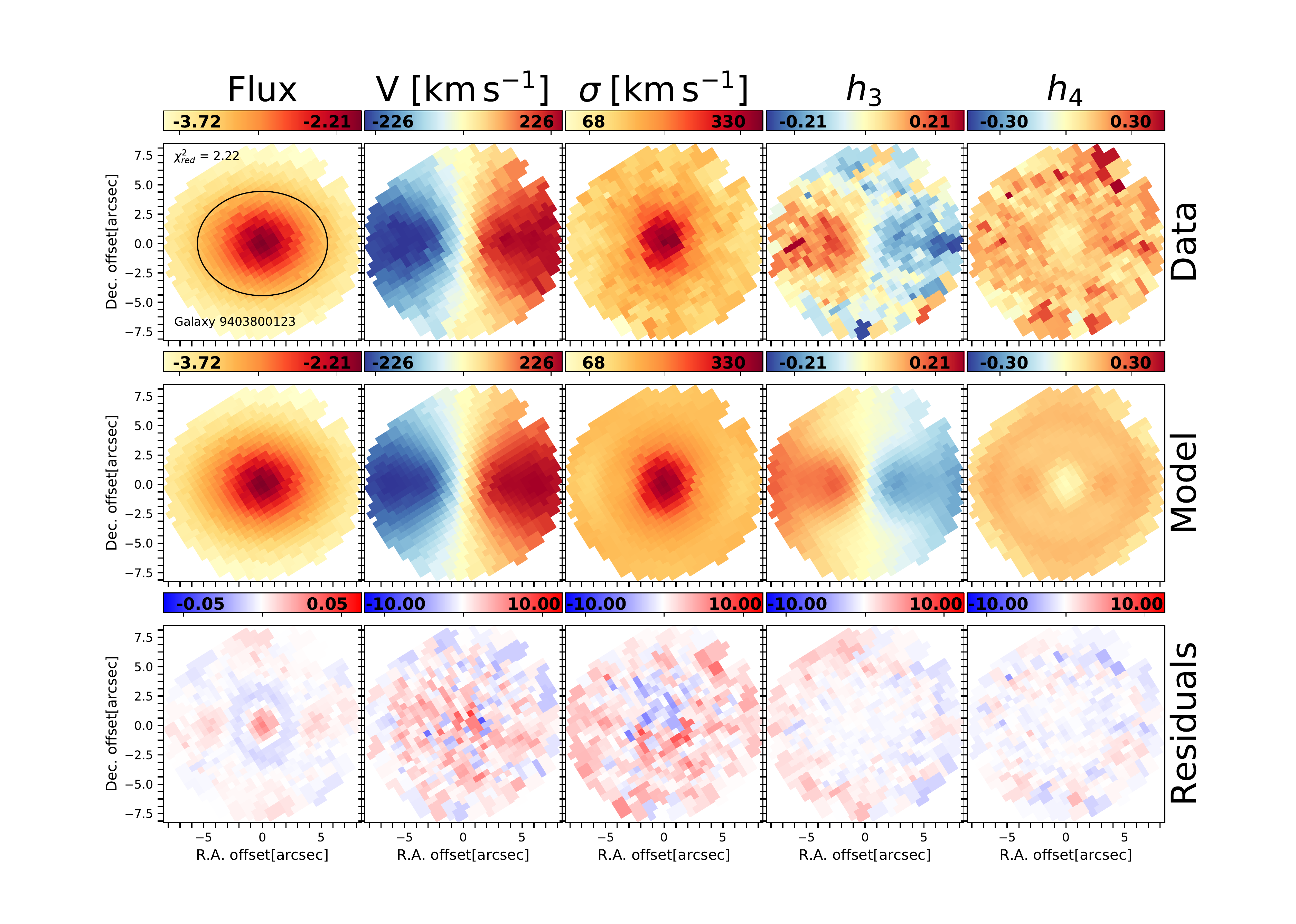}
\caption{Example of a galaxy with excellent spatial sampling: SAMI CATID 9403800123 in the cluster Abell 4038. This galaxy ($\log M_{\star}/M_{\odot} = 11.05$ and $R_{\rm e} = 5.52^{\prime\prime}$) is a non edge-on oblate galaxy and has stellar kinematic measurements up to 1.36 $R_{\rm e}$ and counts 255 spatial bins within 1$R_{\rm e}$ (black ellipse). Columns show 2D maps for, from left to right, flux, velocity, velocity dispersion, $h_3$ and $h_4$. First row shows the observed maps, second row shows the best-fit maps derived from the \schw modelling and the third row shows the residuals, calculated as the difference between the observation and the model, divided by the observational uncertainties. The best-fit model maps ($\chi^2_{red} = 2.22$) accurately reconstruct the structures seen in the observations, not only for the velocity and velocity dispersion maps, but also for $h_3$ and $h_4$. }
\label{fig:123_bestfit}
\end{figure*}

\begin{figure*}
\centering
\includegraphics[scale=0.5, trim=2.2cm 1.8cm 2.5cm 2cm,clip=true]{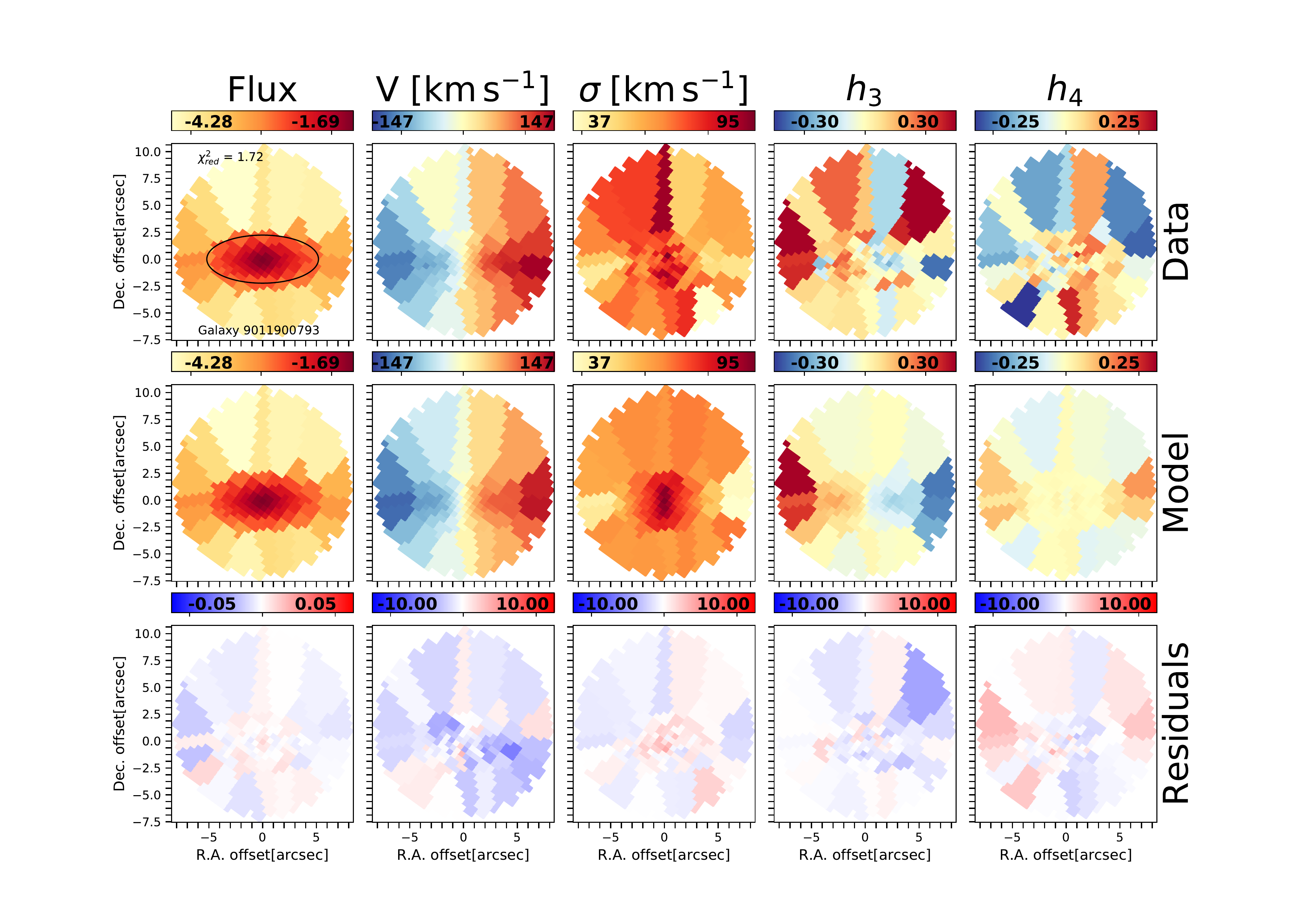}
\caption{Example of a galaxy near the minimum requirement of 85 spatial bins: SAMI CATID 9011900793 in the cluster Abell 119. This galaxy ($\log M_{\star}/M_{\odot} = 10.34$ and $R_{\rm e} = 5.19^{\prime\prime}$) is a edge-on oblate galaxy and has stellar kinematic measurements up to 1.45 $R_{\rm e}$ and 87 spatial bins within 1$R_{\rm e}$. Panels are as in Fig. \ref{fig:123_bestfit}. The best-fit model maps ($\chi^2_{red} = 1.72)$ accurately reconstruct the structures seen in the observations.}
\label{fig:793_bestfit}
\end{figure*}

\begin{figure*}
\centering
\includegraphics[scale=0.5, trim=2.2cm 1.8cm 2.5cm 2cm,clip=true]{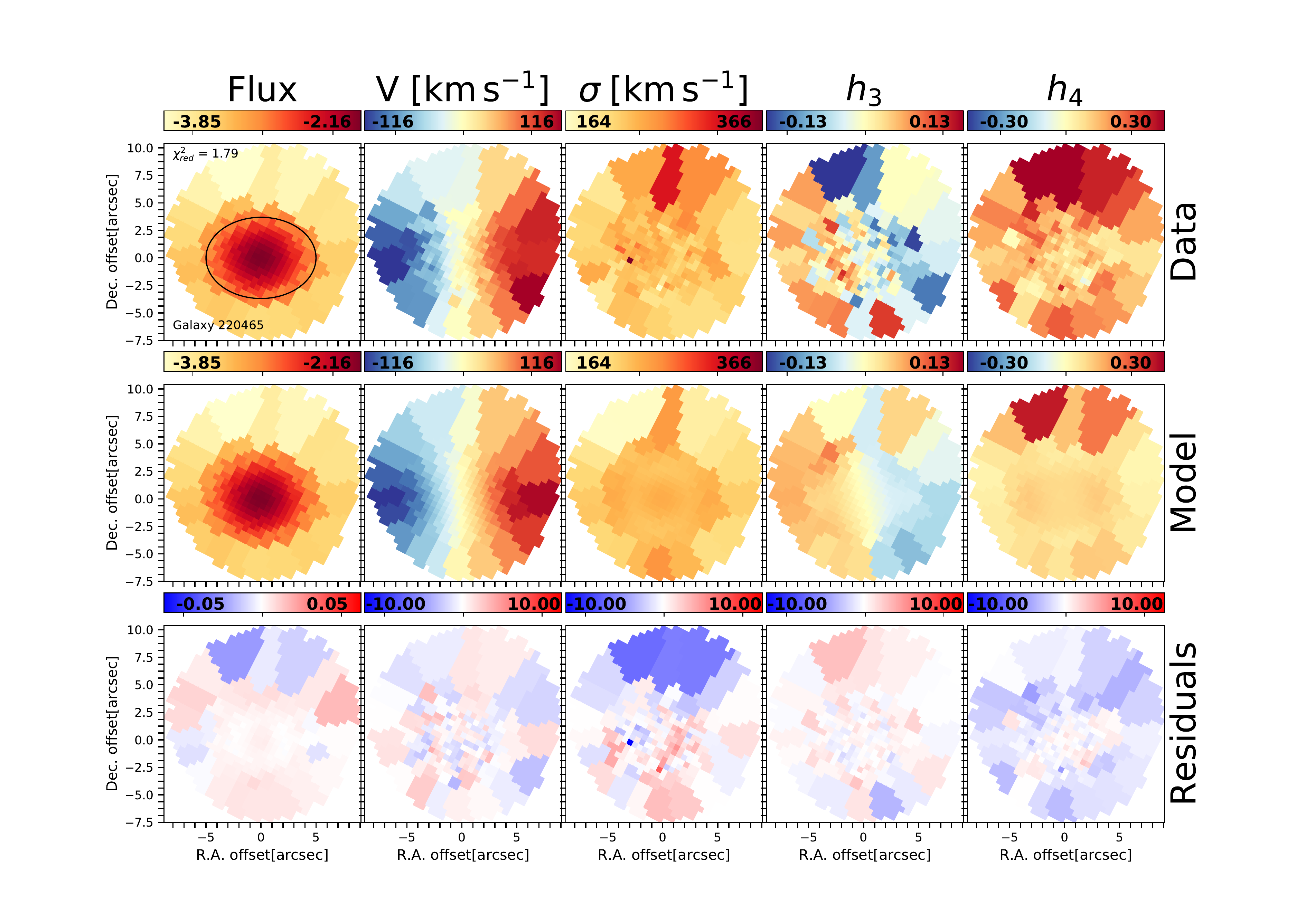}
\caption{Example galaxy SAMI CATID 220465, in the GAMA region. This galaxy ($\log M_{\star}/M_{\odot} = 11.31$ and $R_{\rm e} = 5.00^{\prime\prime}$) is a triaxial galaxy and has stellar kinematic measurements up to 1.5 $R_{\rm e}$ and 142 spatial bins within 1$R_{\rm e}$. Panels are as in Fig. \ref{fig:123_bestfit}. The best-fit model maps ($\chi^2_{red} = 1.79 )$ accurately reconstruct the structures seen in the observations, not only for the velocity and velocity dispersion maps, but also for $h_3$ and $h_4$. }
\label{fig:465_bestfit}
\end{figure*}

\begin{figure*}
\centering
\includegraphics[scale=0.5, trim=2.2cm 1.8cm 2.5cm 2cm,clip=true]{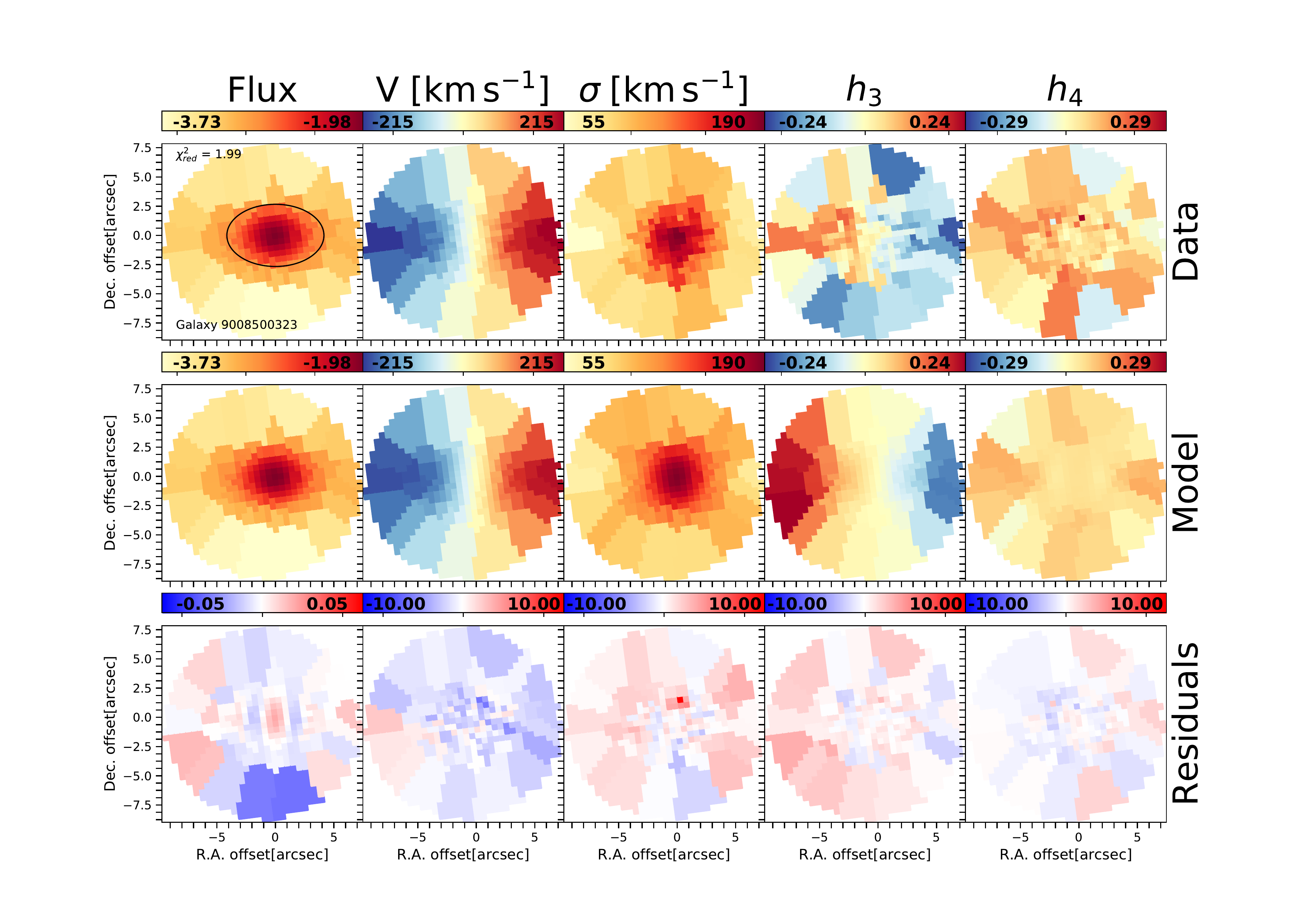}
\caption{Example galaxy SAMI CATID 9008500323, in the cluster Abell 85. This galaxy ($\log M_{\star}/M_{\odot} = 10.78$ and $R_{\rm e} = 4.15^{\prime\prime}$) is a prolate galaxy and has stellar kinematic measurements up to 1.81 $R_{\rm e}$ and 104 spatial bins within 1$R_{\rm e}$. Panels are as in Fig. \ref{fig:123_bestfit}. The best-fit model maps ($\chi^2_{red} = 1.99)$ accurately reconstruct the structures seen in the observations, not only for the velocity and velocity dispersion maps, but also for $h_3$ and $h_4$. }
\label{fig:323_bestfit}
\end{figure*}

The kinematic maps for the best-fit models of example galaxies
9403800123, 9011900793, 220465 and 9008500323 are presented in Fig. \ref{fig:123_bestfit}, Fig. \ref{fig:793_bestfit}, Fig. \ref{fig:465_bestfit} and Fig. \ref{fig:323_bestfit}. We selected these four galaxies as representative of the sample, with 9403800123 being a non edge-on oblate galaxy (with 255 Voronoi bins within 1$R_{\rm e}$), 9011900793 an edge-on oblate galaxy (with 87 Voronoi bins within 1$R_{\rm e}$), 220465 a triaxial galaxy (142 Voronoi bins within 1$R_{\rm e}$) and 9008500323 a prolate galaxy (with 104 Voronoi bins within 1$R_{\rm e}$). Even when the spatial sampling is low, as in the case of 9011900793, the model is able to reproduce the observed kinematic maps well ($\chi^2_{red} = 2.22$ for galaxy 9403800123, $\chi^2_{red} = 1.72$ for galaxy 9011900793, $\chi^2_{red} = 1.79$ for galaxy 220465 and $\chi^2_{red} = 1.99$ for galaxy 9008500323\footnote{The reduced $\chi^2$ is defined as $\chi^2_{red} = \frac{\chi^2}{4N_{kin}-N_{par}}$, with $\chi^2$ calculated following Eq. \ref{eq:chi2}. The values of $\chi^2_{red}$ are not always equal to 1 for the best-fit models of the galaxies in our sample. This is because the input kinematic maps of the galaxies in our sample were not symmetrized. Therefore, comparing the observed maps to the model maps, which are symmetric, can result in values of $\chi^2_{red}$ higher than 1.}).
We also show the explored parameter grids and the obtained internal mass distribution, orbit circularity, triaxiality and tangential anisotropy for the best fits of these four galaxies in Appendix \ref{sec:example_gals}. These parameters are fully described in the following section.


\section{Results}
In this section and the next we present the results we obtain modelling a sample of 161 passive galaxies in the SAMI Galaxy Survey with the \schw orbit-superposition technique. For each galaxy we explore a range in parameter space by building on average 1250 different models. This is consistent with previous analyses that used an iterative grid search in $\sim$6 dimensions. For example \cite{Jin2019,Jin2020} required 1000 to 2000 separate Schwarzschild models per galaxy to be run. By comparing the 2D maps of the flux and kinematic parameters derived from each model and observations we determine the best-fit parameters. From the best-fit model we derive the intrinsic properties of the inner mass distribution (for both stellar and dark matter components), intrinsic stellar shape (axis ratios and ellipticity), velocity anisotropy and the orbit circularity distribution. We take as our best-fit values the parameters calculated at or averaged within an aperture of 1$R_{\rm e}$, depending on the parameter.
Uncertainties on the measured values are calculated using Monte Carlo realizations, as described in Appendix \ref{app:errors}, combined with the 1$\sigma$ confidence levels for the parameters fluctuations from the best-fit model that we describe in Sec. \ref{sec:bfmodel}.

\subsection{Inner mass distribution}
\begin{figure}[!ht]
\centering
\includegraphics[scale=0.43]{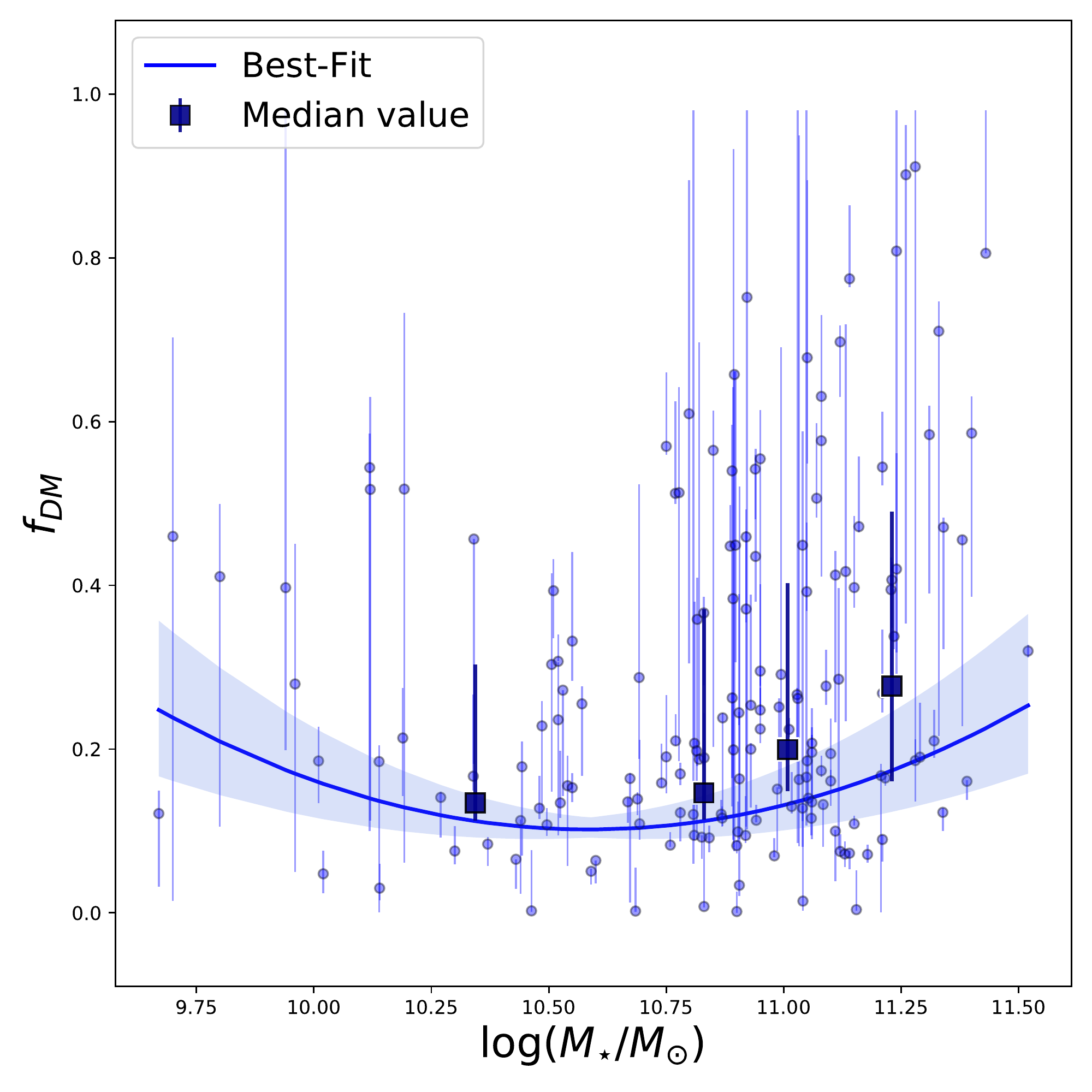}
\caption{Fraction of dark matter to total mass ($f_{\rm{DM}} = M_{dark}/M_{tot}$) within 1$R_{\rm e}$ as a function of stellar mass. The median values of the fraction of dark matter for each of the 4 mass bins are shown as dark blue squares, with the error bars marking the $25^{th}$ and $75^{th}$ percentiles. The blue solid line is the parabolic best-fit to the data - $f_{\rm{DM}} \sim 0.10 + 0.17 \times (\log \frac{M_{\star}}{M_{\odot}}-10.59)^2$. The fit is calculated by taking into account the errors on the data points, so that it is critically determined by the data points with small errors. The shaded region represents the error on the best-fit.}
\label{fig:f_dm}
\end{figure}

The total mass ($M_{tot}$) radial distribution is one of the fundamental parameters of the Schwarzschild model, which includes a stellar component and a dark matter component ($M_{dark}$). A black hole mass component is included as well, but not discussed here as its contribution to the total mass distribution is negligible. The distribution of the fraction of dark matter ($f_{\rm{DM}} = M_{dark}/M_{tot}$) within 1$R_{\rm e}$ for the galaxies in our sample is shown in Fig. \ref{fig:f_dm}. The average value of the dark matter fraction is $ 0.28$, with a standard deviation of $0.20$. Similar to \cite{Cappellari2013a}, we fit a quadratic function to the $f_{\rm{DM}}$ versus stellar mass distribution. The best-fit relation follows $f_{\rm{DM}} \sim 0.10 + 0.17 \times (\log M_{\star}/M_{\odot}-10.59)^2$, although the 1-$\sigma$ scatter along this relation is as high as $\delta f_{\rm{DM}} = 0.24$. 

Above a stellar mass of $\log (M_{\star}/M_{\odot}) \sim 10.75 $ we see a hint of an increasing $f_{\rm{DM}}$ as a function of stellar mass. To test whether this trend is statistically significant, we use the Kendall's correlation coefficient $\tau$, using the Python package \textit{scipy.stats.kendalltau} \citep{Virtanen2019}. This correlation coefficient is robust to small sample sizes. A $\tau$ value close to 1 indicates strong correlation, whereas a value close to $-$1 indicates strong anti-correlation. For galaxies with $\log (M_{\star}/M_{\odot}) \gtrsim 10.75 $ we find a value of $\tau = 0.17$, with a probability of correlation of 99.73\%. While the trend of increasing fraction of dark matter with increasing stellar mass is mild, it is significant at the 3-$\sigma$ level.

\subsection{Intrinsic stellar shape}
Next, we investigate the intrinsic shapes of the galaxies in our sample. As shown in Sec. \ref{sec:param}, three parameters are used to model the dynamically-based intrinsic stellar shape of each galaxy: $p$, $q$ and $u$. The intrinsic shape has been shown to be connected to various other galaxy properties such as: stellar mass \citep{Sanchez-Janssen2010}, luminosity \citep{Sanchez-Janssen2016}, spin parameter \citep[e.g.][]{Foster2017}, mean stellar population age \citep{vandeSande2018} and its environment \citep{Fasano2010, Rodriguez2016}. Furthermore, theoretical simulations suggest that intrinsic shape depends on a galaxy’s formation history \citep{Jesseit2009,Li2018a, Li2018b}.

Here, in particular, we analyse the triaxial parameter $T_{Re}$, calculated at 1$R_{\rm e}$ and defined as: 

\begin{equation}
    T_{Re} = (1 - p_{Re}^2)/(1 - q_{Re}^2).
\end{equation}

\begin{figure}[!ht]
\centering
\includegraphics[scale=0.45]{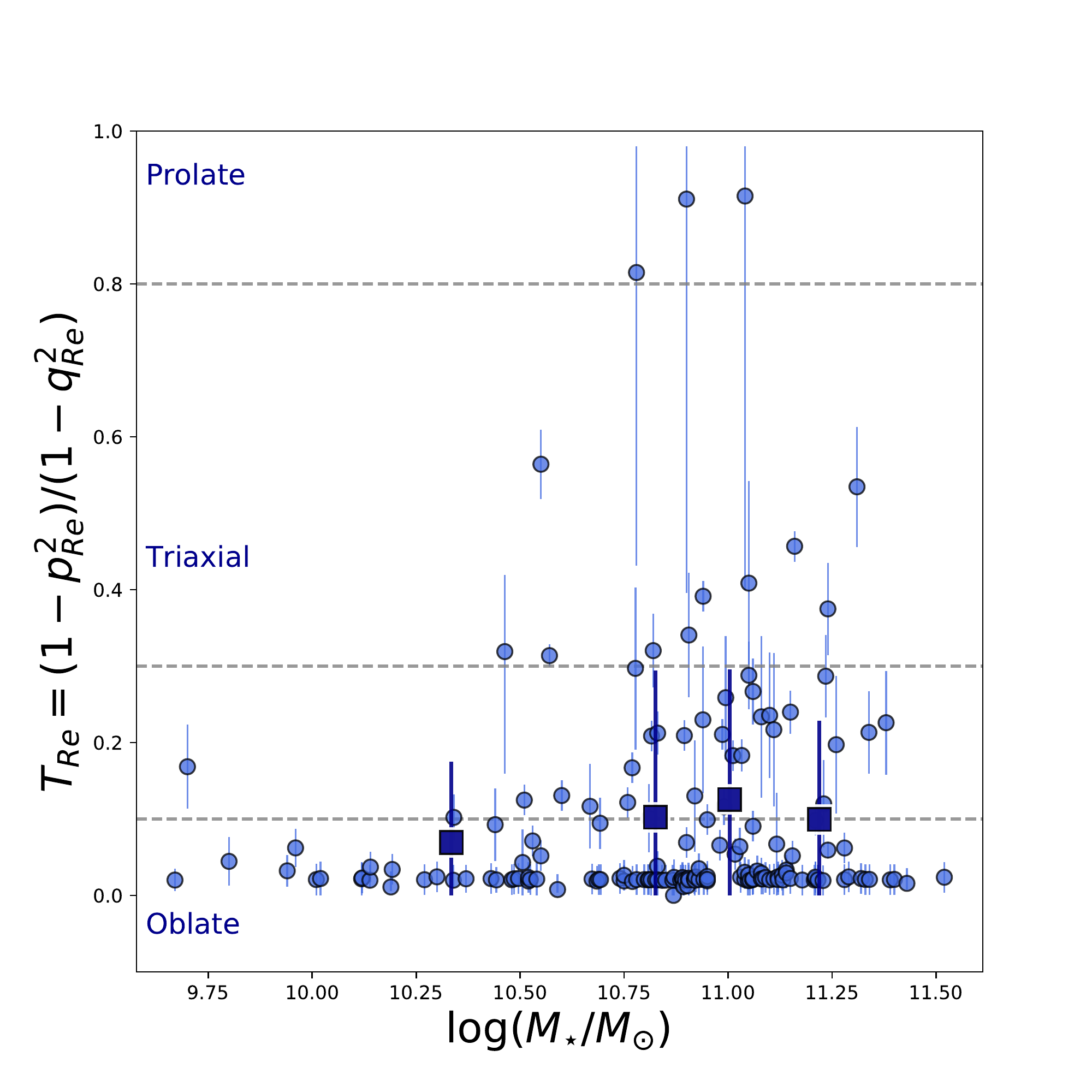}
\caption{Triaxial parameter $T_{Re} = (1 - p_{Re}^2)/(1 - q_{Re}^2)$ as a function of stellar mass. Galaxies with $T_{Re} = 0$ are classified as oblate, galaxies with $T_{Re} = 1$ as prolate and those in-between as triaxial. Grey dashed lines represent $T_{Re} = 0.1$, $T_{Re} = 0.3$ and $T_{Re} = 0.8$. The majority of the galaxies in our sample are oblate, with a few galaxies with non-oblate shape. The average values of the triaxiality parameter for each of the 4 mass bins are shown as dark blue squares, with the error bars marking the 1-$\sigma$ scatter. There is a weak increase in the triaxiality parameter with increasing stellar mass. The percentage of galaxies that are non-oblate ($T_{Re} > 0.1$) increases with increasing stellar mass, going from $12\% \pm 4\%$ below $10^{10.50} M_{\star}/M_{\odot}$ to $29\% \pm 2\%$ above this mass.}
\label{fig:Te_mass}
\end{figure}

We show an example of the best-fit intrinsic shape parameters $p$, $q$ and $T$ as a function of radius in Appendix \ref{sec:example_gals}, Fig. \ref{fig:pqt}. Based on the triaxiallity parameter $T_{Re}$, we separate galaxies into three groups according to their dynamically-based intrisic shape: oblate ($T_{Re} = 0$), prolate ($T_{Re} = 1$) and triaxial ($T_{Re} \neq 0, 1$).
In Fig. \ref{fig:Te_mass} we show the triaxial parameter $T_{Re}$ as a function of stellar mass $\log (M_{\star}/M_{\odot})$. The majority of the galaxies in our sample are close to oblate (118 out of 161 galaxies; $73\% \pm 3\%$), 30 galaxies ($19\% \pm 3\%$) show evidence of being mildly triaxial ($0.1 < T_{Re} \leq 0.3$) and 13 galaxies ($8\% \pm 2\%$) have triaxial/prolate shapes (with $T_{Re} > 0.3$). There is evidence of a slight increase of triaxiality with increasing stellar mass ($\tau= 0.1$), however, this trend is only significant at a 1-$\sigma$ level (with a probability of 82.96\%). However, if we consider the fraction of galaxies that have $T_{Re} > 0.1$ (non-oblate galaxies), we find a clear increase of the fraction with stellar mass, with a sharp change at $\sim 10^{10.50} M_{\star}/M_{\odot}$, with the fraction of non-oblate galaxies increasing from $12\% \pm 4\%$ to $29\% \pm 2\%$ at this mass.

Non-oblate galaxies are often dispersion-dominated, with their shape reflecting the anisotropic velocity dispersion. In contrast, oblate galaxies may have varying degrees of rotation support and anisotropy \citep[e.g.][]{Kireeva2019}. We analyse the distribution of the velocity dispersion anisotropy in the next section.

\subsection{Velocity anisotropy} \label{sec:beta}
Velocity dispersion anisotropy parameters (e.g. $\beta_r, \beta_z$) are widely used as indicators of the underlying orbit distribution of a galaxy. However, various definitions and approaches exist in the literature. The velocity dispersion anisotropy parameter used in more recent literature, $\beta_z$, is in cylindrical coordinates and has been used in particular to describe the global anisotropy in fast-rotating axisymmetric galaxies \citep{Cappellari2007}. This parameter measures the velocity anisotropy along the radius on the disk plane, in cylindrical coordinates, following the idea of cylindrically aligned stellar velocity ellipsoids ellipsoids in oblate galaxies. However, for triaxial galaxies $\beta_z$ ($<R_{\rm e}$) will have a contribution from both circular orbits (which have cylindrically-aligned velocity dispersion ellipsoids) as well as radial and box orbits (which have spherically-aligned velocity dispersion ellipsoids). Recent results \citep{Thater2021} show that the velocity dispersion ellipsoids for the elliptical galaxy NGC 6958 are more closely aligned with spherical coordinates. The misalignment between the measured ellipsoids and the cylindrical coordinates can reach angles as high as $80^{\circ}$. This misalignment can even occur in disk galaxies, most notably, our own Milky Way \citep{Budenbender2015, Hagen2019}.
Following \cite{Thater2021}, we measure the misalignment of the velocity ellipsoids for the galaxies in our sample and find that they are more closely aligned with spherical coordinates. For this reason, we focus on the radial velocity anisotropy parameter, $\beta_r$, in the results presented here. For completeness, we also include the results for $\beta_z$ in Appendix \ref{app:betaz}.

We define the velocity anisotropy parameter $\beta_r$, in spherical coordinates, following \cite{Binney2008}: 
\begin{equation}
\beta_r = 1 - \frac{\Pi_{tt}}{ \Pi_{rr}},
\end{equation}
with 
\begin{equation}
\Pi_{tt} = \frac{\Pi_{\theta\theta}+\Pi_{\phi\phi}}{2},
\end{equation}
($r,\theta,\phi$) the standard spherical coordinates, and
\begin{equation}\label{eq:D_kk}
\Pi_{kk} =  \int \rho \sigma_k^2 \,d^3x  = \sum_{n=1}^{N} M_n \sigma_{k,n}^2 
\end{equation}
with $\sigma_k$ the velocity dispersion along the direction $k$ at a given location inside the galaxy. The summation defines how we computed this quantity from our Schwarzschild models. $M_n$ is the mass contained in each of the $N$ polar grid cells in the meridional plane of the model, and $\sigma_{k,n}$ is the corresponding mean velocity dispersion along the direction $k$.

\begin{figure}[!ht]
\centering
\includegraphics[scale=0.46, trim= 0cm 0cm 2cm 2cm , clip=True]{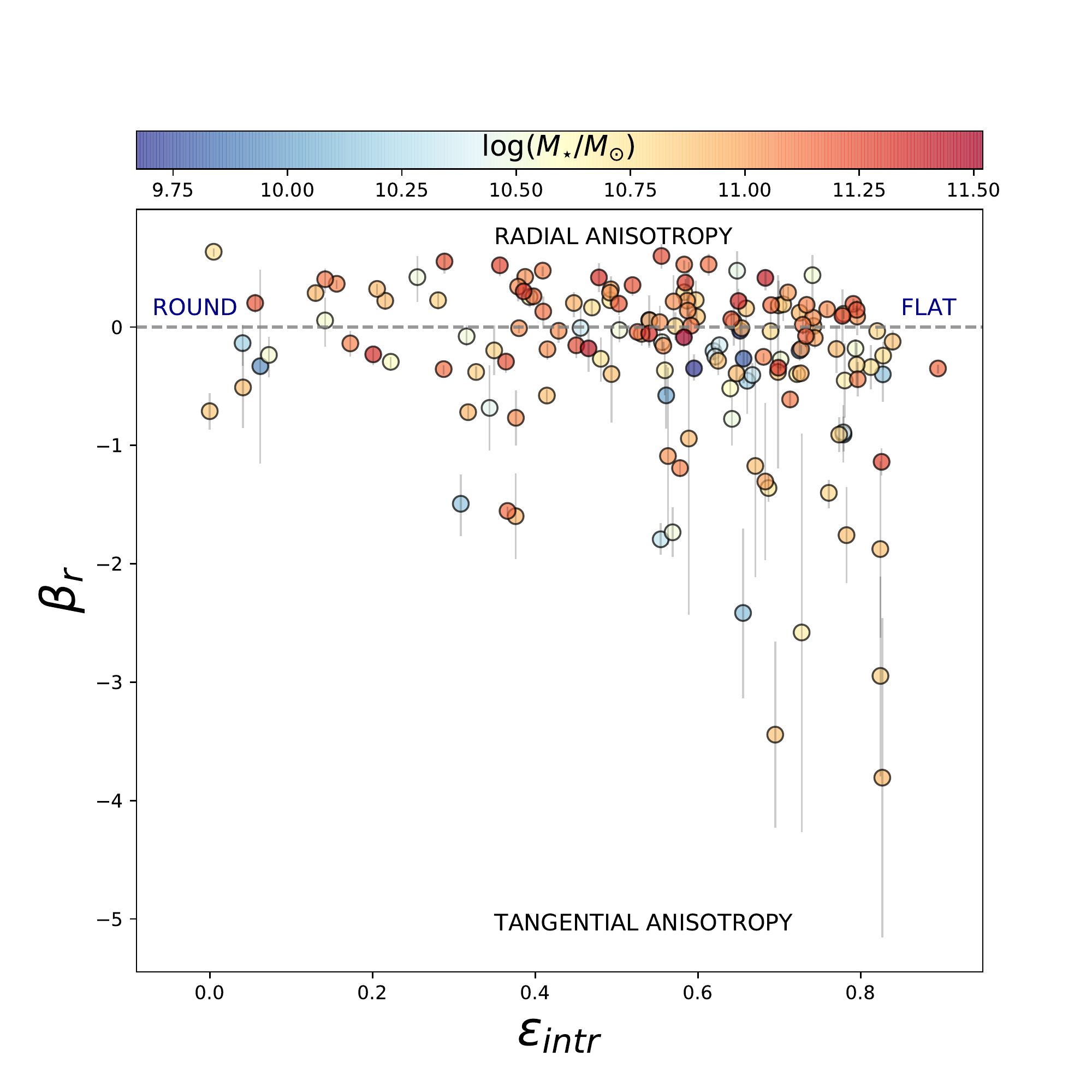}
\caption{Velocity dispersion anisotropy in spherical coordinates $\beta_r$ within 1$R_{\rm e}$ as a function of intrinsic ellipticity ($\varepsilon_{intr} = 1-q$) at 1 $R_{\rm e}$, color-coded by stellar mass. The grey dashed line represents isotropy, $\beta_r = 0$. Negative $\beta_r$ indicates tangentially-anisotropic systems (supported by rotation), while positive $\beta_r$ are radially-anisotropic systems (supported by random motions). Galaxies with very high ellipticity are close to isotropic or tangentially anisotropic. Radially-anisotropic galaxies are generally more massive.}
\label{fig:beta_r_mass}
\end{figure}

We calculate the value of $\beta_r$ within 1$R_{\rm e}$, excluding the inner regions ($r<2^{\prime\prime}$) since this region is affected by atmospheric seeing. $\beta_r > 0$ indicates radial anisotropy, $\beta_r < 0$ indicates tangential anisotropy and $\beta_r = 0$ indicates isotropy. Figure \ref{fig:beta_r_mass} shows the derived values of $\beta_r$, for each galaxy, as a function of intrinsic ellipticity. Here, we derive $\varepsilon$ using the intrinsic flattening, $q_{Re}$, from the best-fit model of the galaxy, measured at 1$R_{\rm e}$; $\varepsilon_{intr} = 1 - q_{Re}$.
In general, galaxies with high ellipticity (flat galaxies, $\varepsilon_{intr}>0.7$) are close to isotropic or tangentially anisotropic (supported by rotation). We also find that radially anisotropic galaxies are typically more massive than tangentially anisotropic galaxies. 

\subsection{Spin Parameter}\label{sec:lambda_r}
The proxy for the spin parameter, $\lambda_{r}$, has previously been used to separate slow-rotating galaxies from fast-rotating galaxies \citep{Emsellem2007, Emsellem2011, Cappellari2016}.
We use the \cite{Cortese2016} definition of the spin parameter to calculate $\lambda_{r}$ for each galaxy: 
\begin{equation}
    \lambda_{r} = \frac{\sum_{i=0}^{N_{spx}} F_i R_i |V_i|}{\sum_{i=0}^{N_{spx}} F_i R_i \sqrt{V_i^2 + \sigma_i^2}}
\label{eq:lambda_r}
\end{equation}
where $i$ refers to each spaxel within the ellipse with semi-major axis $R_{\rm e}$ and ellipticity $\varepsilon$, $F_i$ is the corresponding flux of the $i^{th}$ spaxel, $V_i$ is its stellar velocity, $\sigma_i$ is the velocity dispersion and $R_i$ is the semi-major axis of the ellipse in which the spaxel lies. Since $\lambda_{r}$ is calculated within 1$R_{\rm e}$, it will be referred to as $\lambda_{Re}$ hereafter.

For completeness, we also measure the ratio of ordered to random motion
$V/\sigma$, also measured within 1$R_{\rm e}$, using the definition from \cite{Cappellari2007}:
\begin{equation}
    \left( \frac{V}{\sigma} \right)^2 \equiv \frac{\langle V^2 \rangle }{\langle \sigma^2 \rangle} = \frac{\sum_{i=0}^{N_{spx}} F_i V_i^2}{\sum_{i=0}^{N_{spx}} F_i \sigma_i^2}.
\label{eq:vsigma}
\end{equation}
Results obtained using $V/\sigma$ are similar to those obtained for $\lambda_{Re}$ and are shown in Appendix \ref{app:vsigma}. 

Inclination has a strong impact on the observed $\lambda_{Re}$ and $V/\sigma$ quantities, in particular when the viewing angle is close to face-on \citep[e.g.][]{Binney1990}. While inclination corrections are now commonly applied to $\lambda_{Re}$ measurements \citep[e.g.][]{Querejeta2015, vandeSande2018,Falcon-Barroso2019, delMoral2020, Fraser2021} these methods cannot be applied for slow rotating or triaxial galaxies (for a detailed discussion see \citealt{vandeSande2021b}). Our triaxial Schwarzschild models now allow us, irrespective of galaxy type, to deproject each galaxy to a consistent edge-on view and reconstruct a best-fit internal orbital distribution for that viewing angle. 

In order to reconstruct the edge-on maps, we re-calculate and store the orbit library for each galaxy, with a specific projection. Schwarzschild models take into account the PSF of the observations when reproducing the kinematics. To construct 2D maps without the impact of seeing within the Schwarzschild routine, we set the PSF FWHM to 0.01$^{\prime \prime}$ for the model to use when projecting the galaxy.

\begin{figure*}
\centering
\includegraphics[scale=0.55, trim=2.5cm 0cm 1cm 1cm, clip=True]{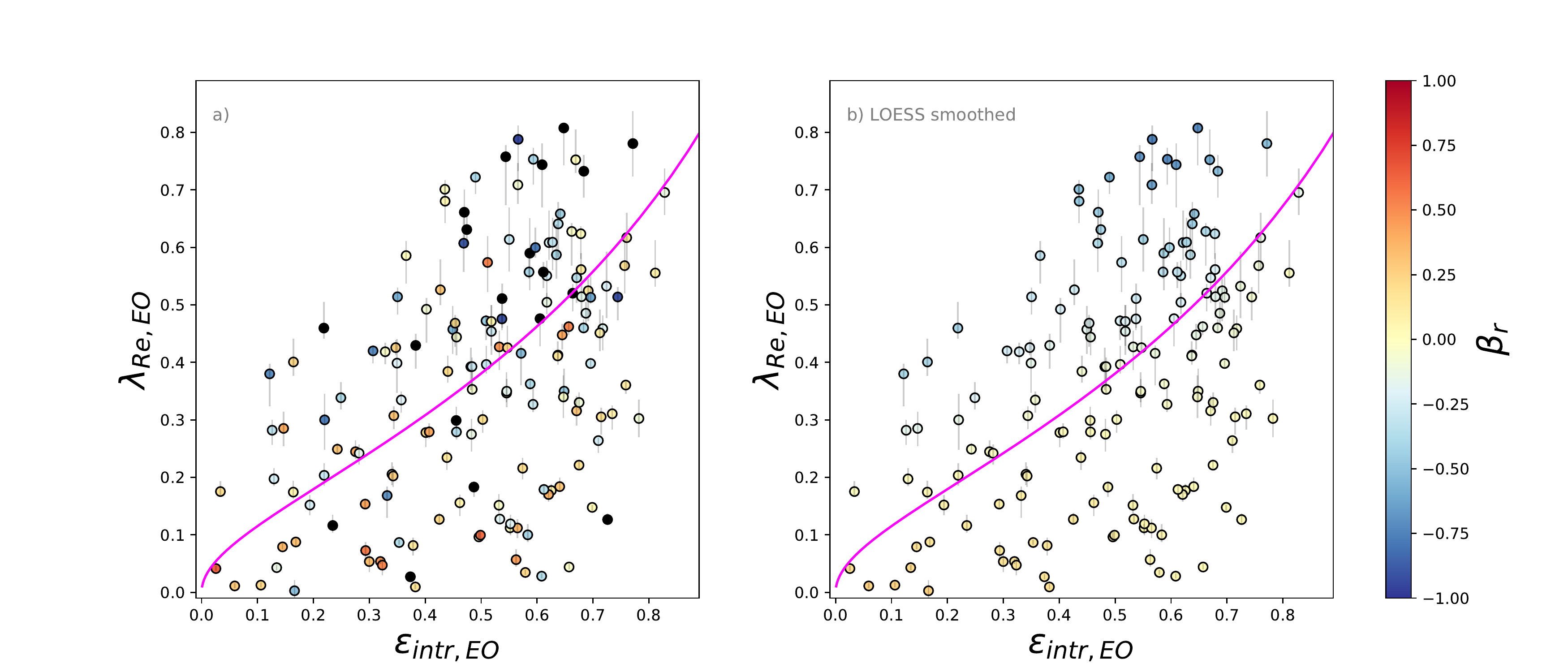}
\caption{$\lambda_{Re, EO}$ as a function of the ellipticity $\varepsilon_{intr, EO}$ derived from MGE fits to the edge-on projected maps, calculated at 1 $R_{\rm e}$. The magenta line corresponds to the relation $\beta_z = 0.65 \times \varepsilon$ for edge-on galaxies as in \cite{Cappellari2007}. Galaxies are colored by their velocity anisotropy $\beta_r$ in panel a and LOESS smoothed in panel b. As expected, $\lambda_{Re, EO}$ increases with increasing intrinsic ellipticity. Galaxies that are radially anisotropic show low- to mid- values of ellipticity and $\lambda_{Re}$, while galaxies with high ellipticity and $\lambda_{Re, EO}$ are more isotropic or tangentially anisotropic.}
\label{fig:lambda_r_eps}
\end{figure*}

Once we have constructed the edge-on projected maps, we measure the spin parameter within 1$R_{\rm e}$ by applying Equation \ref{eq:lambda_r}. In order to produce results comparable to observations, we remeasure the MGE model on our edge-on projected maps to derive $R_{\rm e}$, using the \texttt{MgeFit} python package \citep{Cappellari2002}. We then derive the ellipticity by finding the model isophote with area $A = \pi R_{\rm e}^2$, and use its ellipticity as the galaxy ellipticity \citep{Deugenio2021}.
We show the derived edge-on $\lambda_{Re, EO}$ values as a function of the edge-on intrinsic ellipticity from our MGE fit, $\varepsilon_{intr,EO}$, in Fig. \ref{fig:lambda_r_eps}, color-coded by their velocity anisotropy $\beta_r$. The magenta line corresponds to the relation $\beta_z = 0.65 \times \varepsilon$ for edge-on galaxies as in \cite{Cappellari2007}.

We find that $\lambda_{Re, EO}$ increases with increasing intrinsic ellipticity. In particular, galaxies that have low values of $\lambda_{Re, EO}$ are rounder than galaxies with higher values of \lam.
Moreover, we find that galaxies that are radially anisotropic (positive values of $\beta_r$) show low- to mid- values of ellipticity and $\lambda_{Re, EO}$, while galaxies with high ellipticity and $\lambda_{Re, EO}$ are more isotropic or tangentially anisotropic. This is seen more clearly when a locally weighted regression algorithm (LOESS - \citealt{Cappellari2013a}) is applied to the data to recover any mean underlying trend in $\beta_r$ (Fig. \ref{fig:lambda_r_eps}, panel b). In general, the variation in $\beta_r$ seems to mostly be driven by the spin parameter, $\lambda_{Re, EO}$.

\begin{figure}[!ht]
\centering
\includegraphics[scale=0.42,trim=0cm 0cm 0cm 0cm, clip=True]{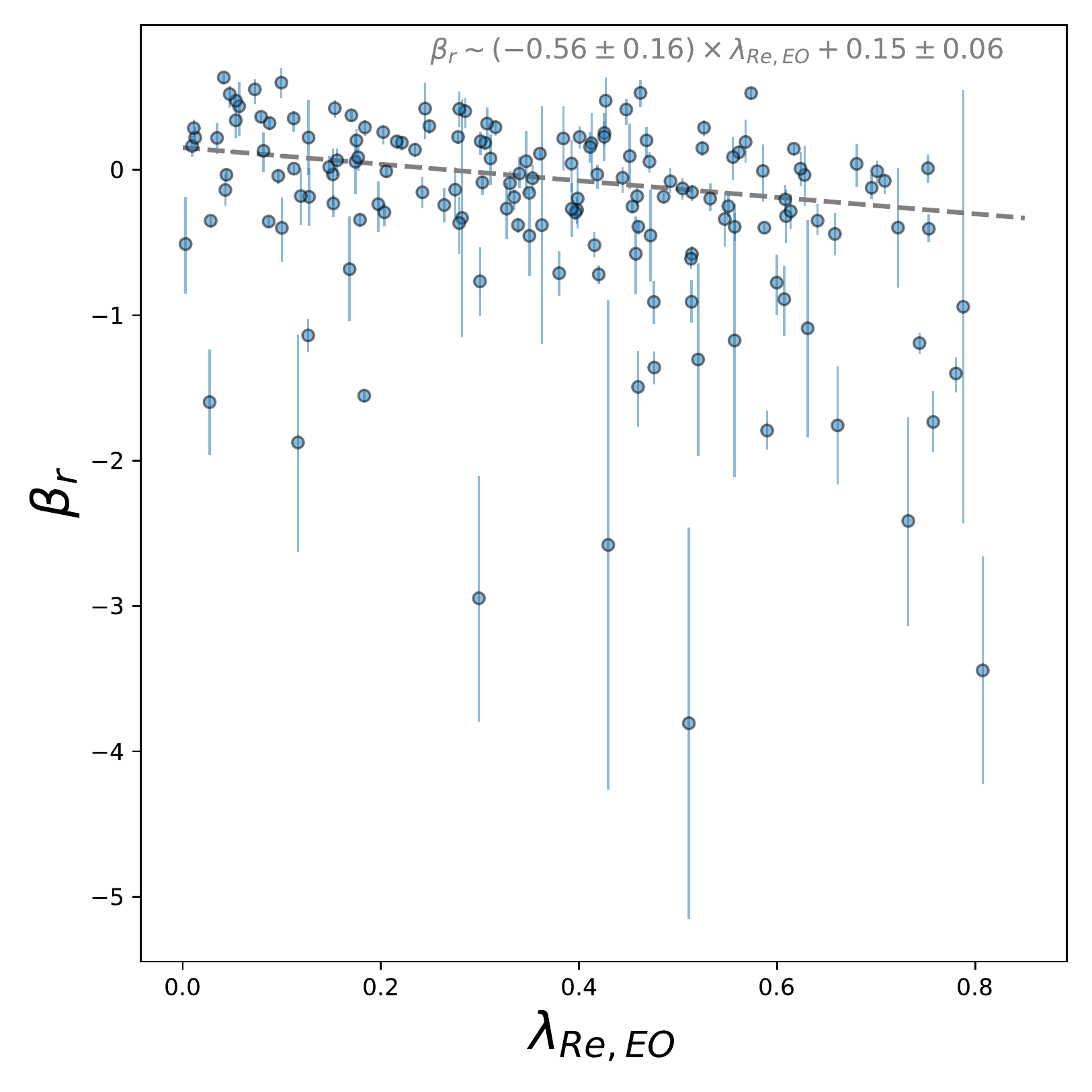}
\caption{Velocity anisotropy $\beta_r$ as a function of the intrinsic spin parameter \lam. The dashed line is the linear best-fit to the data points (shown in the top right-hand corner). The two parameters are anti-correlated, so that $\beta_r$ decreases with increasing \lam. This means that fast-rotating galaxies are more likely to be tangentially anisotropic.}
\label{fig:lambda_r_beta}
\end{figure}

The anti-correlation between $\lambda_{Re, EO}$ and $\beta_r$ can be seen in Fig. \ref{fig:lambda_r_beta}. Testing the correlation using the Kendall's correlation coefficient $\tau$, we find a value of $\tau =-0.27$, with a probability of correlation of 99.99\% that $\beta_r$ decreases with increasing \lam. This means that fast-rotating galaxies are, as expected, more tangentially anisotropic than slow-rotating systems, which are more radially anisotropic.

\subsection{Orbital structure}
Stellar orbits can be characterized by two main properties: the time-averaged radius $r$, representing the size of each orbit, and the circularity $\lambda_z= \overline{L_z}/(r \times \overline{V_c})$, where $\overline{L_{z}}$ is the time averaged z-component of the orbit's angular momentum ($\overline{x v_{y}-y v_{x}}$), $r=\overline{\sqrt{x^{2}+y^{2}+z^{2}}}$, and $\overline{V_{c}}=
\sqrt{\overline{v_{i}^{2}+v_{y}^{2}+v_{z}^{2}+2 v_{x} v_{y}+2 v_{x} v_{z}+2 v_{y} v_{z}}}$. The denominator represents the angular momentum of a typical circular orbit associated with the original orbit. Using the ratio of these two angular momentum terms, we can quantify the orbit circularity. $|\lambda_z| = 1$ represents highly-rotating short-axis tube orbits (circular orbits), while $\lambda_z = 0$ represents mostly box or radial orbits. Taking the radius, $r$, and the circularity, $\lambda_z$, of each orbit, and considering their weights given by the solution from the best-fit model, we can use the orbit circularity distribution in the phase space to obtain the probability density of orbits within 1$R_{\rm e}$, for each galaxy.

\begin{figure*}
\centering
\includegraphics[trim=2cm 0cm 2cm 0cm, clip=True,width=19cm]{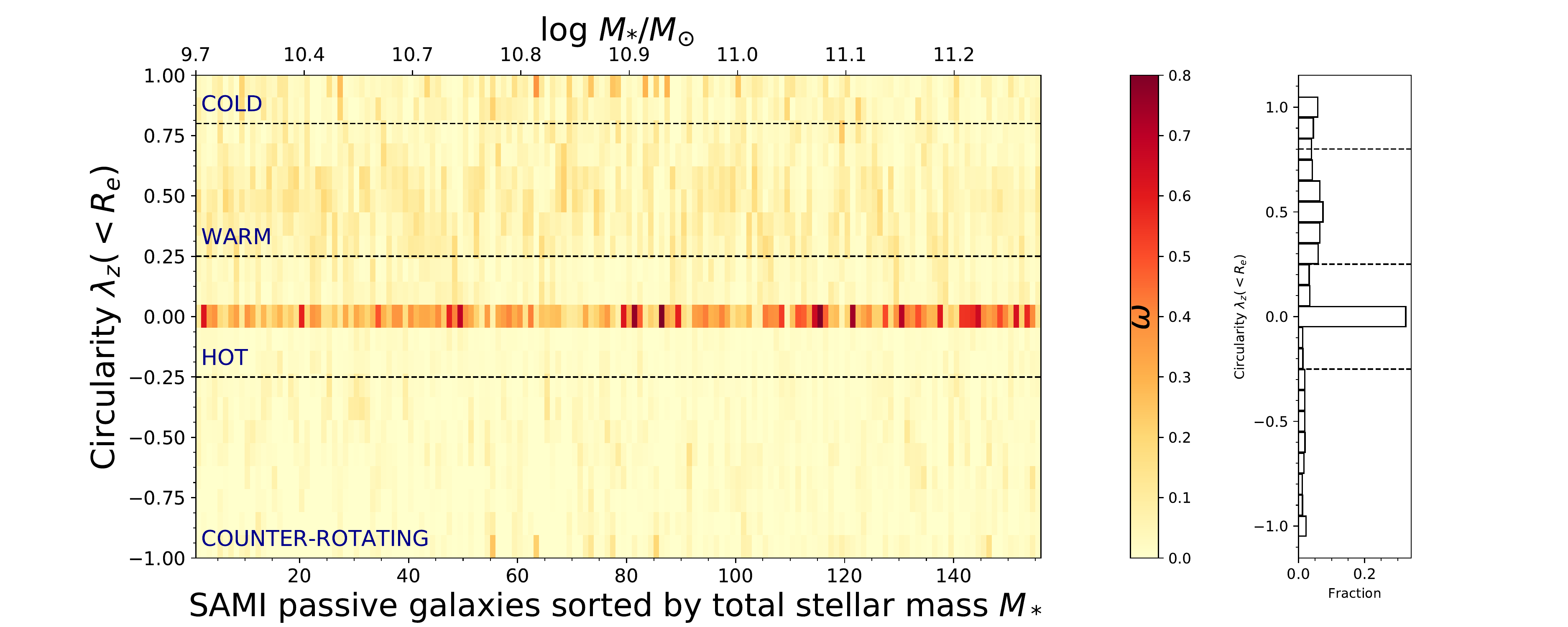}
\caption{Overall orbit circularity distribution (calculated by integrating the probability distribution of $\lambda_z$, over all radii within 1$R_{\rm e}$ and normalizing it to unity), for all the galaxies in our sample, sorted by increasing stellar mass (shown in the top x-axis). The color indicates the normalized density, $\omega$, of the orbits on the phase space. The orbits are divided into four categories: cold orbits ($\lambda_z \geq 0.80$), warm orbits ($ 0.25 < \lambda_z < 0.80$), hot orbits ($-0.25 \geq \lambda_z \leq 0.25$) and counter-rotating orbits ($\lambda_z < -0.25$). Darker colors indicate higher probabilities as illustrated by the color bar. The right-hand panel shows the average orbit-circularity distribution within the mass range. Overall, the fraction of hot orbits seems to increase with increasing stellar mass, while the fraction of warm and cold orbits becomes smaller with increasing mass.}
\label{fig:lambda_z_mass}
\end{figure*}

Figure \ref{fig:lambda_z_mass} shows the overall orbit circularity distribution for all the galaxies in our sample, sorted by increasing stellar mass (shown in the top $x$-axis). The orbit circularity distribution is calculated by integrating the probability distribution of $\lambda_z$ over all radii within 1$R_{\rm e}$ and normalizing it to unity. The color of each square represents the normalized density, $\omega$, of the orbits on the phase space. We divide the orbits into four broad categories (similar to \citealt{Zhu2018c, Zhu2018b}): cold orbits, $\lambda_z \geq 0.80$ (close to circular orbits); warm orbits, $ 0.25 < \lambda_z < 0.80$ (short-axis tube orbits with a component of rotation but also contribution of random motions); hot orbits, $-0.25 \geq \lambda_z \leq 0.25$ (mostly box orbits and long-axis tube orbits); counter-rotating orbits, $\lambda_z < -0.25$, (similar to the warm and cold components, but with opposite rotation). Overall, the amount of hot orbits increases with increasing stellar mass, while the number of warm and cold orbits becomes smaller with increasing mass.

\begin{figure*}[!ht]
\centering
\includegraphics[trim=2cm 2.5cm 2cm 2.5cm,clip=true, width=14cm]{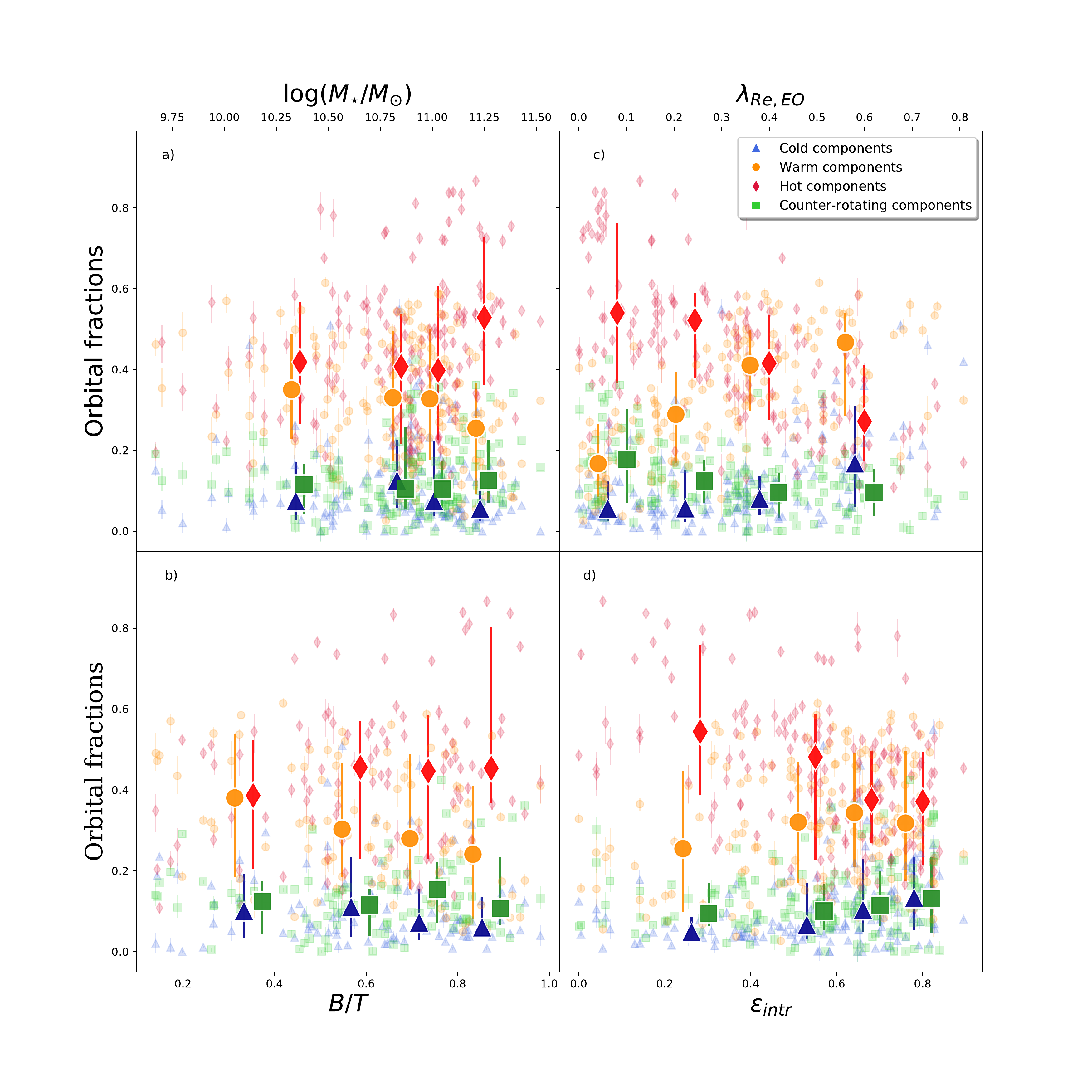}
\caption{Fractions of orbital components as a function of:  a) stellar mass, b) bulge to total flux ratio, B/T, c) $\lambda_{Re,EO}$, d) $\varepsilon_{intr}$. Bold points show the median values for each mass bin, with error bars representing the 1$\sigma$ scatter around the median value. There is a clear increase of hot orbits (red diamonds) with increasing stellar mass (and B/T ratio), while the fraction of warm orbits (orange circles) decreases with increasing stellar mass (and B/T ratio), both of them showing a large scatter. Hot orbits decrease with increasing $\lambda_{Re,EO}$ (and $\varepsilon_{intr}$), while the fraction of warm orbit increases with increasing $\lambda_{Re,EO}$ (and $\varepsilon_{intr}$).The fraction of cold orbits (blue triangles) is also declining towards more massive galaxies and increases with galaxies becoming flatter. The fraction of counter-rotating orbits (green squares) does not show any significant trend with B/T ratio or $\varepsilon_{intr}$, but it does decrease with increasing $\lambda_{Re,EO}$. The correlation between the orbital fractions and $\lambda_{Re,EO}$ shows very little scatter. 
}
\label{fig:fracs_mass}
\end{figure*}

To better visualize these trends with stellar mass, we calculate the luminosity-weighted fractions of each component within 1$R_{\rm e}$ as a function of stellar mass in Fig. \ref{fig:fracs_mass}, panel a. We also divide the sample into 4 mass bins with 29 galaxies each and we show the median values for each mass bin as bold points. We find a clear increase in the fraction of hot orbits with increasing stellar mass ($\tau = 0.16$, with a probability of correlation of 99.71\%), while the fraction of warm orbits decreases with increasing stellar mass ($\tau = -0.19$, with a probability of correlation of 99.95\%), both of them showing a large scatter. In particular, the fraction of hot and warm orbits seem to have a sharp change above $\log M_{\star}/M_{\odot}= 11$. The fraction of cold orbits only have a weak correlation with mass ($\tau = -0.10$, with a probability of correlation of 94.21\%), declining towards more massive galaxies. The fraction of counter-rotating orbits does not seem to depend on stellar mass ($\tau = 0.05$, with a probability of correlation of 61.44\%). 

We also explore the correlation of the fractions of the orbital components with the bulge to total flux ratio, B/T in panel b, with the intrinsic spin parameter $\lambda_{Re,EO}$ in panel c and with the intrinsic ellipticity $\varepsilon_{intr}$ in panel d. B/T ratios are calculated from the $r$-band photometry, performing a 2D photometric bulge-disk decomposition (\citealt{barsanti2021colors} for the decomposition of cluster galaxies and Casura et al., in prep, for the galaxies in the GAMA region). Only 97 galaxies in our sample have reliable B/T values for the 2 component decomposition. The orbital fractions show a correlation with the B/T ratios similar to that with stellar mass. 

Looking at $\lambda_{Re,EO}$ and $\varepsilon_{intr}$ the orbital fractions have similar trends: hot orbits decrease with increasing $\lambda_{Re,EO}$ and $\varepsilon_{intr}$, warm orbits increase with increasing $\lambda_{Re,EO}$ and $\varepsilon_{intr}$ and cold orbits show an increase in the fractions, while there is a significant change ($\tau=-0.21$, with a probability of correlation of 99.99\%) in the fraction of counter-rotating orbits only with $\lambda_{Re,EO}$, so that the fraction decreases with increasing $\lambda_{Re,EO}$. In particular, we note that the trends with $\lambda_{Re,EO}$ are tighter than those with stellar mass (average 1-$\sigma$ scatter $\sim$ 0.09 compared to the average 1-$\sigma$ scatter $\sim$ 0.12 with stellar mass).

\subsection{Higher-order stellar kinematics and orbital components}

\cite{vandeSande2017a} used the higher-order stellar kinematic moments ($h_3$ and $h_4$) to classify galaxies in the SAMI Galaxy Survey into 5 distinct classes based on each galaxy's individual $h_3$ versus $V/\sigma$ signature. 
Galaxies belonging to Class 1 are typically the most massive, large and red. Most of Class 1 galaxies are also classified as slow rotators, indicating that they have more complex dynamical structures as compared to fast rotators. Galaxies in Class 2-5 are all consistent with being oblate rotating axisymmetric spheroids as based on $\lambda_{Re}$ and $\varepsilon$, but have a range of higher-order kinematic signatures. Galaxies in Class 2 are less massive, but still red, and reside in between slow and fast rotators. True fast rotators are in Class 3 and 4, with galaxies showing a strong anti-correlation of $V/\sigma$ and $h_3$. Galaxies in Class 5 have very high $V/\sigma$ and ellipticity, but they do not show any anti-correlation with $h_3$. Here, we examine the connection between the distributions of these classes and the orbital components of the galaxies in our sample.

\begin{figure*}

\centering

\includegraphics[trim=4cm 0.5cm 2cm 1cm, clip=True,width=19cm]{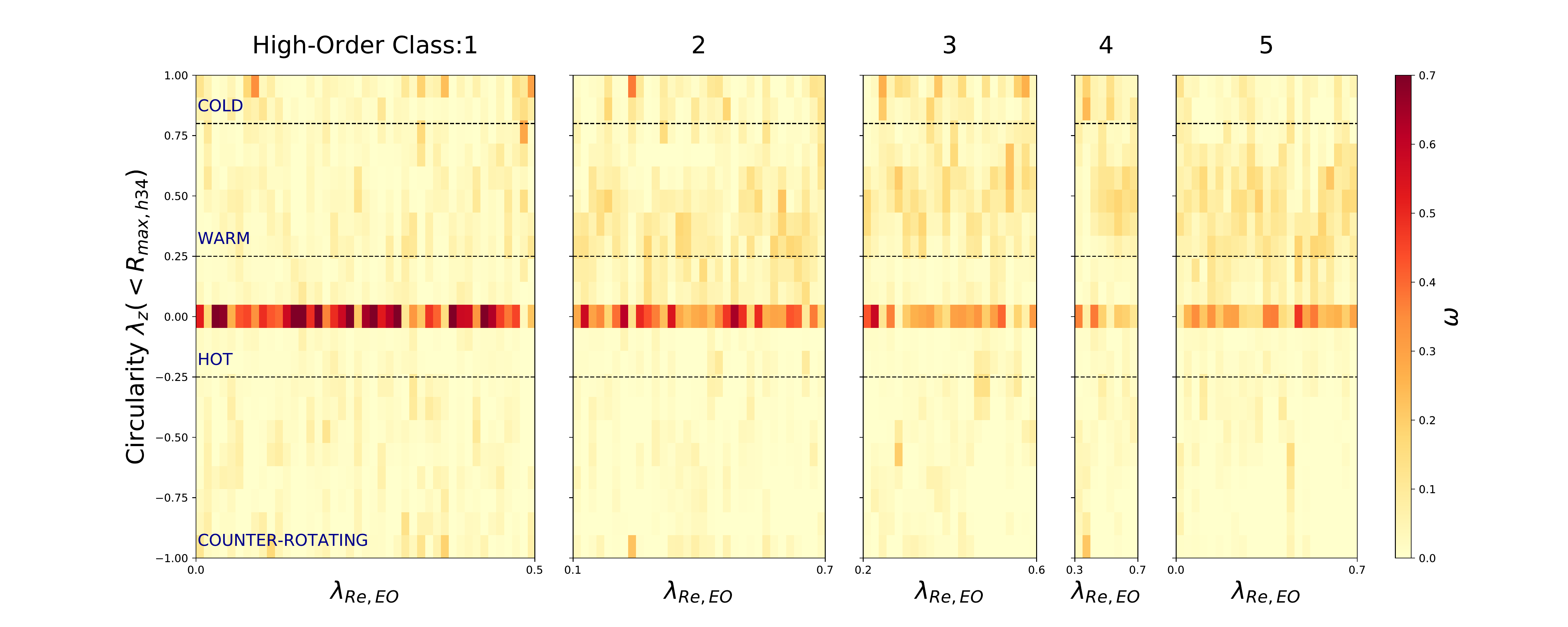}

\caption{Orbit circularity distribution calculated by integrating the probability distribution of $\lambda_z$ over all radii within $R_{\rm max,h3h4}$ and normalizing it to unity, for all the galaxies in our sample, grouped by their kinematic classes from \cite{vandeSande2017a} based on the higher-order ($V/\sigma$ - $h_3$) signatures. Each class has been ordered by the intrinsic $\lambda_{Re,EO}$ values. The color indicates the normalized density, $\omega$, of the orbits on the phase space.
Galaxies in Class 1 are dominated by hot orbits. Warm orbits become important for galaxies in Class 2, in particular at higher values of $\lambda_{Re,EO}$, with the warm orbits contribution increasing for Classes 3, 4 and 5. Hot orbits become less important with increasing $\lambda_{Re,EO}$.}

\label{fig:orb_lambda}

\end{figure*}

In Fig. \ref{fig:orb_lambda} we show the overall orbit circularity distribution for all the galaxies in our sample, grouped by their kinematic classes. The orbit circularity distribution is calculated by integrating the probability distribution of $\lambda_z$ over all radii within $R_{\rm max,h3h4}$ and normalizing it to unity, similarly to Fig. \ref{fig:lambda_z_mass}. $R_{\rm max,h3h4}$ is the radius within which the $h_3$ versus $V/\sigma$ signatures were derived for each galaxy, due to S/N restrictions \citep{vandeSande2017a}. Within each subpanel in Fig. \ref{fig:orb_lambda}, we have ordered the galaxies by their intrinsic $\lambda_{Re,EO}$ values. The color indicates the normalized density, $\omega$, of the orbits on the phase space.
There is a clear distinction between the orbital distributions, depending on the galaxy kinematic class. In general, hot orbits are more dominant in galaxies belonging to Class 1, and they decrease going towards Class 5, with Class 4 showing the lowest values. The contribution of cold orbits becomes more important in Classes 3, 4 and 5, while warm orbits can also be a significant fraction for galaxies in Class 2. Counter-rotating orbits do not have any significant contribution for Class 3 and 5. 

\begin{figure}
\centering
\includegraphics[trim=1.5cm 1.5cm 2cm 2cm, clip=True,width=8.5cm]{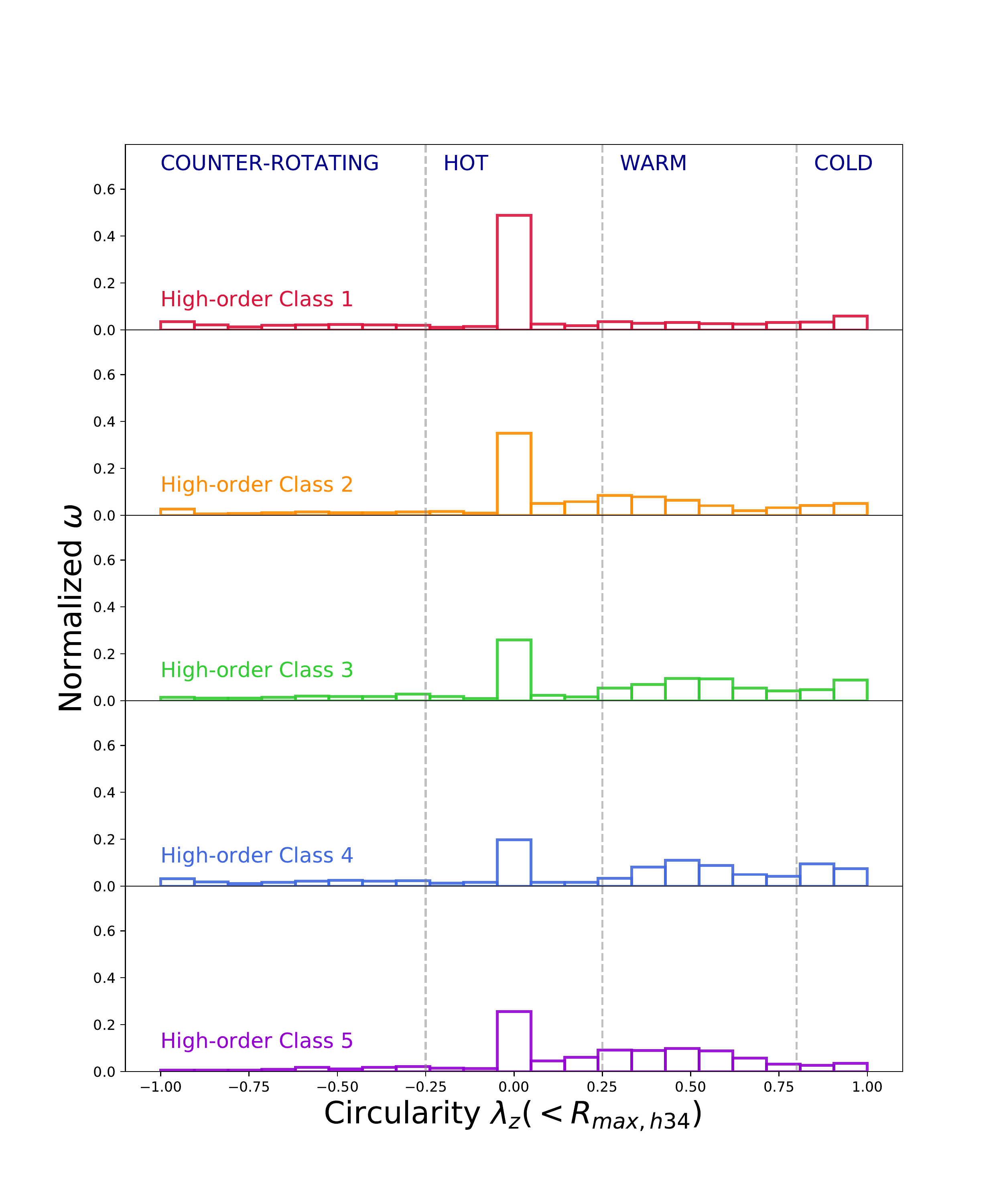}
\caption{Normalized density of orbits as a function of the orbit circularity $\lambda_z$ for each kinematic class from \cite{vandeSande2017a} based on the higher-order ($V/\sigma$ - $h_3$) signatures. There is a decrease in the contribution from hot orbits going from Class 1 to 5, with Class 4 having the lowest value. Warm orbits become more important from Class 2 to Class 5. Counter-rotating orbits do not have any significant contribution for Class 3 and 5.}
\label{fig:orb_fracs}
\end{figure}

The distribution of orbits in each class is clearer if we look at their integrated distributions, shown in Fig. \ref{fig:orb_fracs}. Within each class, there are also clear trends of the orbital components with $\lambda_{Re}$, so that, as expected, cold orbits are increasing with increasing $\lambda_{Re}$ (rotationally supported galaxies). Similarly, warm orbits also increase with increasing $\lambda_{Re, EO}$. In contrast, the hot component becomes less important with increasing $\lambda_{Re, EO}$, while the counter-rotating orbits do not show any particular trend. In particular, in slow-rotating galaxies, the main contribution is given by hot orbits. This is not unexpected, since these galaxies are expected to be pressure-supported. The warm component starts to become important for galaxies in Class 2, with its contribution increasing with increasing $\lambda_{Re, EO}$. Galaxies in Class 3, 4 and 5 show higher contributions from warm and cold orbits for all the galaxies (compared to Class 1 and 2). We do not find strong evidence for a difference in the orbital distribution between the higher-order kinematic Classes 3-5 as derived from the circularity diagram. Nonetheless, the existence of the different signatures in the higher-order moment maps points to kinematic features that are not captured in the $\lambda_z$--r space. This will be explored further in future work, but is beyond the scope of this paper.

\section{Discussion}
We have constructed Schwarzschild orbit-superposition models of 161 passive galaxies from the SAMI Galaxy Survey in order to derive intrinsic properties such as the internal mass distribution, intrinsic stellar shape, velocity anisotropy and orbit circularity distribution. We find that changes in the internal structures are mostly correlated with the stellar mass of the galaxies.

\subsection{Comparison with previous studies}
\subsubsection{Fractions of dark matter}
\begin{figure}
\centering
\includegraphics[scale=0.43, clip=True]{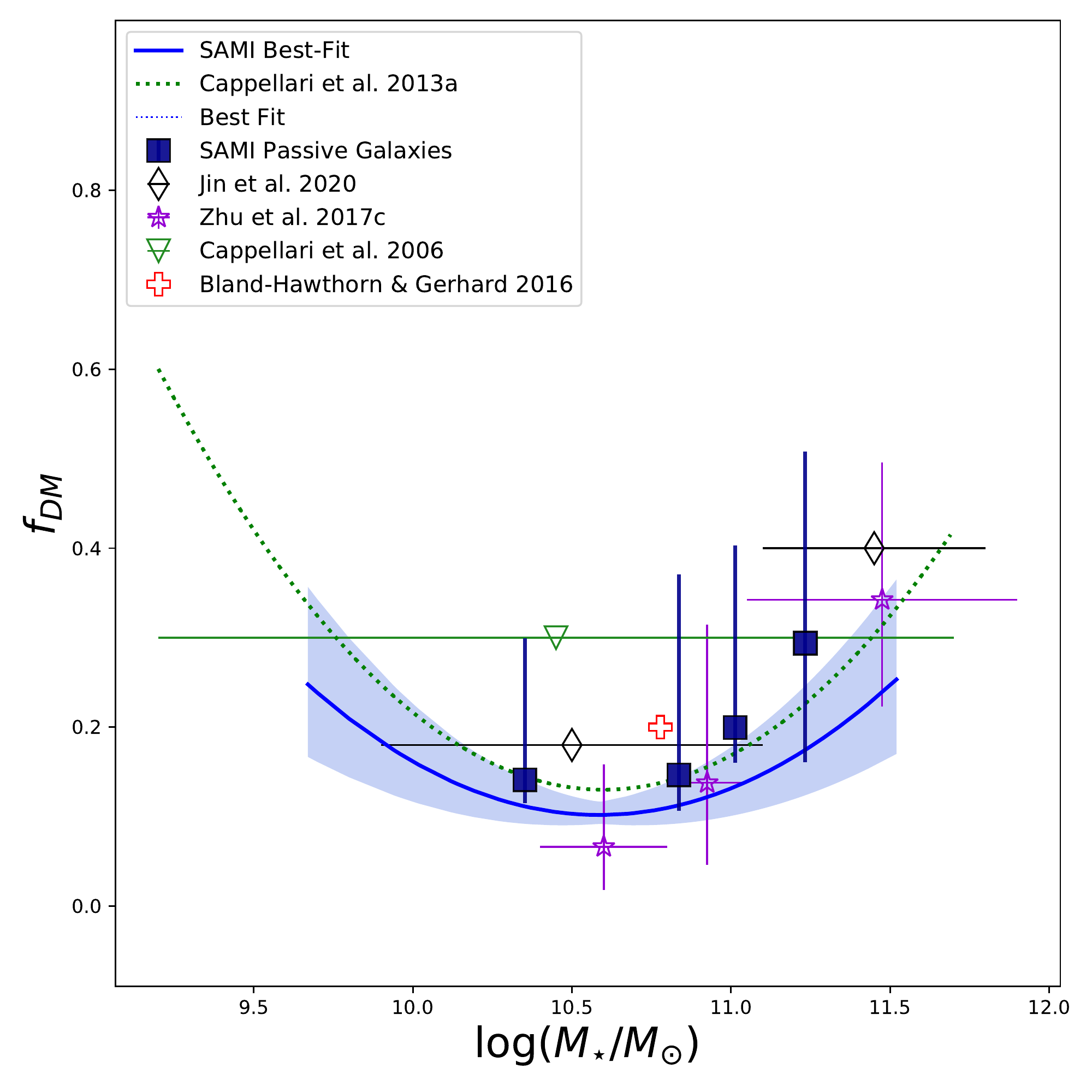}
\caption{Median values of the fractions of dark matter ($f_{\rm{DM}}=M_{dark}/M_{tot}$) within 1 $R_e$ as a function of stellar mass for: SAURON (green triangle, \citealt{Cappellari2006}), ATLAS 3D (green dotted line, \citealt{Cappellari2013a} - derived with a cosmologically-motivated NFW halo) galaxies, the Milky Way (red cross, \citealt{Bland2016}), CALIFA (purple stars, \citealt{Zhu2018a}), (MaNGA (black diamonds, \citealt{Jin2020}) and SAMI (dark blue squares and blue solid line). Horizontal error bars delimit the mass range covered by each study. Vertical error bars mark the $25^{th}$ and $75^{th}$ percentiles, when available. The shaded region represents the error on the best-fit line for SAMI galaxies. Our results are in good agreement with the results presented in the literature. }
\label{fig:fdm_comp}
\end{figure}

We find an average value of the dark matter fraction of $f_{\rm{DM}} = 0.28$, with a standard deviation of $0.20$, within 1$R_e$. 
In general, our results for \fdm are broadly consistent with previous stellar dynamic determinations within 1$R_{\rm e}$ found in the literature which also all assume a NFW dark matter halo distribution (Fig. \ref{fig:fdm_comp}). For example, \cite{Gerhard2001} found \fdm = $0.1-0.4$ from spherical dynamical modelling of 21 ETGs, \cite{Cappellari2006} inferred a median \fdm $\approx 0.3$ by comparing dynamics and population masses of 25 ETGs, and assuming a universal IMF, \cite{Thomas2007, Thomas2011} measured \fdm $= 0.23 \pm 0.17 $ via axisymmetric dynamical models of 17 ETGs, \cite{Cappellari2013a} measured a \fdm of 0.15 for early-type galaxies in ATLAS$^{3D}$ using Jeans Anisotropic Modelling (JAM), with galaxies showing an increasing fraction of dark matter with increasing mass for masses $\log (M_{\star}/M_{\odot}) > 10.6$, consistent with our findings here. Similar results were also found by \cite{Posacki2015}, $f_{\rm{DM}} = 0.14$ for 55 early-type galaxies from stellar dynamics and lensing, and by \cite{Poci2017} - $f_{\rm{DM}} = 0.19$ using JAM to model a sample of 258 early-type galaxies in ATLAS$^{3D}$.
For the Milky Way, \cite{Bland2016} found a $f_{\rm{DM}}= 0.3$. Overall, these studies show that baryons dominate the centers of galaxies, especially in our mass range, where the efficiency of galaxy building is peaking. 

\cite{Jin2020} found a similar trend for early-type galaxies in the MaNGA sample, with the \fdm for the most massive galaxies ($11.0 < \log (M_{\star}/M_{\odot}) < 11.5$) generally above 0.4, similar to what we see for galaxies in the same mass bin. However, we note that, as suggested by model tests with mock data from the Illustris simulations \citep{Jin2019}, estimations of \fdm can have a systematic offset as a result of modelling the dark matter halos assuming that galaxies follow a NFW profile, which may not be correct. This is an interesting aspect that will need to be explored further and tested with a range of simulations and datasets.
The trend we see in the $f_{\rm{DM}}$ with stellar mass is also consistent with predictions from simulations, where galaxies with $\log (M_{\star}/M_{\odot}) \sim 10.6$ are the most efficient at forming stars \citep[e.g.][]{Behroozi2010, Behroozi2013,Henriques2019}. The physical interpretation of this behavior is the interplay between the feedback processes that impact star formation efficiency at different galaxy masses. Supernova feedback is more effective at reheating and expelling gas in low-mass galaxies, while AGN feedback is more effective in high-mass galaxies.
\subsubsection{Intrinsic shape distribution}

As seen in Fig. \ref{fig:Te_mass}, the majority of our galaxies are very close to oblate axisymmetric ($73\% \pm 3\%$), with $T_{\rm Re} \leq 0.1$ , with varying degrees of intrinsic flattening, with $19\% \pm 3\%$ being mildly triaxial ($ 0.1< T_{\rm Re} \geq 0.3$) and a small percentage ($8\% \pm 2\%$) being triaxial/prolate ($T_{\rm Re} > 0.3$). There is a weak increase in the triaxiality parameter with increasing stellar mass. The percentage of galaxies that are non-oblate ($T_{Re} > 0.1$) increases with increasing stellar mass, going from $12\% \pm 4\%$ below $10^{10.50} M_{\star}/M_{\odot}$ to $29\% \pm 2\%$ above this mass.

Triaxial Schwarzschild orbit-superposition dynamical models allow to measure intrinsic shapes directly. Previous studies used statistical methods to derive intrinsic shape properties; for example
\cite{Kimm2007} studied a sample of 3922 galaxies from SDSS \citep{Adelman-McCarthy2006} and found that more massive galaxies are more likely to be triaxial than lower-mass galaxies. \cite{Foster2017} derived the intrinsic shape of 845 galaxies in the SAMI Galaxy Survey using an algorithm to simultaneously invert the distributions of apparent ellipticities and kinematic misalignments using the methodology of \cite{Weijmans2014}. They find the majority ($\sim 85\%$) of the galaxies in their sample to be oblate axisymmetric, in good agreement with \cite{Weijmans2014} and our results. 
Our result is also in agreement with previous results from the Illustris simulations, where only a very small fraction of galaxies are found to have prolate shapes, with the fraction decreasing to zero prolate galaxies below $\log (M_{\star}/M_{\odot}) = 11.48$ \citep{Li2018a}.
\cite{Jin2020} found higher fractions of triaxial and prolate galaxies in a sample of 149 early-type galaxies from the MaNGA survey. This discrepancy is partly explained by their higher stellar mass range analysed (their stellar masses ranged between $10^{9.9}$ and $10^{11.8}  M_{\odot}$), and their different sample selection. 
\cite{Jin2020} also find an increase of the fraction of non-oblate galaxies with increasing stellar mass, in agreement with our results.

\subsubsection{Velocity Anisotropy}
We find that galaxies with higher ellipticities have, in general, more negative values of $\beta_r$. This means that flatter galaxies are more tangentially anisotropic than rounder galaxies, while the latter are more likely to be supported by radial anisotropy. Moreover, we find a tight relationship of $\beta_r$ with \lam. This is not unexpected, since both parameters are a measure of rotation. The idea that the most giant early-type galaxies are not flattened by rotation but by anisotropy was proposed in the late 1970s \citep{Bertola1977,Illingworth1977,Binney1978}, however most of the dynamical modelling methods available to date do not allow for triaxiality, which is needed for a significant fraction of massive galaxies in order to construct accurate models.

Our results are also in agreement with more recent studies. For example, \cite{Gerhard2001} found that most of the galaxies in their sample of 21 ETGs were moderately radially anisotropic ($\beta_r \approx 0.2$), in agreement with the values we find in this study. 

\subsubsection{Orbital structures}

We find that the hot orbital component generally dominates within $R_{\rm e}$, becoming the most prevalent component among galaxies with total stellar mass $\log (M_{\star}/M_{\odot}) > 11$. As expected, bulge-dominated galaxies have high fractions of hot orbits (consistent with a pressure-supported bulge). In most galaxies a substantial number of stars within $R_{\rm e}$ are on warm orbits, with the contribution becoming more important at lower stellar masses. The cold component rarely dominates within $R_{\rm e}$ and its importance decreases with increasing stellar mass. The counter-rotating component is roughly constant for galaxies at all masses.

\begin{figure}
\centering
\includegraphics[scale=0.4, trim= 1.3cm 1cm 2cm 2cm, clip=True]{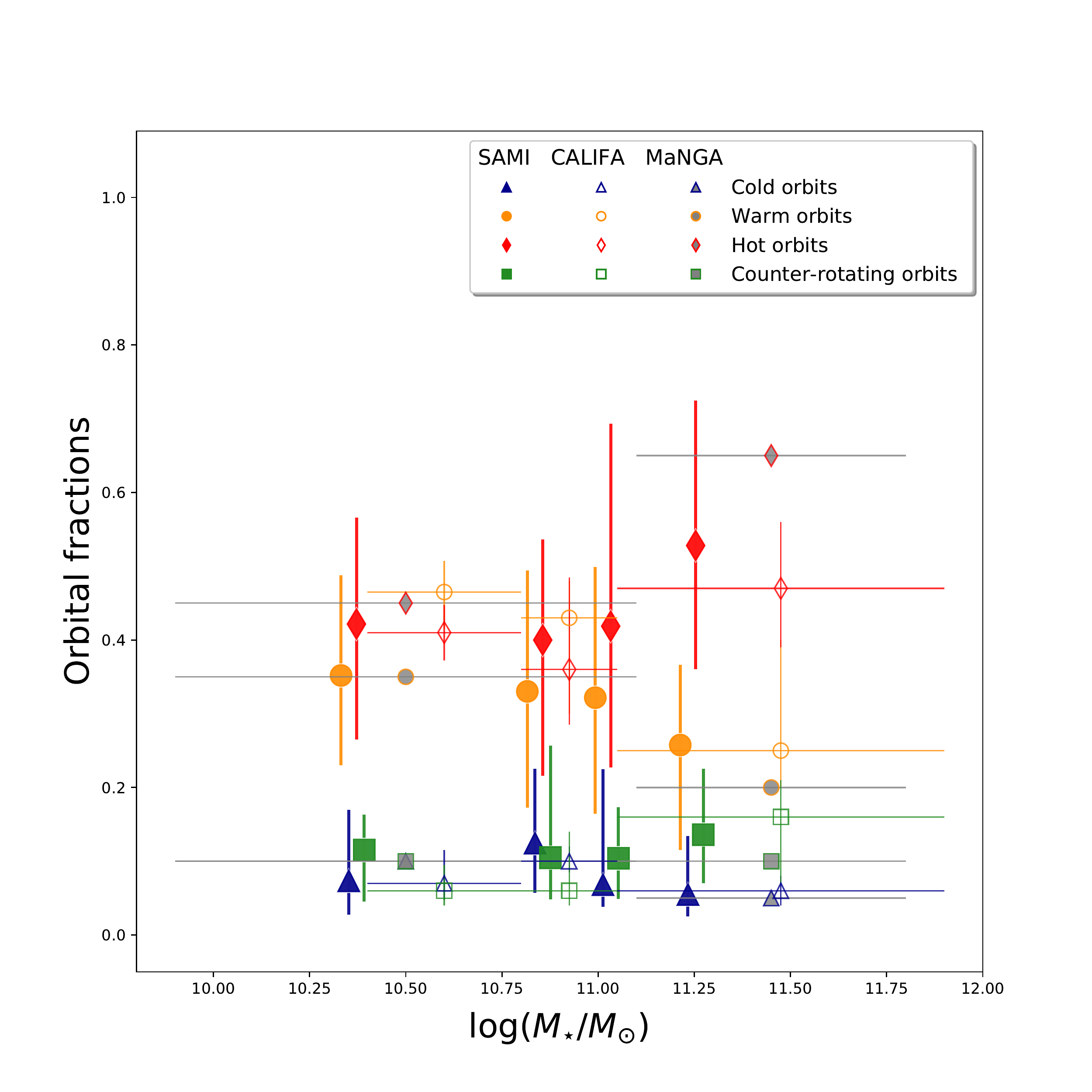}
\caption{Median values of the fractions of orbital components as a function of stellar mass. The cold component is shown in blue, the warm component in orange, hot component in red and counter rotating in green. SAMI passive galaxies are shown as filled points, MaNGA early-type galaxies as open points \citep{Jin2020} and the shaded areas represent the median values of early-type galaxies in the CALIFA sample \citep{Zhu2018b}. Horizontal error bars delimit the mass range covered. Vertical error bars mark the $25^{th}$ and $75^{th}$ percentiles, when available. The distribution of the fractions of orbits in SAMI and MaNGA are similar. All three samples show similar trends of orbital fractions with stellar mass. }
\label{fig:comparison_orbs}
\end{figure}

Stellar orbit distributions have only been derived explicitly before for two large (N$>$100) samples of galaxies, in the CALIFA \citep{Zhu2018b} and MaNGA \citep{Jin2020} surveys. We show the orbital fractions derived for early-type CALIFA and MaNGA galaxies, as well as the results from this work, in Fig. \ref{fig:comparison_orbs}. 
The variations of the fraction of orbits is in good agreement with the general trends with stellar mass seen by \cite{Zhu2018b} and \cite{Jin2020}. \cite{Jin2020} also found an increase in the fraction of hot orbits for massive ($\log (M_{\star}/M_{\odot}) > 11$) galaxies, similar to what we find.

Previous studies that did not have access to stellar orbit modelling, commonly used the proxy for the spin parameter $\lambda_{Re}$, and the flattening of galaxies, to shed light on galaxy intrinsic properties. Schwarzschild dynamical models allow us to explain the trends in $\lambda_{Re}$ by showing the contributions from different orbital components, providing a new insight into how $\lambda_{Re}$ is built-up. We measured the edge-on $\lambda_{Re, EO}$ from our model fits and compared it to the orbital fractions, shown in Fig. \ref{fig:orb_lambda}. 
We find a clear trend of the fractions of orbits with $\lambda_{Re, EO}$: hot orbits show a rapid decrease in fraction with increasing $\lambda_{Re, EO}$, while warm orbits have the opposite behaviour (increasing rapidly with increasing $\lambda_{Re, EO}$). Counter-rotating orbits have slightly lower fractions for galaxies with higher spin parameter, while cold orbits show low fractions up to $\lambda_{Re, EO} \approx 0.3$, after which their importance starts to increase. This confirms that $\lambda_{Re, EO}$ is a good indicator of the underlying orbit distribution of a galaxy.
The observed spin parameter used in the literature \citep[e.g.][]{Cappellari2007, Emsellem2007, Emsellem2011, vandeSande2017a} is a projected quantity along an often-unknown line-of-sight viewing angle. Slow rotators are found to be more massive, dominating above $2 \times 10^{11} M_{\odot}$ \citep[e.g.][]{Emsellem2011,Cappellari2016, Brough2017, Veale2017, Greene2017, vandeSande2017b,vandeSande2021a}. This is in agreement with our more direct orbit-based finding of an increase of the hot component with increasing galaxy mass and the hot component starting to be dominating for galaxies with $\log (M_{\star}/M_{\odot}) > 10.75$.

\subsection{Implications for galaxy formation}
While the degeneracy due to deprojection impacts the reliability of the recovered shape \citep{Rybicki1987,Krajnovic2005,deNicola2020}, the Schwarzschild orbit-superposition method is still the best method that exists to derive the true three-dimensional structure of individual galaxies. In this paper we find that the changes of internal structures within 1$R_{\rm e}$ are correlated with the total stellar mass of individual galaxies. 
In particular, we find a rapid change in structure for galaxies above a stellar mass $\log (M_{\star}/M_{\odot}) \sim 11$. Below this stellar mass, galaxies tend to be oblate and with a substantial number of stars within $R_{\rm e}$ on warm orbits, while higher-mass galaxies with $\log (M_{\star}/M_{\odot}) > 11$ tend to be more triaxial and dominated by hot orbits. A similar change is also seen in the fraction of dark matter (Fig \ref{fig:f_dm}). The change in the hot and warm orbital fractions that we observe in Fig. \ref{fig:fracs_mass} at stellar masses higher than $\sim 10^{11} M_{\star}$ and the change in intrinsic shape at similar mass that we see in Fig. \ref{fig:Te_mass} could be interpreted as an indication of different formation channels. In particular, major and minor mergers are found to be the main driver of triaxial and prolate shapes, while exclusively very minor mergers are largely
associated with triaxial systems and oblate slow rotators are formed in the absence of mergers \citep{Lagos2020}. The increasing fractions of hot orbits with increasing stellar mass supports a scenario where the most massive slow rotators form via gas-poor major mergers \citep{Li2018a}.

The trends we observe in the inner parts of passive galaxies (within 1$R_{\rm e}$) are generally consistent with the two formation paths of early-type galaxies \citep[e.g. see recent review by][]{Cappellari2016}. In this picture slow-rotating ETGs assemble near the center of massive dark matter halos via intense star formation at high redshift, and their evolution is dominated by gas-poor mergers. These galaxies are more likely to be triaxial and more massive, in agreement with what we find. By comparison, low-mass fast-rotating ETGs grow via gas accretion and their structures show similarities with that of spiral galaxies. Moreover, since the warm component can be interpreted as being similar to a thick disk, the increasing contribution that we see from warm orbits in fast-rotating galaxies provides further evidence for disk-like components in these systems as indicated by \cite{Krajnovic2008}.

Simulations suggest that stars on different orbits have different formation paths. The cold components are mostly young stars formed in-situ, the warm component likely traces old stars formed in-situ, or stars being heated from cold disks via secular evolution, and a small fraction of the warm component stars could be accreted \citep{Gomez2017,Park2021}. The stars on hot orbits in the outer regions should mostly be accreted \citep{Gomez2017, Tissera2017} via minor or major mergers, while stars on hot orbits in the inner regions are predicted to have formed at high-redshift. Further comparison with simulations will help us to understand the physical processes that lead to the orbit distribution observed at present times.

\subsubsection{Evidence of early accretion from stellar populations}

Resolved stellar dynamics trace the change in angular momentum and orbital distribution of stars due to mergers, but major mergers are likely to have obscured the effects of earlier interactions. However, evidence of these earlier interactions can be found in the stellar populations. In particular, a galaxy's mean stellar age provides information on when the stars were formed \citep[e.g.][]{Tinsley1980,Bender1993,Park2021}.
So combining stellar population and stellar kinematic studies can provide unique but complementary insights into how galaxies build-up their stellar mass and angular momentum. 

\cite{vandeSande2018} studied a sample of galaxies in the SAMI Galaxy Survey and found that there is a strong relation between $V/\sigma_{Re}$ and mean stellar age, such that galaxies with young stellar populations are predominantly rotationally supported, whereas galaxies with old stellar populations are more pressure supported by random orbital motion of stars. For the large majority of galaxies that are oblate-rotating spheroids, they found that characteristic stellar age is related to the intrinsic ellipticity of galaxies. They studied a full range of morphologies, but showed that this trend is still observed when galaxies are in early-type or late-type subsamples.

\begin{figure}[!ht]
\centering
\includegraphics[scale=0.6, trim=0cm 0cm 1.5cm 1cm, clip=True]{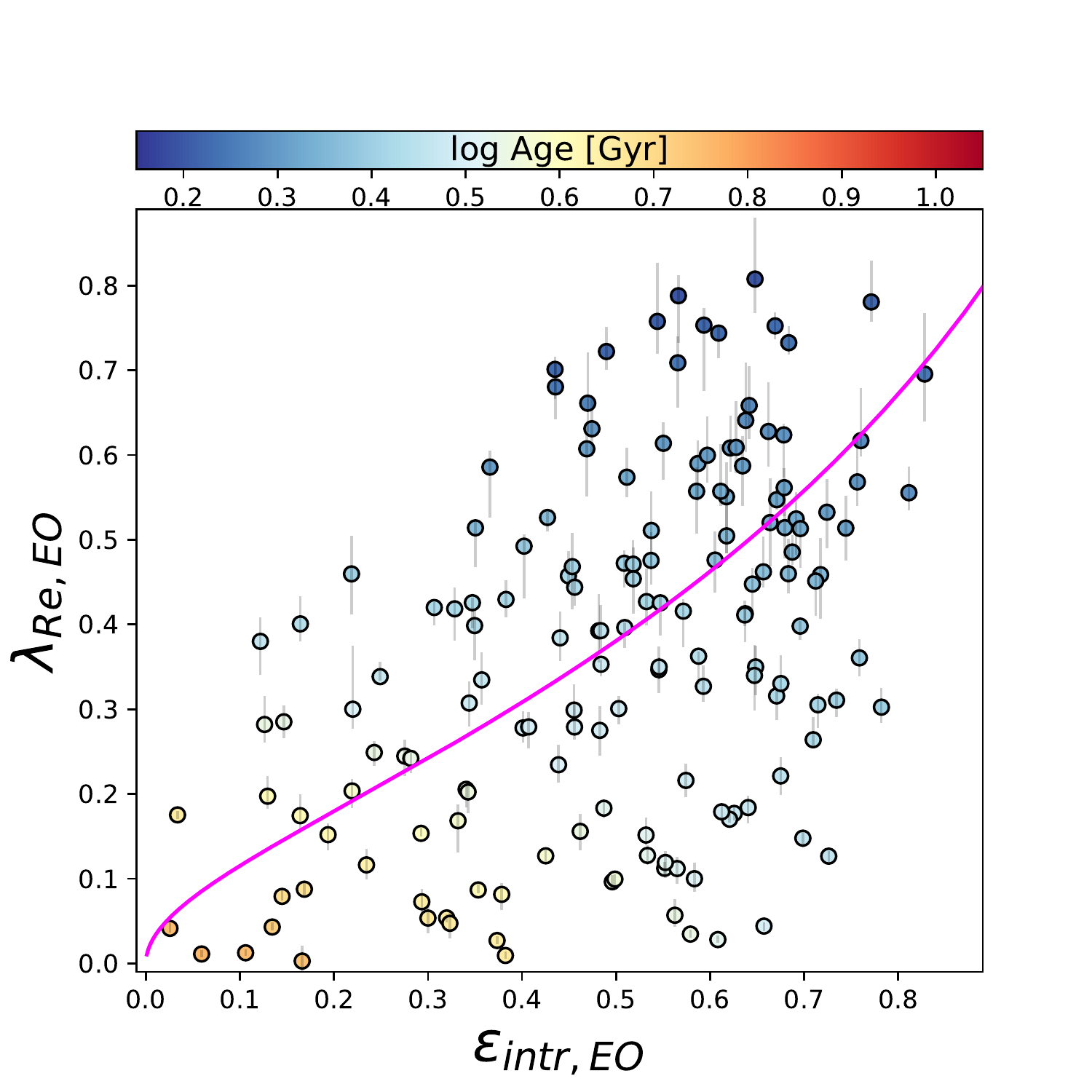}
\caption{$\lambda_{Re, EO}$ as a function of the intrinsic ellipticity $\varepsilon_{intr, EO}$, calculated at 1$R_{\rm e}$. The magenta line represents the relation between the anisotropy parameter $\beta_z$ and the intrinsic ellipticity $\varepsilon_{intr, EO}$, for galaxies viewed edge-on, that bounds all regular rotating galaxies. Galaxies are color-coded by $\log$ Age, and LOESS smoothed to recover any mean underlying trend. Older galaxies are generally slow-rotating and rounder than younger systems.}
\label{fig:lambda_r_age}
\end{figure}

To check whether this relation holds for our parameters derived using \schw models, we color-code our data in the $\lambda_{Re, EO} - \varepsilon_{intr, EO}$ plot by luminosity-weighted, mean stellar population age (see \citealt{Scott2017}) in Fig. \ref{fig:lambda_r_age} and use LOESS smoothing to recover any mean underlying trend. We find a good match to the trends as found by \cite{vandeSande2018}, with slow-rotating galaxies being generally older and rounder than fast-rotating galaxies.
This relationship is consistent with predictions from hydrodynamical cosmological simulations and observations, where slow-rotating galaxies form via intense star formation at high redshift, and evolve from a set of processes dominated by gas-poor mergers \citep{Cappellari2016}.

All the results presented here are in agreement with a formation scenario in which passive galaxies form through two main channels and where the changes of internal structures within 1$R_{\rm e}$ are generally correlated with the total stellar mass of the individual galaxies.

\vspace*{1cm}
\section{Conclusion}
We constructed Schwarzschild orbit-superposition models of 161 passive galaxies, from the SAMI Galaxy Survey, with stellar masses raging from $9.5 < \log (M_{\star}/ M_{\odot}) < 11.4$. We derived intrinsic properties such as the internal mass distribution (for both stellar and dark matter), intrinsic stellar shape (axis ratios and ellipticity), velocity anisotropy and orbit circularity distribution which gives us the most detailed insight into their assembly history.
We draw the following conclusions:
\begin{itemize}
	\item Passive galaxies have an average dark matter fraction $f_{\rm{DM}} = 0.28 \pm 0.20$, consistent with previous results (Fig. \ref{fig:fdm_comp}).
	\item The majority of our galaxies are consistent with having oblate axisymmetry ($73\% \pm 3\%$), with $T_{\rm Re} \leq 0.1$ , with varying degrees of intrinsic flattening, with $19\% \pm 3\%$ being mildly triaxial ($ 0.1< T_{\rm Re} \geq 0.3$) and a small percentage ($8\% \pm 2\%$) being triaxial/prolate ($T_{\rm Re} > 0.3$). The fraction of non-oblate galaxies increases with increasing stellar mass, with a sudden change at $\sim 10^{10.50} M_{\star}/M_{\odot}$ (Fig. \ref{fig:Te_mass}).
	\item Galaxies with high intrinsic ellipticity (flat galaxies, $\varepsilon > 0.7$) are found to be more isotropic ($\beta_r \sim 0$) or more tangentially anisotropic ($\beta_r < 0$; Fig. \ref{fig:beta_r_mass}). $\beta_r$ is anti-correlated with the spin parameter \lam, so that $\beta_r$ decreases with increasing \lam, consistent with slow-rotating galaxies being more radially anisotropic and fast-rotating galaxies being more tangentially anisotropic (Fig. \ref{fig:lambda_r_beta}). 
	\item By dividing the stellar orbital distribution into cold, warm, hot, and counter-rotating components, we find that the hot component generally dominates within $R_{\rm e}$, becoming the most prevalent component among galaxies with total stellar mass $\log (M_{\star}/M_{\odot}) > 11$. In most galaxies a substantial number ($\sim 40\%$ of stars within $R_{\rm e}$) are on warm orbits, with the warm contribution becoming more important at lower stellar masses. The contribution from the cold orbital components is small across stellar mass, with its contribution decreasing further with increasing mass. The counter-rotating component is roughly constant for galaxies at all masses (Fig. \ref{fig:fracs_mass}). 
	\item The changes of internal structures (fraction of dark matter, $f_{\rm DM}$, intrinsic shape and orbital distribution) within 1$R_{\rm e}$ are correlated with the total stellar mass of the individual galaxies.
	\item The fractions of orbits show tight correlations with the intrinsic $\lambda_{Re,EO}$, with hot orbits being dominant for slow-rotating galaxies and contributions from warm and cold orbits becoming more important with increasing $\lambda_{Re,EO}$. We also find a clear distinction between the orbital distributions of galaxies, depending on their kinematic class (from \cite{vandeSande2017a} based on the higher-order $V/\sigma$ - $h_3$ signatures). Class 1 is dominated by hot orbits, with little contribution from other components. The contribution of warm orbits increases from Class 2 to 5, while the contribution from hot orbits become less important. Class 4 and 5 also show contributions from cold and counter rotating components (Fig. \ref{fig:orb_lambda} and \ref{fig:orb_fracs}).
	\item These results are in agreement with a formation scenario in which galaxies form through two main different channels. Slow-rotating ETGs assemble near the center of massive dark matter halos via intense star formation at high redshift, and their evolution is dominated by gas-poor mergers. These galaxies are more likely to be triaxial and more massive, dominated by radial anisotropy, in agreement with what we find. By comparison, low-mass fast-rotating ETGs grow via gas accretion and their structures show similarities with that of spiral galaxies. Moreover, the intrinsic shapes of slow rotators could point to different type of mergers in their evolutionary history.
\end{itemize} 

The work presented here expands on previous analyses by including the higher-order stellar kinematic moments. We found that including the higher-order kinematic moments $h_3$ and $h_4$ can improve the model fits, even if the $h_3$ and $h_4$ measurements have high uncertainties. We therefore recommend the inclusion of $h_3$ and $h_4$ in future works. Moreover, since $h_3$ and $h_4$ are quantities that are predicted to be connected with a galaxy’s assembly history \citep{Naab2014}, studying their relation to the internal orbital structure of galaxies provides an extra tool to help disentangle the different possible formation scenarios. We did not find a significant difference between the orbital components of fast-rotating galaxies in the different high-order classes from \cite{vandeSande2017a} (determined using the V/sigma- h3 correlation) in the $\lambda_z - r$ space (Fig. \ref{fig:orb_lambda}), but this is an interesting aspect that should be further explored in future works with larger samples (e.g. the forthcoming Hector survey; \citealt{Bryant2016}).

\section*{Acknowledgements}
We thank the anonymous referee for their comments that helped to improve this manuscript.
We thank the DYNAMITE team for their invaluable support with the code and optimisation and Michele Cappellari for useful discussions. The SAMI Galaxy Survey is based on observations made at the Anglo-Australian Telescope. The Sydney-AAO Multi-object Integral field spectrograph (SAMI) was developed jointly by the University of Sydney and the Australian Astronomical Observatory. The SAMI input catalogue is based on data taken from the Sloan Digital Sky Survey, the GAMA Survey and the VST ATLAS Survey. The SAMI Galaxy Survey is supported by the Australian Research Council center of Excellence for All Sky Astrophysics in 3 Dimensions (ASTRO 3D), through project number CE170100013, the Australian Research Council center of Excellence for All-sky Astrophysics (CAASTRO), through project number CE110001020, and other participating institutions. The SAMI Galaxy Survey website is \href{http://sami-survey.org/}{http://sami-survey.org/}.\\
\\
This research is supported by an Australian Government Research Training Program (RTP) Scholarship.
SB acknowledges funding support from the Australian Research Council
through a Future Fellowship (FT140101166).
JvdS acknowledges support of an Australian Research Council Discovery Early Career Research Award (project number DE200100461) funded by the Australian Government.
RMcD acknowledges funding support via an Australian Research Council Future Fellowship (project number FT150100333).
GvdV acknowledges funding from the European Research Council (ERC) under the European Union's Horizon 2020 research and innovation programme under grant agreement No 724857 (Consolidator Grant ArcheoDyn).
LZ acknowledges the support from National Natural Science Foundation of China under grant No. Y945271001.
FDE acknowledges funding through the ERC Advanced grant 695671 “QUENCH”, the H2020 ERC Consolidator Grant 683184 and support by the Science and Technology Facilities Council (STFC).
JBH is supported by an ARC Laureate Fellowship FL140100278. The SAMI instrument was funded by Bland-Hawthorn's former Federation Fellowship FF0776384, an ARC LIEF grant LE130100198 (PI Bland-Hawthorn) and funding from the Anglo-Australian Observatory.
JJB acknowledges support of an Australian Research Council Future Fellowship (FT180100231).
M.S.O. acknowledges the funding support from the Australian Research Council through a Future Fellowship (FT140100255).
S.K.Y. acknowledges support from the Korean National Research Foundation (NRF-2020R1A2C3003769).

\software{pPXF \citep{Cappellari2004, Cappellari2017},  MgeFit \citep{Cappellari2002}, Voronoi binning code \citep{Cappellari2003}, Scipy \citep{2020SciPy-NMeth}, DYNAMITE \citep{DYN2020},  UNSW Katana computational cluster \citep{Katana2010}.}
\bibliography{bib_schw}

\begin{thebibliography}{}
\expandafter\ifx\csname natexlab\endcsname\relax\def\natexlab#1{#1}\fi
\providecommand{\url}[1]{\href{#1}{#1}}
\providecommand{\dodoi}[1]{doi:~\href{http://doi.org/#1}{\nolinkurl{#1}}}
\providecommand{\doeprint}[1]{\href{http://ascl.net/#1}{\nolinkurl{http://ascl.net/#1}}}
\providecommand{\doarXiv}[1]{\href{https://arxiv.org/abs/#1}{\nolinkurl{https://arxiv.org/abs/#1}}}

\bibitem[{{Adelman-McCarthy} {et~al.}(2006){Adelman-McCarthy}, {Ag{\"u}eros},
  {Allam}, {Anderson}, {Anderson}, {Annis}, {Bahcall}, {Baldry}, {Barentine},
  {Berlind}, {Bernardi}, {Blanton}, {Boroski}, {Brewington}, {Brinchmann},
  {Brinkmann}, {Brunner}, {Budav{\'a}ri}, {Carey}, {Carr}, {Castander},
  {Connolly}, {Csabai}, {Czarapata}, {Dalcanton}, {Doi}, {Dong}, {Eisenstein},
  {Evans}, {Fan}, {Finkbeiner}, {Friedman}, {Frieman}, {Fukugita}, {Gillespie},
  {Glazebrook}, {Gray}, {Grebel}, {Gunn}, {Gurbani}, {de Haas}, {Hall},
  {Harris}, {Harvanek}, {Hawley}, {Hayes}, {Hendry}, {Hennessy}, {Hindsley},
  {Hirata}, {Hogan}, {Hogg}, {Holmgren}, {Holtzman}, {Ichikawa}, {Ivezi{\'c}},
  {Jester}, {Johnston}, {Jorgensen}, {Juri{\'c}}, {Kent}, {Kleinman}, {Knapp},
  {Kniazev}, {Kron}, {Krzesinski}, {Kuropatkin}, {Lamb}, {Lampeitl}, {Lee},
  {Leger}, {Lin}, {Long}, {Loveday}, {Lupton}, {Margon},
  {Mart{\'\i}nez-Delgado}, {Mandelbaum}, {Matsubara}, {McGehee}, {McKay},
  {Meiksin}, {Munn}, {Nakajima}, {Nash}, {Neilsen}, {Newberg}, {Newman},
  {Nichol}, {Nicinski}, {Nieto-Santisteban}, {Nitta}, {O'Mullane}, {Okamura},
  {Owen}, {Padmanabhan}, {Pauls}, {Peoples}, {Pier}, {Pope}, {Pourbaix},
  {Quinn}, {Richards}, {Richmond}, {Rockosi}, {Schlegel}, {Schneider},
  {Schroeder}, {Scranton}, {Seljak}, {Sheldon}, {Shimasaku}, {Smith},
  {Smol{\v{c}}i{\'c}}, {Snedden}, {Stoughton}, {Strauss}, {SubbaRao}, {Szalay},
  {Szapudi}, {Szkody}, {Tegmark}, {Thakar}, {Tucker}, {Uomoto}, {Vanden Berk},
  {Vandenberg}, {Vogeley}, {Voges}, {Vogt}, {Walkowicz}, {Weinberg}, {West},
  {White}, {Xu}, {Yanny}, {Yocum}, {York}, {Zehavi}, {Zibetti}, \&
  {Zucker}}]{Adelman-McCarthy2006}
{Adelman-McCarthy}, J.~K., {Ag{\"u}eros}, M.~A., {Allam}, S.~S., {et~al.} 2006,
  \apjs, 162, 38, \dodoi{10.1086/497917}

\bibitem[{{Ahn} {et~al.}(2012){Ahn}, {Alexandroff}, {Allende Prieto},
  {Anderson}, {Anderton}, {Andrews}, {Aubourg}, {Bailey}, {Balbinot}, {Barnes},
  \& et~al.}]{Ahn2012}
{Ahn}, C.~P., {Alexandroff}, R., {Allende Prieto}, C., {et~al.} 2012, \apjs,
  203, 21, \dodoi{10.1088/0067-0049/203/2/21}

\bibitem[{{Aquino-Ort{\'\i}z} {et~al.}(2020){Aquino-Ort{\'\i}z}, {S{\'a}nchez},
  {Valenzuela}, {Hern{\'a}ndez-Toledo}, {Jin}, {Zhu}, {van de Ven},
  {Barrera-Ballesteros}, {Avila-Reese}, {Rodr{\'\i}guez-Puebla}, \&
  {Tissera}}]{Aquino-Ortiz2020}
{Aquino-Ort{\'\i}z}, E., {S{\'a}nchez}, S.~F., {Valenzuela}, O., {et~al.} 2020,
  arXiv e-prints, arXiv:2005.09149.
\newblock \doarXiv{2005.09149}

\bibitem[{Barsanti {et~al.}(2021)Barsanti, Owers, McDermid, Bekki, Bryant,
  Croom, Oh, Robotham, Scott, \& van~de Sande}]{barsanti2021colors}
Barsanti, S., Owers, M.~S., McDermid, R.~M., {et~al.} 2021, The colors of
  bulges and disks in the core and outskirts of galaxy clusters.
\newblock \doarXiv{2012.12480}

\bibitem[{{Behroozi} {et~al.}(2010){Behroozi}, {Conroy}, \&
  {Wechsler}}]{Behroozi2010}
{Behroozi}, P.~S., {Conroy}, C., \& {Wechsler}, R.~H. 2010, \apj, 717, 379,
  \dodoi{10.1088/0004-637X/717/1/379}

\bibitem[{{Behroozi} {et~al.}(2013){Behroozi}, {Wechsler}, \&
  {Conroy}}]{Behroozi2013}
{Behroozi}, P.~S., {Wechsler}, R.~H., \& {Conroy}, C. 2013, \apj, 770, 57,
  \dodoi{10.1088/0004-637X/770/1/57}

\bibitem[{{Bender}(1988)}]{Bender1988}
{Bender}, R. 1988, \aap, 202, L5

\bibitem[{{Bender} {et~al.}(1993){Bender}, {Burstein}, \& {Faber}}]{Bender1993}
{Bender}, R., {Burstein}, D., \& {Faber}, S.~M. 1993, \apj, 411, 153,
  \dodoi{10.1086/172815}

\bibitem[{{Bertola} \& {Capaccioli}(1977)}]{Bertola1977}
{Bertola}, F., \& {Capaccioli}, M. 1977, \apj, 211, 697, \dodoi{10.1086/154980}

\bibitem[{{Binney}(1978)}]{Binney1978}
{Binney}, J. 1978, \mnras, 183, 501, \dodoi{10.1093/mnras/183.3.501}

\bibitem[{{Binney} \& {Tremaine}(2008)}]{Binney2008}
{Binney}, J., \& {Tremaine}, S. 2008, {Galactic Dynamics: Second Edition}

\bibitem[{{Binney} {et~al.}(1990){Binney}, {Davies}, \&
  {Illingworth}}]{Binney1990}
{Binney}, J.~J., {Davies}, R.~L., \& {Illingworth}, G.~D. 1990, \apj, 361, 78,
  \dodoi{10.1086/169169}

\bibitem[{{Bland-Hawthorn} \& {Gerhard}(2016)}]{Bland2016}
{Bland-Hawthorn}, J., \& {Gerhard}, O. 2016, \araa, 54, 529,
  \dodoi{10.1146/annurev-astro-081915-023441}

\bibitem[{{Bland-Hawthorn} {et~al.}(2011){Bland-Hawthorn}, {Bryant},
  {Robertson}, {Gillingham}, {O'Byrne}, {Cecil}, {Haynes}, {Croom}, {Ellis},
  {Maack}, {Skovgaard}, \& {Noordegraaf}}]{Bland-Hawthorn2011}
{Bland-Hawthorn}, J., {Bryant}, J., {Robertson}, G., {et~al.} 2011, Optics
  Express, 19, 2649, \dodoi{10.1364/OE.19.002649}

\bibitem[{{Breddels} \& {Helmi}(2014)}]{Breddels2014}
{Breddels}, M.~A., \& {Helmi}, A. 2014, \apjl, 791, L3,
  \dodoi{10.1088/2041-8205/791/1/L3}

\bibitem[{{Brough} {et~al.}(2017){Brough}, {van de Sande}, {Owers},
  {d'Eugenio}, {Sharp}, {Cortese}, {Scott}, {Croom}, {Bassett}, {Bekki},
  {Bland-Hawthorn}, {Bryant}, {Davies}, {Drinkwater}, {Driver}, {Foster},
  {Goldstein}, {L{\'o}pez-S{\'a}nchez}, {Medling}, {Sweet}, {Taranu}, {Tonini},
  {Yi}, {Goodwin}, {Lawrence}, \& {Richards}}]{Brough2017}
{Brough}, S., {van de Sande}, J., {Owers}, M.~S., {et~al.} 2017, \apj, 844, 59,
  \dodoi{10.3847/1538-4357/aa7a11}

\bibitem[{{Bryant} {et~al.}(2014){Bryant}, {Bland-Hawthorn}, {Fogarty},
  {Lawrence}, \& {Croom}}]{Bryant2014}
{Bryant}, J.~J., {Bland-Hawthorn}, J., {Fogarty}, L.~M.~R., {Lawrence}, J.~S.,
  \& {Croom}, S.~M. 2014, \mnras, 438, 869, \dodoi{10.1093/mnras/stt2254}

\bibitem[{{Bryant} {et~al.}(2015){Bryant}, {Owers}, {Robotham}, {Croom},
  {Driver}, {Drinkwater}, {Lorente}, {Cortese}, {Scott}, {Colless}, {Schaefer},
  {Taylor}, {Konstantopoulos}, {Allen}, {Baldry}, {Barnes}, {Bauer},
  {Bland-Hawthorn}, {Bloom}, {Brooks}, {Brough}, {Cecil}, {Couch}, {Croton},
  {Davies}, {Ellis}, {Fogarty}, {Foster}, {Glazebrook}, {Goodwin}, {Green},
  {Gunawardhana}, {Hampton}, {Ho}, {Hopkins}, {Kewley}, {Lawrence},
  {Leon-Saval}, {Leslie}, {McElroy}, {Lewis}, {Liske}, {L{\'o}pez-S{\'a}nchez},
  {Mahajan}, {Medling}, {Metcalfe}, {Meyer}, {Mould}, {Obreschkow}, {O'Toole},
  {Pracy}, {Richards}, {Shanks}, {Sharp}, {Sweet}, {Thomas}, {Tonini}, \&
  {Walcher}}]{Bryant2015}
{Bryant}, J.~J., {Owers}, M.~S., {Robotham}, A.~S.~G., {et~al.} 2015, \mnras,
  447, 2857, \dodoi{10.1093/mnras/stu2635}

\bibitem[{{Bryant} {et~al.}(2016){Bryant}, {Bland-Hawthorn}, {Lawrence},
  {Croom}, {Brown}, {Venkatesan}, {Gillingham}, {Zhelem}, {Content},
  {Saunders}, {Staszak}, {van de Sande}, {Couch}, {Leon-Saval}, {Tims},
  {McDermid}, \& {Schaefer}}]{Bryant2016}
{Bryant}, J.~J., {Bland-Hawthorn}, J., {Lawrence}, J., {et~al.} 2016, in
  Society of Photo-Optical Instrumentation Engineers (SPIE) Conference Series,
  Vol. 9908, Ground-based and Airborne Instrumentation for Astronomy VI, ed.
  C.~J. {Evans}, L.~{Simard}, \& H.~{Takami}, 99081F,
  \dodoi{10.1117/12.2230740}

\bibitem[{{B{\"u}denbender} {et~al.}(2015){B{\"u}denbender}, {van de Ven}, \&
  {Watkins}}]{Budenbender2015}
{B{\"u}denbender}, A., {van de Ven}, G., \& {Watkins}, L.~L. 2015, \mnras, 452,
  956, \dodoi{10.1093/mnras/stv1314}

\bibitem[{{Bundy} {et~al.}(2015){Bundy}, {Bershady}, {Law}, {Yan}, {Drory},
  {MacDonald}, {Wake}, {Cherinka}, {S{\'a}nchez-Gallego}, {Weijmans}, {Thomas},
  {Tremonti}, {Masters}, {Coccato}, {Diamond-Stanic}, {Arag{\'o}n-Salamanca},
  {Avila-Reese}, {Badenes}, {Falc{\'o}n-Barroso}, {Belfiore}, {Bizyaev},
  {Blanc}, {Bland-Hawthorn}, {Blanton}, {Brownstein}, {Byler}, {Cappellari},
  {Conroy}, {Dutton}, {Emsellem}, {Etherington}, {Frinchaboy}, {Fu}, {Gunn},
  {Harding}, {Johnston}, {Kauffmann}, {Kinemuchi}, {Klaene}, {Knapen},
  {Leauthaud}, {Li}, {Lin}, {Maiolino}, {Malanushenko}, {Malanushenko}, {Mao},
  {Maraston}, {McDermid}, {Merrifield}, {Nichol}, {Oravetz}, {Pan}, {Parejko},
  {Sanchez}, {Schlegel}, {Simmons}, {Steele}, {Steinmetz}, {Thanjavur},
  {Thompson}, {Tinker}, {van den Bosch}, {Westfall}, {Wilkinson}, {Wright},
  {Xiao}, \& {Zhang}}]{Bundy2015}
{Bundy}, K., {Bershady}, M.~A., {Law}, D.~R., {et~al.} 2015, \apj, 798, 7,
  \dodoi{10.1088/0004-637X/798/1/7}

\bibitem[{{Cappellari}(2002)}]{Cappellari2002}
{Cappellari}, M. 2002, \mnras, 333, 400,
  \dodoi{10.1046/j.1365-8711.2002.05412.x}

\bibitem[{{Cappellari}(2016)}]{Cappellari2016}
---. 2016, \araa, 54, 597, \dodoi{10.1146/annurev-astro-082214-122432}

\bibitem[{{Cappellari}(2017)}]{Cappellari2017}
---. 2017, \mnras, 466, 798, \dodoi{10.1093/mnras/stw3020}

\bibitem[{{Cappellari} \& {Copin}(2003)}]{Cappellari2003}
{Cappellari}, M., \& {Copin}, Y. 2003, \mnras, 342, 345,
  \dodoi{10.1046/j.1365-8711.2003.06541.x}

\bibitem[{{Cappellari} \& {Emsellem}(2004)}]{Cappellari2004}
{Cappellari}, M., \& {Emsellem}, E. 2004, \pasp, 116, 138,
  \dodoi{10.1086/381875}

\bibitem[{{Cappellari} {et~al.}(2006){Cappellari}, {Bacon}, {Bureau}, {Damen},
  {Davies}, {de Zeeuw}, {Emsellem}, {Falc{\'o}n-Barroso}, {Krajnovi{\'c}},
  {Kuntschner}, {McDermid}, {Peletier}, {Sarzi}, {van den Bosch}, \& {van de
  Ven}}]{Cappellari2006}
{Cappellari}, M., {Bacon}, R., {Bureau}, M., {et~al.} 2006, \mnras, 366, 1126,
  \dodoi{10.1111/j.1365-2966.2005.09981.x}

\bibitem[{{Cappellari} {et~al.}(2007){Cappellari}, {Emsellem}, {Bacon},
  {Bureau}, {Davies}, {de Zeeuw}, {Falc{\'o}n-Barroso}, {Krajnovi{\'c}},
  {Kuntschner}, {McDermid}, {Peletier}, {Sarzi}, {van den Bosch}, \& {van de
  Ven}}]{Cappellari2007}
{Cappellari}, M., {Emsellem}, E., {Bacon}, R., {et~al.} 2007, \mnras, 379, 418,
  \dodoi{10.1111/j.1365-2966.2007.11963.x}

\bibitem[{{Cappellari} {et~al.}(2011){Cappellari}, {Emsellem}, {Krajnovi{\'c}},
  {McDermid}, {Scott}, {Verdoes Kleijn}, {Young}, {Alatalo}, {Bacon}, {Blitz},
  {Bois}, {Bournaud}, {Bureau}, {Davies}, {Davis}, {de Zeeuw}, {Duc},
  {Khochfar}, {Kuntschner}, {Lablanche}, {Morganti}, {Naab}, {Oosterloo},
  {Sarzi}, {Serra}, \& {Weijmans}}]{Cappellari2011}
{Cappellari}, M., {Emsellem}, E., {Krajnovi{\'c}}, D., {et~al.} 2011, \mnras,
  413, 813, \dodoi{10.1111/j.1365-2966.2010.18174.x}

\bibitem[{{Cappellari} {et~al.}(2013){Cappellari}, {McDermid}, {Alatalo},
  {Blitz}, {Bois}, {Bournaud}, {Bureau}, {Crocker}, {Davies}, {Davis}, {de
  Zeeuw}, {Duc}, {Emsellem}, {Khochfar}, {Krajnovi{\'c}}, {Kuntschner},
  {Morganti}, {Naab}, {Oosterloo}, {Sarzi}, {Scott}, {Serra}, {Weijmans}, \&
  {Young}}]{Cappellari2013a}
{Cappellari}, M., {McDermid}, R.~M., {Alatalo}, K., {et~al.} 2013, \mnras, 432,
  1862, \dodoi{10.1093/mnras/stt644}

\bibitem[{{Chabrier}(2003)}]{Chabrier2003}
{Chabrier}, G. 2003, \apjl, 586, L133, \dodoi{10.1086/374879}

\bibitem[{{Chandrasekhar}(1969)}]{Chandrasekhar1969}
{Chandrasekhar}, S. 1969, {Ellipsoidal figures of equilibrium}

\bibitem[{{Cortese} {et~al.}(2016){Cortese}, {Fogarty}, {Bekki}, {van de Sand
  e}, {Couch}, {Catinella}, {Colless}, {Obreschkow}, {Taranu}, \&
  {Tescari}}]{Cortese2016}
{Cortese}, L., {Fogarty}, L.~M.~R., {Bekki}, K., {et~al.} 2016, \mnras, 463,
  170, \dodoi{10.1093/mnras/stw1891}

\bibitem[{{Cretton} {et~al.}(1999){Cretton}, {de Zeeuw}, {van der Marel}, \&
  {Rix}}]{Cretton1999}
{Cretton}, N., {de Zeeuw}, P.~T., {van der Marel}, R.~P., \& {Rix}, H.-W. 1999,
  \apjs, 124, 383, \dodoi{10.1086/313264}

\bibitem[{{Croom} {et~al.}(2012){Croom}, {Lawrence}, {Bland-Hawthorn},
  {Bryant}, {Fogarty}, {Richards}, {Goodwin}, {Farrell}, {Miziarski}, {Heald},
  {Jones}, {Lee}, {Colless}, {Brough}, {Hopkins}, {Bauer}, {Birchall}, {Ellis},
  {Horton}, {Leon-Saval}, {Lewis}, {L{\'o}pez-S{\'a}nchez}, {Min}, {Trinh}, \&
  {Trowland}}]{Croom2012}
{Croom}, S.~M., {Lawrence}, J.~S., {Bland-Hawthorn}, J., {et~al.} 2012, Monthly
  Notices of the Royal Astronomical Society, 421, 872,
  \dodoi{10.1111/j.1365-2966.2011.20365.x}

\bibitem[{{Croom} {et~al.}(2021){Croom}, {Owers}, {Scott}, {Poetrodjojo},
  {Groves}, {van de Sande}, {Barone}, {Cortese}, {D'Eugenio}, {Bland-Hawthorn},
  {Bryant}, {Oh}, {Brough}, {Agostino}, {Casura}, {Catinella}, {Colless},
  {Cecil}, {Davies}, {Drinkwater}, {Driver}, {Ferreras}, {Foster},
  {Fraser-McKelvie}, {Lawrence}, {Leslie}, {Liske}, {L{\'o}pez-S{\'a}nchez},
  {Lorente}, {McElroy}, {Medling}, {Obreschkow}, {Richards}, {Sharp}, {Sweet},
  {Taranu}, {Taylor}, {Tescari}, {Thomas}, {Tocknell}, \&
  {Vaughan}}]{Croom2021}
{Croom}, S.~M., {Owers}, M.~S., {Scott}, N., {et~al.} 2021, \mnras,
  \dodoi{10.1093/mnras/stab229}

\bibitem[{{de Nicola} {et~al.}(2020){de Nicola}, {Saglia}, {Thomas}, {Dehnen},
  \& {Bender}}]{deNicola2020}
{de Nicola}, S., {Saglia}, R.~P., {Thomas}, J., {Dehnen}, W., \& {Bender}, R.
  2020, \mnras, 496, 3076, \dodoi{10.1093/mnras/staa1703}

\bibitem[{{de Zeeuw} {et~al.}(2002){de Zeeuw}, {Bureau}, {Emsellem}, {Bacon},
  {Carollo}, {Copin}, {Davies}, {Kuntschner}, {Miller}, {Monnet}, {Peletier},
  \& {Verolme}}]{deZeeuw2002}
{de Zeeuw}, P.~T., {Bureau}, M., {Emsellem}, E., {et~al.} 2002, \mnras, 329,
  513, \dodoi{10.1046/j.1365-8711.2002.05059.x}

\bibitem[{{del Moral-Castro} {et~al.}(2020){del Moral-Castro},
  {Garc{\'\i}a-Lorenzo}, {Ramos Almeida}, {Ruiz-Lara}, {Falc{\'o}n-Barroso},
  {S{\'a}nchez}, {S{\'a}nchez-Bl{\'a}zquez}, {M{\'a}rquez}, \&
  {Masegosa}}]{delMoral2020}
{del Moral-Castro}, I., {Garc{\'\i}a-Lorenzo}, B., {Ramos Almeida}, C.,
  {et~al.} 2020, \aap, 639, L9, \dodoi{10.1051/0004-6361/202038091}

\bibitem[{{den Brok} {et~al.}(2021){den Brok}, {Krajnovi{\'c}}, {Emsellem},
  {Brinchmann}, \& {Maseda}}]{denBrok2021}
{den Brok}, M., {Krajnovi{\'c}}, D., {Emsellem}, E., {Brinchmann}, J., \&
  {Maseda}, M. 2021, arXiv e-prints, arXiv:2109.14640.
\newblock \doarXiv{2109.14640}

\bibitem[{{D'Eugenio} {et~al.}(2021){D'Eugenio}, {Colless}, {Scott}, {van der
  Wel}, {Davies}, {van de Sande}, {Sweet}, {Oh}, {Groves}, {Sharp}, {Owers},
  {Bland-Hawthorn}, {Croom}, {Brough}, {Bryant}, {Goodwin}, {Lawrence},
  {Lorente}, \& {Richards}}]{Deugenio2021}
{D'Eugenio}, F., {Colless}, M., {Scott}, N., {et~al.} 2021, \mnras,
  \dodoi{10.1093/mnras/stab1146}

\bibitem[{{Driver} {et~al.}(2011){Driver}, {Hill}, {Kelvin}, {Robotham},
  {Liske}, {Norberg}, {Baldry}, {Bamford}, {Hopkins}, {Loveday}, {Peacock},
  {Andrae}, {Bland-Hawthorn}, {Brough}, {Brown}, {Cameron}, {Ching}, {Colless},
  {Conselice}, {Croom}, {Cross}, {de Propris}, {Dye}, {Drinkwater}, {Ellis},
  {Graham}, {Grootes}, {Gunawardhana}, {Jones}, {van Kampen}, {Maraston},
  {Nichol}, {Parkinson}, {Phillipps}, {Pimbblet}, {Popescu}, {Prescott},
  {Roseboom}, {Sadler}, {Sansom}, {Sharp}, {Smith}, {Taylor}, {Thomas},
  {Tuffs}, {Wijesinghe}, {Dunne}, {Frenk}, {Jarvis}, {Madore}, {Meyer},
  {Seibert}, {Staveley-Smith}, {Sutherland}, \& {Warren}}]{Driver2011}
{Driver}, S.~P., {Hill}, D.~T., {Kelvin}, L.~S., {et~al.} 2011, \mnras, 413,
  971, \dodoi{10.1111/j.1365-2966.2010.18188.x}

\bibitem[{{Emsellem} {et~al.}(1994){Emsellem}, {Monnet}, \&
  {Bacon}}]{Emsellem1994}
{Emsellem}, E., {Monnet}, G., \& {Bacon}, R. 1994, \aap, 285, 723

\bibitem[{{Emsellem} {et~al.}(2004){Emsellem}, {Cappellari}, {Peletier},
  {McDermid}, {Bacon}, {Bureau}, {Copin}, {Davies}, {Krajnovi{\'c}},
  {Kuntschner}, {Miller}, \& {de Zeeuw}}]{Emsellem2004}
{Emsellem}, E., {Cappellari}, M., {Peletier}, R.~F., {et~al.} 2004, \mnras,
  352, 721, \dodoi{10.1111/j.1365-2966.2004.07948.x}

\bibitem[{{Emsellem} {et~al.}(2007){Emsellem}, {Cappellari}, {Krajnovi{\'c}},
  {van de Ven}, {Bacon}, {Bureau}, {Davies}, {de Zeeuw}, {Falc{\'o}n-Barroso},
  {Kuntschner}, {McDermid}, {Peletier}, \& {Sarzi}}]{Emsellem2007}
{Emsellem}, E., {Cappellari}, M., {Krajnovi{\'c}}, D., {et~al.} 2007, \mnras,
  379, 401, \dodoi{10.1111/j.1365-2966.2007.11752.x}

\bibitem[{{Emsellem} {et~al.}(2011){Emsellem}, {Cappellari}, {Krajnovi{\'c}},
  {Alatalo}, {Blitz}, {Bois}, {Bournaud}, {Bureau}, {Davies}, {Davis}, {de
  Zeeuw}, {Khochfar}, {Kuntschner}, {Lablanche}, {McDermid}, {Morganti},
  {Naab}, {Oosterloo}, {Sarzi}, {Scott}, {Serra}, {van de Ven}, {Weijmans}, \&
  {Young}}]{Emsellem2011}
---. 2011, \mnras, 414, 888, \dodoi{10.1111/j.1365-2966.2011.18496.x}

\bibitem[{{Fahrion} {et~al.}(2019){Fahrion}, {Lyubenova}, {van de Ven},
  {Leaman}, {Hilker}, {Mart{\'\i}n-Navarro}, {Zhu}, {Alfaro-Cuello}, {Coccato},
  {Corsini}, {Falc{\'o}n-Barroso}, {Iodice}, {McDermid}, {Sarzi}, \& {de
  Zeeuw}}]{Fahrion2019}
{Fahrion}, K., {Lyubenova}, M., {van de Ven}, G., {et~al.} 2019, \aap, 628,
  A92, \dodoi{10.1051/0004-6361/201935832}

\bibitem[{{Falc{\'o}n-Barroso} {et~al.}(2017){Falc{\'o}n-Barroso}, {Lyubenova},
  {van de Ven}, {Mendez-Abreu}, {Aguerri}, {Garc{\'\i}a-Lorenzo},
  {Bekerait{\'e}}, {S{\'a}nchez}, {Husemann}, {Garc{\'\i}a-Benito}, {Mast},
  {Walcher}, {Zibetti}, {Barrera-Ballesteros}, {Galbany},
  {S{\'a}nchez-Bl{\'a}zquez}, {Singh}, {van den Bosch}, {Wild}, {Zhu}, {Bland
  -Hawthorn}, {Cid Fernandes}, {de Lorenzo-C{\'a}ceres}, {Gallazzi},
  {Gonz{\'a}lez Delgado}, {Marino}, {M{\'a}rquez}, {P{\'e}rez}, {P{\'e}rez},
  {Roth}, {Rosales-Ortega}, {Ruiz-Lara}, {Wisotzki}, {Ziegler}, \& {Califa
  Collaboration}}]{Falcon-Barroso2017}
{Falc{\'o}n-Barroso}, J., {Lyubenova}, M., {van de Ven}, G., {et~al.} 2017,
  \aap, 597, A48, \dodoi{10.1051/0004-6361/201628625}

\bibitem[{{Falc{\'o}n-Barroso} {et~al.}(2019){Falc{\'o}n-Barroso}, {van de
  Ven}, {Lyubenova}, {Mendez-Abreu}, {Aguerri}, {Garc{\'\i}a-Lorenzo},
  {Bekerait{\'e}}, {S{\'a}nchez}, {Husemann}, {Garc{\'\i}a-Benito},
  {Gonz{\'a}lez Delgado}, {Mast}, {Walcher}, {Zibetti}, {Zhu},
  {Barrera-Ballesteros}, {Galbany}, {S{\'a}nchez-Bl{\'a}zquez}, {Singh}, {van
  den Bosch}, {Wild}, {Bland-Hawthorn}, {Cid Fernandes}, {de
  Lorenzo-C{\'a}ceres}, {Gallazzi}, {Marino}, {M{\'a}rquez}, {Peletier},
  {P{\'e}rez}, {P{\'e}rez}, {Roth}, {Rosales-Ortega}, {Ruiz-Lara}, {Wisotzki},
  \& {Ziegler}}]{Falcon-Barroso2019}
{Falc{\'o}n-Barroso}, J., {van de Ven}, G., {Lyubenova}, M., {et~al.} 2019,
  \aap, 632, A59, \dodoi{10.1051/0004-6361/201936413}

\bibitem[{{Fall} \& {Efstathiou}(1980)}]{Fall1980}
{Fall}, S.~M., \& {Efstathiou}, G. 1980, \mnras, 193, 189,
  \dodoi{10.1093/mnras/193.2.189}

\bibitem[{{Fasano} {et~al.}(2010){Fasano}, {Bettoni}, {Ascaso}, {Tormen},
  {Poggianti}, {Valentinuzzi}, {D'Onofrio}, {Fritz}, {Moretti}, {Omizzolo},
  {Cava}, {Moles}, {Dressler}, {Couch}, {Kj{\ae}rgaard}, \&
  {Varela}}]{Fasano2010}
{Fasano}, G., {Bettoni}, D., {Ascaso}, B., {et~al.} 2010, \mnras, 404, 1490,
  \dodoi{10.1111/j.1365-2966.2010.16361.x}

\bibitem[{{Feldmeier-Krause} {et~al.}(2017){Feldmeier-Krause}, {Zhu},
  {Neumayer}, {van de Ven}, {de Zeeuw}, \& {Sch{\"o}del}}]{Feldmeier2017}
{Feldmeier-Krause}, A., {Zhu}, L., {Neumayer}, N., {et~al.} 2017, \mnras, 466,
  4040, \dodoi{10.1093/mnras/stw3377}

\bibitem[{{Foster} {et~al.}(2017){Foster}, {van de Sande}, {D'Eugenio},
  {Cortese}, {McDermid}, {Bland-Hawthorn}, {Brough}, {Bryant}, {Croom},
  {Goodwin}, {Konstantopoulos}, {Lawrence}, {L{\'o}pez-S{\'a}nchez}, {Medling},
  {Owers}, {Richards}, {Scott}, {Taranu}, {Tonini}, \& {Zafar}}]{Foster2017}
{Foster}, C., {van de Sande}, J., {D'Eugenio}, F., {et~al.} 2017, \mnras, 472,
  966, \dodoi{10.1093/mnras/stx1869}

\bibitem[{{Fraser-McKelvie} {et~al.}(2021){Fraser-McKelvie}, {Cortese}, {van de
  Sande}, {Bryant}, {Catinella}, {Colless}, {Croom}, {Groves}, {Medling},
  {Scott}, {Sweet}, {Bland-Hawthorn}, {Goodwin}, {Lawrence}, {Lorente},
  {Owers}, \& {Richards}}]{Fraser2021}
{Fraser-McKelvie}, A., {Cortese}, L., {van de Sande}, J., {et~al.} 2021,
  \mnras, 503, 4992, \dodoi{10.1093/mnras/stab573}

\bibitem[{{Gebhardt} {et~al.}(2003){Gebhardt}, {Richstone}, {Tremaine},
  {Lauer}, {Bender}, {Bower}, {Dressler}, {Faber}, {Filippenko}, {Green},
  {Grillmair}, {Ho}, {Kormendy}, {Magorrian}, \& {Pinkney}}]{Gebhardt2003}
{Gebhardt}, K., {Richstone}, D., {Tremaine}, S., {et~al.} 2003, \apj, 583, 92,
  \dodoi{10.1086/345081}

\bibitem[{{Gerhard} {et~al.}(2001){Gerhard}, {Kronawitter}, {Saglia}, \&
  {Bender}}]{Gerhard2001}
{Gerhard}, O., {Kronawitter}, A., {Saglia}, R.~P., \& {Bender}, R. 2001, \aj,
  121, 1936, \dodoi{10.1086/319940}

\bibitem[{{Gerhard}(1993)}]{Gerhard1993}
{Gerhard}, O.~E. 1993, \mnras, 265, 213, \dodoi{10.1093/mnras/265.1.213}

\bibitem[{{G{\'o}mez} {et~al.}(2017){G{\'o}mez}, {Grand}, {Monachesi}, {White},
  {Bustamante}, {Marinacci}, {Pakmor}, {Simpson}, {Springel}, \&
  {Frenk}}]{Gomez2017}
{G{\'o}mez}, F.~A., {Grand}, R. J.~J., {Monachesi}, A., {et~al.} 2017, \mnras,
  472, 3722, \dodoi{10.1093/mnras/stx2149}

\bibitem[{{Greene} {et~al.}(2017){Greene}, {Leauthaud}, {Emsellem}, {Goddard},
  {Ge}, {Andrews}, {Brinkman}, {Brownstein}, {Greco}, {Law}, {Lin}, {Masters},
  {Merrifield}, {More}, {Okabe}, {Schneider}, {Thomas}, {Wake}, {Yan}, \&
  {Drory}}]{Greene2017}
{Greene}, J.~E., {Leauthaud}, A., {Emsellem}, E., {et~al.} 2017, \apjl, 851,
  L33, \dodoi{10.3847/2041-8213/aa8ace}

\bibitem[{{Hagen} {et~al.}(2019){Hagen}, {Helmi}, {de Zeeuw}, \&
  {Posti}}]{Hagen2019}
{Hagen}, J.~H.~J., {Helmi}, A., {de Zeeuw}, P.~T., \& {Posti}, L. 2019, \aap,
  629, A70, \dodoi{10.1051/0004-6361/201935264}

\bibitem[{{Henriques} {et~al.}(2019){Henriques}, {White}, {Lilly}, {Bell},
  {Bluck}, \& {Terrazas}}]{Henriques2019}
{Henriques}, B. M.~B., {White}, S. D.~M., {Lilly}, S.~J., {et~al.} 2019,
  \mnras, 485, 3446, \dodoi{10.1093/mnras/stz577}

\bibitem[{{Hill} {et~al.}(2011){Hill}, {Kelvin}, {Driver}, {Robotham},
  {Cameron}, {Cross}, {Andrae}, {Baldry}, {Bamford}, {Bland-Hawthorn},
  {Brough}, {Conselice}, {Dye}, {Hopkins}, {Liske}, {Loveday}, {Norberg},
  {Peacock}, {Croom}, {Frenk}, {Graham}, {Jones}, {Kuijken}, {Madore},
  {Nichol}, {Parkinson}, {Phillipps}, {Pimbblet}, {Popescu}, {Prescott},
  {Seibert}, {Sharp}, {Sutherland}, {Thomas}, {Tuffs}, \& {van
  Kampen}}]{Hill2011}
{Hill}, D.~T., {Kelvin}, L.~S., {Driver}, S.~P., {et~al.} 2011, \mnras, 412,
  765, \dodoi{10.1111/j.1365-2966.2010.17950.x}

\bibitem[{{Illingworth}(1977)}]{Illingworth1977}
{Illingworth}, G. 1977, \apjl, 218, L43, \dodoi{10.1086/182572}

\bibitem[{{Jesseit} {et~al.}(2009){Jesseit}, {Cappellari}, {Naab}, {Emsellem},
  \& {Burkert}}]{Jesseit2009}
{Jesseit}, R., {Cappellari}, M., {Naab}, T., {Emsellem}, E., \& {Burkert}, A.
  2009, \mnras, 397, 1202, \dodoi{10.1111/j.1365-2966.2009.14984.x}

\bibitem[{{Jethwa} {et~al.}(2020){Jethwa}, {Thater}, {Maindl}, \& {Van de
  Ven}}]{DYN2020}
{Jethwa}, P., {Thater}, S., {Maindl}, T., \& {Van de Ven}, G. 2020, {DYNAMITE:
  DYnamics, Age and Metallicity Indicators Tracing Evolution}.
\newblock \doeprint{2011.007}

\bibitem[{{Jin} {et~al.}(2020){Jin}, {Zhu}, {Long}, {Mao}, {Wang}, \& {van de
  Ven}}]{Jin2020}
{Jin}, Y., {Zhu}, L., {Long}, R.~J., {et~al.} 2020, \mnras, 491, 1690,
  \dodoi{10.1093/mnras/stz3072}

\bibitem[{{Jin} {et~al.}(2019){Jin}, {Zhu}, {Long}, {Mao}, {Xu}, {Li}, \& {van
  de Ven}}]{Jin2019}
---. 2019, \mnras, 486, 4753, \dodoi{10.1093/mnras/stz1170}

\bibitem[{{Kimm} \& {Yi}(2007)}]{Kimm2007}
{Kimm}, T., \& {Yi}, S.~K. 2007, \apj, 670, 1048, \dodoi{10.1086/522573}

\bibitem[{{Kireeva} \& {Kondratyev}(2019)}]{Kireeva2019}
{Kireeva}, E.~N., \& {Kondratyev}, B.~P. 2019, Astronomy Reports, 63, 713,
  \dodoi{10.1134/S106377291909004X}

\bibitem[{{Krajnovi{\'c}} {et~al.}(2005){Krajnovi{\'c}}, {Cappellari},
  {Emsellem}, {McDermid}, \& {de Zeeuw}}]{Krajnovic2005}
{Krajnovi{\'c}}, D., {Cappellari}, M., {Emsellem}, E., {McDermid}, R.~M., \&
  {de Zeeuw}, P.~T. 2005, \mnras, 357, 1113,
  \dodoi{10.1111/j.1365-2966.2005.08715.x}

\bibitem[{{Krajnovi{\'c}} {et~al.}(2009){Krajnovi{\'c}}, {McDermid},
  {Cappellari}, \& {Davies}}]{Krajnovic2009}
{Krajnovi{\'c}}, D., {McDermid}, R.~M., {Cappellari}, M., \& {Davies}, R.~L.
  2009, \mnras, 399, 1839, \dodoi{10.1111/j.1365-2966.2009.15415.x}

\bibitem[{{Krajnovi{\'c}} {et~al.}(2008){Krajnovi{\'c}}, {Bacon}, {Cappellari},
  {Davies}, {de Zeeuw}, {Emsellem}, {Falc{\'o}n-Barroso}, {Kuntschner},
  {McDermid}, {Peletier}, {Sarzi}, {van den Bosch}, \& {van de
  Ven}}]{Krajnovic2008}
{Krajnovi{\'c}}, D., {Bacon}, R., {Cappellari}, M., {et~al.} 2008, \mnras, 390,
  93, \dodoi{10.1111/j.1365-2966.2008.13712.x}

\bibitem[{{Krajnovi{\'c}} {et~al.}(2015){Krajnovi{\'c}}, {Weilbacher},
  {Urrutia}, {Emsellem}, {Carollo}, {Shirazi}, {Bacon}, {Contini}, {Epinat},
  {Kamann}, {Martinsson}, \& {Steinmetz}}]{Krajnovic2015}
{Krajnovi{\'c}}, D., {Weilbacher}, P.~M., {Urrutia}, T., {et~al.} 2015, \mnras,
  452, 2, \dodoi{10.1093/mnras/stv958}

\bibitem[{{Kuijken} {et~al.}(1996){Kuijken}, {Fisher}, \&
  {Merrifield}}]{Kuijken1996}
{Kuijken}, K., {Fisher}, D., \& {Merrifield}, M.~R. 1996, \mnras, 283, 543,
  \dodoi{10.1093/mnras/283.2.543}

\bibitem[{{Lagos} {et~al.}(2020){Lagos}, {Emsellem}, {van de Sande},
  {Harborne}, {Cortese}, {Davison}, {Foster}, \& {Wright}}]{Lagos2020}
{Lagos}, C. d.~P., {Emsellem}, E., {van de Sande}, J., {et~al.} 2020, arXiv
  e-prints, arXiv:2012.08060.
\newblock \doarXiv{2012.08060}

\bibitem[{{Lawson} \& {Hanson}(1974)}]{Lawson1974}
{Lawson}, C.~L., \& {Hanson}, R.~J. 1974, {Solving least squares problems}

\bibitem[{{Li} {et~al.}(2018{\natexlab{a}}){Li}, {Mao}, {Cappellari}, {Graham},
  {Emsellem}, \& {Long}}]{Li2018b}
{Li}, H., {Mao}, S., {Cappellari}, M., {et~al.} 2018{\natexlab{a}}, \apjl, 863,
  L19, \dodoi{10.3847/2041-8213/aad54b}

\bibitem[{{Li} {et~al.}(2018{\natexlab{b}}){Li}, {Mao}, {Emsellem}, {Xu},
  {Springel}, \& {Krajnovi{\'c}}}]{Li2018a}
{Li}, H., {Mao}, S., {Emsellem}, E., {et~al.} 2018{\natexlab{b}}, \mnras, 473,
  1489, \dodoi{10.1093/mnras/stx2374}

\bibitem[{{Liepold} {et~al.}(2020){Liepold}, {Quenneville}, {Ma}, {Walsh},
  {McConnell}, {Greene}, \& {Blakeslee}}]{Liepold2020}
{Liepold}, C.~M., {Quenneville}, M.~E., {Ma}, C.-P., {et~al.} 2020, \apj, 891,
  4, \dodoi{10.3847/1538-4357/ab6f71}

\bibitem[{{Lipka} \& {Thomas}(2021)}]{Lipka2021}
{Lipka}, M., \& {Thomas}, J. 2021, \mnras, 504, 4599,
  \dodoi{10.1093/mnras/stab1092}

\bibitem[{{Lyubenova} {et~al.}(2013){Lyubenova}, {van den Bosch},
  {C{\^o}t{\'e}}, {Kuntschner}, {van de Ven}, {Ferrarese}, {Jord{\'a}n},
  {Infante}, \& {Peng}}]{Lyubenova2013}
{Lyubenova}, M., {van den Bosch}, R. C.~E., {C{\^o}t{\'e}}, P., {et~al.} 2013,
  \mnras, 431, 3364, \dodoi{10.1093/mnras/stt414}

\bibitem[{{Ma} {et~al.}(2014){Ma}, {Greene}, {McConnell}, {Janish},
  {Blakeslee}, {Thomas}, \& {Murphy}}]{Ma2014}
{Ma}, C.-P., {Greene}, J.~E., {McConnell}, N., {et~al.} 2014, \apj, 795, 158,
  \dodoi{10.1088/0004-637X/795/2/158}

\bibitem[{{McConnell} {et~al.}(2011){McConnell}, {Ma}, {Gebhardt}, {Wright},
  {Murphy}, {Lauer}, {Graham}, \& {Richstone}}]{McConnell2011}
{McConnell}, N.~J., {Ma}, C.-P., {Gebhardt}, K., {et~al.} 2011, \nat, 480, 215,
  \dodoi{10.1038/nature10636}

\bibitem[{{Naab} {et~al.}(2009){Naab}, {Johansson}, \& {Ostriker}}]{Naab2009}
{Naab}, T., {Johansson}, P.~H., \& {Ostriker}, J.~P. 2009, \apjl, 699, L178,
  \dodoi{10.1088/0004-637X/699/2/L178}

\bibitem[{{Naab} {et~al.}(2014){Naab}, {Oser}, {Emsellem}, {Cappellari},
  {Krajnovi{\'c}}, {McDermid}, {Alatalo}, {Bayet}, {Blitz}, \&
  {Bois}}]{Naab2014}
{Naab}, T., {Oser}, L., {Emsellem}, E., {et~al.} 2014, \mnras, 444, 3357,
  \dodoi{10.1093/mnras/stt1919}

\bibitem[{{Navarro} {et~al.}(1996){Navarro}, {Frenk}, \& {White}}]{Navarro1996}
{Navarro}, J.~F., {Frenk}, C.~S., \& {White}, S. D.~M. 1996, \apj, 462, 563,
  \dodoi{10.1086/177173}

\bibitem[{{Neureiter} {et~al.}(2021){Neureiter}, {Thomas}, {Saglia}, {Bender},
  {Finozzi}, {Krukau}, {Naab}, {Rantala}, \& {Frigo}}]{Neureiter2021}
{Neureiter}, B., {Thomas}, J., {Saglia}, R., {et~al.} 2021, \mnras, 500, 1437,
  \dodoi{10.1093/mnras/staa3014}

\bibitem[{{Oser} {et~al.}(2010){Oser}, {Ostriker}, {Naab}, {Johansson}, \&
  {Burkert}}]{Oser2010}
{Oser}, L., {Ostriker}, J.~P., {Naab}, T., {Johansson}, P.~H., \& {Burkert}, A.
  2010, \apj, 725, 2312, \dodoi{10.1088/0004-637X/725/2/2312}

\bibitem[{{Owers} {et~al.}(2017){Owers}, {Allen}, {Baldry}, {Bryant}, {Cecil},
  {Cortese}, {Croom}, {Driver}, {Fogarty}, {Green}, {Helmich}, {de Jong},
  {Kuijken}, {Mahajan}, {McFarland}, {Pracy}, {Robotham}, {Sikkema}, {Sweet},
  {Taylor}, {Verdoes Kleijn}, {Bauer}, {Bland-Hawthorn}, {Brough}, {Colless},
  {Couch}, {Davies}, {Drinkwater}, {Goodwin}, {Hopkins}, {Konstantopoulos},
  {Foster}, {Lawrence}, {Lorente}, {Medling}, {Metcalfe}, {Richards}, {van de
  Sande}, {Scott}, {Shanks}, {Sharp}, {Thomas}, \& {Tonini}}]{Owers2017}
{Owers}, M.~S., {Allen}, J.~T., {Baldry}, I., {et~al.} 2017, \mnras, 468, 1824,
  \dodoi{10.1093/mnras/stx562}

\bibitem[{{Owers} {et~al.}(2019){Owers}, {Hudson}, {Oman}, {Bland-Hawthorn},
  {Brough}, {Bryant}, {Cortese}, {Couch}, {Croom}, {van de Sande}, {Federrath},
  {Groves}, {Hopkins}, {Lawrence}, {Lorente}, {McDermid}, {Medling},
  {Richards}, {Scott}, {Taranu}, {Welker}, \& {Yi}}]{Owers2019}
{Owers}, M.~S., {Hudson}, M.~J., {Oman}, K.~A., {et~al.} 2019, arXiv e-prints.
\newblock \doarXiv{1901.08185}

\bibitem[{{Park} {et~al.}(2019){Park}, {Yi}, {Dubois}, {Pichon}, {Kimm},
  {Devriendt}, {Choi}, {Volonteri}, {Kaviraj}, \& {Peirani}}]{Park2019}
{Park}, M.-J., {Yi}, S.~K., {Dubois}, Y., {et~al.} 2019, \apj, 883, 25,
  \dodoi{10.3847/1538-4357/ab3afe}

\bibitem[{{Park} {et~al.}(2021){Park}, {Yi}, {Peirani}, {Pichon}, {Dubois},
  {Choi}, {Devriendt}, {Kaviraj}, {Kimm}, {Kraljic}, \& {Volonteri}}]{Park2021}
{Park}, M.~J., {Yi}, S.~K., {Peirani}, S., {et~al.} 2021, \apjs, 254, 2,
  \dodoi{10.3847/1538-4365/abe937}

\bibitem[{{Poci} {et~al.}(2017){Poci}, {Cappellari}, \& {McDermid}}]{Poci2017}
{Poci}, A., {Cappellari}, M., \& {McDermid}, R.~M. 2017, \mnras, 467, 1397,
  \dodoi{10.1093/mnras/stx101}

\bibitem[{{Poci} {et~al.}(2019){Poci}, {McDermid}, {Zhu}, \& {van de
  Ven}}]{Poci2019}
{Poci}, A., {McDermid}, R.~M., {Zhu}, L., \& {van de Ven}, G. 2019, \mnras,
  487, 3776, \dodoi{10.1093/mnras/stz1154}

\bibitem[{{Posacki} {et~al.}(2015){Posacki}, {Cappellari}, {Treu},
  {Pellegrini}, \& {Ciotti}}]{Posacki2015}
{Posacki}, S., {Cappellari}, M., {Treu}, T., {Pellegrini}, S., \& {Ciotti}, L.
  2015, \mnras, 446, 493, \dodoi{10.1093/mnras/stu2098}

\bibitem[{{Quenneville} {et~al.}(2021){Quenneville}, {Liepold}, \&
  {Ma}}]{Quenneville2021}
{Quenneville}, M.~E., {Liepold}, C.~M., \& {Ma}, C.-P. 2021, \apjs, 254, 25,
  \dodoi{10.3847/1538-4365/abe6a0}

\bibitem[{{Quenneville} {et~al.}(2022){Quenneville}, {Liepold}, \&
  {Ma}}]{Quenneville2021b}
---. 2022, \apj, 926, 30, \dodoi{10.3847/1538-4357/ac3e68}

\bibitem[{{Querejeta} {et~al.}(2015){Querejeta}, {Eliche-Moral}, {Tapia},
  {Borlaff}, {van de Ven}, {Lyubenova}, {Martig}, {Falc{\'o}n-Barroso}, \&
  {M{\'e}ndez-Abreu}}]{Querejeta2015}
{Querejeta}, M., {Eliche-Moral}, M.~C., {Tapia}, T., {et~al.} 2015, \aap, 579,
  L2, \dodoi{10.1051/0004-6361/201526354}

\bibitem[{{Rodr{\'\i}guez} {et~al.}(2016){Rodr{\'\i}guez}, {Padilla}, \&
  {Garc{\'\i}a Lambas}}]{Rodriguez2016}
{Rodr{\'\i}guez}, S., {Padilla}, N.~D., \& {Garc{\'\i}a Lambas}, D. 2016,
  \mnras, 456, 571, \dodoi{10.1093/mnras/stv2660}

\bibitem[{{Rusli} {et~al.}(2013){Rusli}, {Thomas}, {Saglia}, {Fabricius},
  {Erwin}, {Bender}, {Nowak}, {Lee}, {Riffeser}, \& {Sharp}}]{Rusli2013}
{Rusli}, S.~P., {Thomas}, J., {Saglia}, R.~P., {et~al.} 2013, \aj, 146, 45,
  \dodoi{10.1088/0004-6256/146/3/45}

\bibitem[{{Rybicki}(1987)}]{Rybicki1987}
{Rybicki}, G.~B. 1987, in Structure and Dynamics of Elliptical Galaxies, ed.
  P.~T. {de Zeeuw}, Vol. 127, 397, \dodoi{10.1007/978-94-009-3971-4\_41}

\bibitem[{{S{\'a}nchez} {et~al.}(2012){S{\'a}nchez}, {Kennicutt}, {Gil de Paz},
  {van de Ven}, {V{\'{\i}}lchez}, {Wisotzki}, {Walcher}, {Mast}, {Aguerri},
  {Albiol-P{\'e}rez}, {Alonso-Herrero}, {Alves}, {Bakos}, {Bart{\'a}kov{\'a}},
  {Bland-Hawthorn}, {Boselli}, {Bomans}, {Castillo-Morales}, {Cortijo-Ferrero},
  {de Lorenzo-C{\'a}ceres}, {Del Olmo}, {Dettmar}, {D{\'{\i}}az}, {Ellis},
  {Falc{\'o}n-Barroso}, {Flores}, {Gallazzi}, {Garc{\'{\i}}a-Lorenzo},
  {Gonz{\'a}lez Delgado}, {Gruel}, {Haines}, {Hao}, {Husemann},
  {Igl{\'e}sias-P{\'a}ramo}, {Jahnke}, {Johnson}, {Jungwiert}, {Kalinova},
  {Kehrig}, {Kupko}, {L{\'o}pez-S{\'a}nchez}, {Lyubenova}, {Marino},
  {M{\'a}rmol-Queralt{\'o}}, {M{\'a}rquez}, {Masegosa}, {Meidt},
  {Mendez-Abreu}, {Monreal-Ibero}, {Montijo}, {Mour{\~a}o}, {Palacios-Navarro},
  {Papaderos}, {Pasquali}, {Peletier}, {P{\'e}rez}, {P{\'e}rez}, {Quirrenbach},
  {Rela{\~n}o}, {Rosales-Ortega}, {Roth}, {Ruiz-Lara},
  {S{\'a}nchez-Bl{\'a}zquez}, {Sengupta}, {Singh}, {Stanishev}, {Trager},
  {Vazdekis}, {Viironen}, {Wild}, {Zibetti}, \& {Ziegler}}]{Sanchez2012}
{S{\'a}nchez}, S.~F., {Kennicutt}, R.~C., {Gil de Paz}, A., {et~al.} 2012,
  \aap, 538, A8, \dodoi{10.1051/0004-6361/201117353}

\bibitem[{{S{\'a}nchez-Janssen} {et~al.}(2010){S{\'a}nchez-Janssen},
  {M{\'e}ndez-Abreu}, \& {Aguerri}}]{Sanchez-Janssen2010}
{S{\'a}nchez-Janssen}, R., {M{\'e}ndez-Abreu}, J., \& {Aguerri}, J.~A.~L. 2010,
  \mnras, 406, L65, \dodoi{10.1111/j.1745-3933.2010.00883.x}

\bibitem[{{S{\'a}nchez-Janssen} {et~al.}(2016){S{\'a}nchez-Janssen},
  {Ferrarese}, {MacArthur}, {C{\^o}t{\'e}}, {Blakeslee}, {Cuillandre}, {Duc},
  {Durrell}, {Gwyn}, {McConnacchie}, {Boselli}, {Courteau}, {Emsellem}, {Mei},
  {Peng}, {Puzia}, {Roediger}, {Simard}, {Boyer}, \&
  {Santos}}]{Sanchez-Janssen2016}
{S{\'a}nchez-Janssen}, R., {Ferrarese}, L., {MacArthur}, L.~A., {et~al.} 2016,
  \apj, 820, 69, \dodoi{10.3847/0004-637X/820/1/69}

\bibitem[{{Sarzi} {et~al.}(2018){Sarzi}, {Iodice}, {Coccato}, {Corsini}, {de
  Zeeuw}, {Falc{\'o}n-Barroso}, {Gadotti}, {Lyubenova}, {McDermid}, {van de
  Ven}, {Fahrion}, {Pizzella}, \& {Zhu}}]{Sarzi2018}
{Sarzi}, M., {Iodice}, E., {Coccato}, L., {et~al.} 2018, \aap, 616, A121,
  \dodoi{10.1051/0004-6361/201833137}

\bibitem[{{Schwarzschild}(1979)}]{Schwarzschild1979}
{Schwarzschild}, M. 1979, \apj, 232, 236, \dodoi{10.1086/157282}

\bibitem[{{Scott} {et~al.}(2009){Scott}, {Cappellari}, {Davies}, {Bacon}, {de
  Zeeuw}, {Emsellem}, {Falc{\'o}n-Barroso}, {Krajnovi{\'c}}, {Kuntschner},
  {McDermid}, {Peletier}, {Pipino}, {Sarzi}, {van den Bosch}, {van de Ven}, \&
  {van Scherpenzeel}}]{Scott2009}
{Scott}, N., {Cappellari}, M., {Davies}, R.~L., {et~al.} 2009, \mnras, 398,
  1835, \dodoi{10.1111/j.1365-2966.2009.15275.x}

\bibitem[{{Scott} {et~al.}(2017){Scott}, {Brough}, {Croom}, {Davies}, {van de
  Sande}, {Allen}, {Bland-Hawthorn}, {Bryant}, {Cortese}, {D'Eugenio},
  {Federrath}, {Ferreras}, {Goodwin}, {Groves}, {Konstantopoulos}, {Lawrence},
  {Medling}, {Moffett}, {Owers}, {Richards}, {Robotham}, {Tonini}, \&
  {Yi}}]{Scott2017}
{Scott}, N., {Brough}, S., {Croom}, S.~M., {et~al.} 2017, \mnras, 472, 2833,
  \dodoi{10.1093/mnras/stx2166}

\bibitem[{{Seth} {et~al.}(2014){Seth}, {van den Bosch}, {Mieske}, {Baumgardt},
  {Brok}, {Strader}, {Neumayer}, {Chilingarian}, {Hilker}, {McDermid},
  {Spitler}, {Brodie}, {Frank}, \& {Walsh}}]{Seth2014}
{Seth}, A.~C., {van den Bosch}, R., {Mieske}, S., {et~al.} 2014, \nat, 513,
  398, \dodoi{10.1038/nature13762}

\bibitem[{{Shanks} {et~al.}(2015){Shanks}, {Metcalfe}, {Chehade}, {Findlay},
  {Irwin}, {Gonzalez-Solares}, {Lewis}, {Yoldas}, {Mann}, {Read}, {Sutorius},
  \& {Voutsinas}}]{Shanks2015}
{Shanks}, T., {Metcalfe}, N., {Chehade}, B., {et~al.} 2015, Monthly Notices of
  the Royal Astronomical Society, 451, 4238, \dodoi{10.1093/mnras/stv1130}

\bibitem[{{Sharp} {et~al.}(2006){Sharp}, {Saunders}, {Smith}, {Churilov},
  {Correll}, {Dawson}, {Farrel}, {Frost}, {Haynes}, {Heald}, {Lankshear},
  {Mayfield}, {Waller}, \& {Whittard}}]{Sharp2006}
{Sharp}, R., {Saunders}, W., {Smith}, G., {et~al.} 2006, in Society of
  Photo-Optical Instrumentation Engineers (SPIE) Conference Series, Vol. 6269,
  Society of Photo-Optical Instrumentation Engineers (SPIE) Conference Series,
  ed. I.~S. {McLean} \& M.~{Iye}, 62690G, \dodoi{10.1117/12.671022}

\bibitem[{{Smith} \& {Betbeder-Matibet}(2010)}]{Katana2010}
{Smith}, D., \& {Betbeder-Matibet}, L. 2010, {Katana},
  \dodoi{https://doi.org/10.26190/669x-a286}

\bibitem[{{Taylor} {et~al.}(2011){Taylor}, {Hopkins}, {Baldry}, {Brown},
  {Driver}, {Kelvin}, {Hill}, {Robotham}, {Bland-Hawthorn}, {Jones}, {Sharp},
  {Thomas}, {Liske}, {Loveday}, {Norberg}, {Peacock}, {Bamford}, {Brough},
  {Colless}, {Cameron}, {Conselice}, {Croom}, {Frenk}, {Gunawardhana},
  {Kuijken}, {Nichol}, {Parkinson}, {Phillipps}, {Pimbblet}, {Popescu},
  {Prescott}, {Sutherland}, {Tuffs}, {van Kampen}, \&
  {Wijesinghe}}]{Taylor2011}
{Taylor}, E.~N., {Hopkins}, A.~M., {Baldry}, I.~K., {et~al.} 2011, \mnras, 418,
  1587, \dodoi{10.1111/j.1365-2966.2011.19536.x}

\bibitem[{{Thater} {et~al.}(2019){Thater}, {Krajnovi{\'c}}, {Cappellari},
  {Davis}, {de Zeeuw}, {McDermid}, \& {Sarzi}}]{Thater2019}
{Thater}, S., {Krajnovi{\'c}}, D., {Cappellari}, M., {et~al.} 2019, \aap, 625,
  A62, \dodoi{10.1051/0004-6361/201834808}

\bibitem[{{Thater} {et~al.}(2017){Thater}, {Krajnovi{\'c}}, {Bourne},
  {Cappellari}, {de Zeeuw}, {Emsellem}, {Magorrian}, {McDermid}, {Sarzi}, \&
  {van de Ven}}]{Thater2017}
{Thater}, S., {Krajnovi{\'c}}, D., {Bourne}, M.~A., {et~al.} 2017, \aap, 597,
  A18, \dodoi{10.1051/0004-6361/201629480}

\bibitem[{{Thater} {et~al.}(2022){Thater}, {Krajnovi{\'c}}, {Weilbacher},
  {Nguyen}, {Bureau}, {Cappellari}, {Davis}, {Iguchi}, {McDermid}, {Onishi},
  {Sarzi}, \& {van de Ven}}]{Thater2021}
{Thater}, S., {Krajnovi{\'c}}, D., {Weilbacher}, P.~M., {et~al.} 2022, \mnras,
  509, 5416, \dodoi{10.1093/mnras/stab3210}

\bibitem[{{Thomas} {et~al.}(2011){Thomas}, {Maraston}, \&
  {Johansson}}]{Thomas2011}
{Thomas}, D., {Maraston}, C., \& {Johansson}, J. 2011, \mnras, 412, 2183,
  \dodoi{10.1111/j.1365-2966.2010.18049.x}

\bibitem[{{Thomas} {et~al.}(2014){Thomas}, {Saglia}, {Bender}, {Erwin}, \&
  {Fabricius}}]{Thomas2014}
{Thomas}, J., {Saglia}, R.~P., {Bender}, R., {Erwin}, P., \& {Fabricius}, M.
  2014, \apj, 782, 39, \dodoi{10.1088/0004-637X/782/1/39}

\bibitem[{{Thomas} {et~al.}(2007){Thomas}, {Saglia}, {Bender}, {Thomas},
  {Gebhardt}, {Magorrian}, {Corsini}, \& {Wegner}}]{Thomas2007}
{Thomas}, J., {Saglia}, R.~P., {Bender}, R., {et~al.} 2007, \mnras, 382, 657,
  \dodoi{10.1111/j.1365-2966.2007.12434.x}

\bibitem[{{Tinsley}(1980)}]{Tinsley1980}
{Tinsley}, B.~M. 1980, \fcp, 5, 287

\bibitem[{{Tissera} {et~al.}(2017){Tissera}, {Machado}, {Vilchez}, {Pedrosa},
  {Sanchez-Blazquez}, \& {Varela}}]{Tissera2017}
{Tissera}, P.~B., {Machado}, R. E.~G., {Vilchez}, J.~M., {et~al.} 2017, \aap,
  604, A118, \dodoi{10.1051/0004-6361/201628915}

\bibitem[{{Valluri} {et~al.}(2004){Valluri}, {Merritt}, \&
  {Emsellem}}]{Valluri2004}
{Valluri}, M., {Merritt}, D., \& {Emsellem}, E. 2004, \apj, 602, 66,
  \dodoi{10.1086/380896}

\bibitem[{{van de Sande} {et~al.}(2017{\natexlab{a}}){van de Sande},
  {Bland-Hawthorn}, {Fogarty}, {Cortese}, {d'Eugenio}, {Croom}, {Scott},
  {Allen}, {Brough}, {Bryant}, {Cecil}, {Colless}, {Couch}, {Davies}, {Elahi},
  {Foster}, {Goldstein}, {Goodwin}, {Groves}, {Ho}, {Jeong}, {Jones},
  {Konstantopoulos}, {Lawrence}, {Leslie}, {L{\'o}pez-S{\'a}nchez}, {McDermid},
  {McElroy}, {Medling}, {Oh}, {Owers}, {Richards}, {Schaefer}, {Sharp},
  {Sweet}, {Taranu}, {Tonini}, {Walcher}, \& {Yi}}]{vandeSande2017a}
{van de Sande}, J., {Bland-Hawthorn}, J., {Fogarty}, L. M.~R., {et~al.}
  2017{\natexlab{a}}, \apj, 835, 104, \dodoi{10.3847/1538-4357/835/1/104}

\bibitem[{{van de Sande} {et~al.}(2017{\natexlab{b}}){van de Sande},
  {Bland-Hawthorn}, {Brough}, {Croom}, {Cortese}, {Foster}, {Scott}, {Bryant},
  {d'Eugenio}, {Tonini}, {Goodwin}, {Konstantopoulos}, {Lawrence}, {Medling},
  {Owers}, {Richards}, {Schaefer}, \& {Yi}}]{vandeSande2017b}
{van de Sande}, J., {Bland-Hawthorn}, J., {Brough}, S., {et~al.}
  2017{\natexlab{b}}, \mnras, 472, 1272, \dodoi{10.1093/mnras/stx1751}

\bibitem[{{van de Sande} {et~al.}(2018){van de Sande}, {Scott},
  {Bland-Hawthorn}, {Brough}, {Bryant}, {Colless}, {Cortese}, {Croom},
  {d'Eugenio}, {Foster}, {Goodwin}, {Konstantopoulos}, {Lawrence}, {McDermid},
  {Medling}, {Owers}, {Richards}, \& {Sharp}}]{vandeSande2018}
{van de Sande}, J., {Scott}, N., {Bland-Hawthorn}, J., {et~al.} 2018, Nature
  Astronomy, 2, 483, \dodoi{10.1038/s41550-018-0436-x}

\bibitem[{{van de Sande} {et~al.}(2021{\natexlab{a}}){van de Sande}, {Croom},
  {Bland-Hawthorn}, {Cortese}, {Scott}, {Lagos}, {D'Eugenio}, {Bryant},
  {Brough}, {Catinella}, {Foster}, {Groves}, {Harborne},
  {L{\'o}pez-S{\'a}nchez}, {McDermid}, {Medling}, {Owers}, {Richards}, {Sweet},
  \& {Vaughan}}]{vandeSande2021b}
{van de Sande}, J., {Croom}, S.~M., {Bland-Hawthorn}, J., {et~al.}
  2021{\natexlab{a}}, \mnras, \dodoi{10.1093/mnras/stab2647}

\bibitem[{{van de Sande} {et~al.}(2021{\natexlab{b}}){van de Sande}, {Vaughan},
  {Cortese}, {Scott}, {Bland-Hawthorn}, {Croom}, {Lagos}, {Brough}, {Bryant},
  {Devriendt}, {Dubois}, {D'Eugenio}, {Foster}, {Fraser-McKelvie}, {Harborne},
  {Lawrence}, {Oh}, {Owers}, {Poci}, {Remus}, {Richards}, {Schulze}, {Sweet},
  {Varidel}, \& {Welker}}]{vandeSande2021a}
{van de Sande}, J., {Vaughan}, S.~P., {Cortese}, L., {et~al.}
  2021{\natexlab{b}}, \mnras, 505, 3078, \dodoi{10.1093/mnras/stab1490}

\bibitem[{{van de Ven} {et~al.}(2008){van de Ven}, {de Zeeuw}, \& {van den
  Bosch}}]{vandeVen2008}
{van de Ven}, G., {de Zeeuw}, P.~T., \& {van den Bosch}, R.~C.~E. 2008, \mnras,
  385, 614, \dodoi{10.1111/j.1365-2966.2008.12873.x}

\bibitem[{{van de Ven} {et~al.}(2006){van de Ven}, {van den Bosch}, {Verolme},
  \& {de Zeeuw}}]{vandeVen2006}
{van de Ven}, G., {van den Bosch}, R.~C.~E., {Verolme}, E.~K., \& {de Zeeuw},
  P.~T. 2006, \aap, 445, 513, \dodoi{10.1051/0004-6361:20053061}

\bibitem[{{van den Bosch} {et~al.}(2008){van den Bosch}, {van de Ven},
  {Verolme}, {Cappellari}, \& {de Zeeuw}}]{vandenBosch2008}
{van den Bosch}, R.~C.~E., {van de Ven}, G., {Verolme}, E.~K., {Cappellari},
  M., \& {de Zeeuw}, P.~T. 2008, \mnras, 385, 647,
  \dodoi{10.1111/j.1365-2966.2008.12874.x}

\bibitem[{{van der Marel} {et~al.}(1998){van der Marel}, {Cretton}, {de Zeeuw},
  \& {Rix}}]{vanderMarel1998}
{van der Marel}, R.~P., {Cretton}, N., {de Zeeuw}, P.~T., \& {Rix}, H.-W. 1998,
  \apj, 493, 613, \dodoi{10.1086/305147}

\bibitem[{{van der Marel} \& {Franx}(1993)}]{vanderMarel1993}
{van der Marel}, R.~P., \& {Franx}, M. 1993, \apj, 407, 525,
  \dodoi{10.1086/172534}

\bibitem[{{Vasiliev} \& {Athanassoula}(2015)}]{Vasiliev2015}
{Vasiliev}, E., \& {Athanassoula}, E. 2015, \mnras, 450, 2842,
  \dodoi{10.1093/mnras/stv805}

\bibitem[{{Vasiliev} \& {Valluri}(2020)}]{Vasiliev2020}
{Vasiliev}, E., \& {Valluri}, M. 2020, \apj, 889, 39,
  \dodoi{10.3847/1538-4357/ab5fe0}

\bibitem[{{Veale} {et~al.}(2017){Veale}, {Ma}, {Greene}, {Thomas}, {Blakeslee},
  {McConnell}, {Walsh}, \& {Ito}}]{Veale2017}
{Veale}, M., {Ma}, C.-P., {Greene}, J.~E., {et~al.} 2017, \mnras, 471, 1428,
  \dodoi{10.1093/mnras/stx1639}

\bibitem[{{Verolme} {et~al.}(2002){Verolme}, {Cappellari}, {Copin}, {van der
  Marel}, {Bacon}, {Bureau}, {Davies}, {Miller}, \& {de Zeeuw}}]{Verolme2002}
{Verolme}, E.~K., {Cappellari}, M., {Copin}, Y., {et~al.} 2002, \mnras, 335,
  517, \dodoi{10.1046/j.1365-8711.2002.05664.x}

\bibitem[{{Virtanen} {et~al.}(2019){Virtanen}, {Gommers}, {Oliphant},
  {Haberland}, {Reddy}, {Cournapeau}, {Burovski}, {Peterson}, {Weckesser},
  {Bright}, {van der Walt}, {Brett}, {Wilson}, {Jarrod Millman}, {Mayorov},
  {Nelson}, {Jones}, {Kern}, {Larson}, {Carey}, {Polat}, {Feng}, {Moore}, {Vand
  erPlas}, {Laxalde}, {Perktold}, {Cimrman}, {Henriksen}, {Quintero}, {Harris},
  {Archibald}, {Ribeiro}, {Pedregosa}, {van Mulbregt}, \&
  {Contributors}}]{Virtanen2019}
{Virtanen}, P., {Gommers}, R., {Oliphant}, T.~E., {et~al.} 2019, arXiv
  e-prints, arXiv:1907.10121.
\newblock \doarXiv{1907.10121}

\bibitem[{Virtanen {et~al.}(2020)Virtanen, Gommers, Oliphant, Haberland, Reddy,
  Cournapeau, Burovski, Peterson, Weckesser, Bright, {van der Walt}, Brett,
  Wilson, Millman, Mayorov, Nelson, Jones, Kern, Larson, Carey, Polat, Feng,
  Moore, {VanderPlas}, Laxalde, Perktold, Cimrman, Henriksen, Quintero, Harris,
  Archibald, Ribeiro, Pedregosa, {van Mulbregt}, \& {SciPy 1.0
  Contributors}}]{2020SciPy-NMeth}
Virtanen, P., Gommers, R., Oliphant, T.~E., {et~al.} 2020, Nature Methods, 17,
  261, \dodoi{10.1038/s41592-019-0686-2}

\bibitem[{{Weijmans} {et~al.}(2014){Weijmans}, {de Zeeuw}, {Emsellem},
  {Krajnovi{\'c}}, {Lablanche}, {Alatalo}, {Blitz}, {Bois}, {Bournaud},
  {Bureau}, {Cappellari}, {Crocker}, {Davies}, {Davis}, {Duc}, {Khochfar},
  {Kuntschner}, {McDermid}, {Morganti}, {Naab}, {Oosterloo}, {Sarzi}, {Scott},
  {Serra}, {Verdoes Kleijn}, \& {Young}}]{Weijmans2014}
{Weijmans}, A.-M., {de Zeeuw}, P.~T., {Emsellem}, E., {et~al.} 2014, \mnras,
  444, 3340, \dodoi{10.1093/mnras/stu1603}

\bibitem[{{White}(1979)}]{White1979}
{White}, S.~D.~M. 1979, \mnras, 186, 145, \dodoi{10.1093/mnras/186.2.145}

\bibitem[{{York} {et~al.}(2000){York}, {Adelman}, {Anderson}, {Anderson},
  {Annis}, {Bahcall}, {Bakken}, {Barkhouser}, {Bastian}, {Berman}, {Boroski},
  {Bracker}, {Briegel}, {Briggs}, {Brinkmann}, {Brunner}, {Burles}, {Carey},
  {Carr}, {Castander}, {Chen}, {Colestock}, {Connolly}, {Crocker}, {Csabai},
  {Czarapata}, {Davis}, {Doi}, {Dombeck}, {Eisenstein}, {Ellman}, {Elms},
  {Evans}, {Fan}, {Federwitz}, {Fiscelli}, {Friedman}, {Frieman}, {Fukugita},
  {Gillespie}, {Gunn}, {Gurbani}, {de Haas}, {Haldeman}, {Harris}, {Hayes},
  {Heckman}, {Hennessy}, {Hindsley}, {Holm}, {Holmgren}, {Huang}, {Hull},
  {Husby}, {Ichikawa}, {Ichikawa}, {Ivezi{\'c}}, {Kent}, {Kim}, {Kinney},
  {Klaene}, {Kleinman}, {Kleinman}, {Knapp}, {Korienek}, {Kron}, {Kunszt},
  {Lamb}, {Lee}, {Leger}, {Limmongkol}, {Lindenmeyer}, {Long}, {Loomis},
  {Loveday}, {Lucinio}, {Lupton}, {MacKinnon}, {Mannery}, {Mantsch}, {Margon},
  {McGehee}, {McKay}, {Meiksin}, {Merelli}, {Monet}, {Munn}, {Narayanan},
  {Nash}, {Neilsen}, {Neswold}, {Newberg}, {Nichol}, {Nicinski}, {Nonino},
  {Okada}, {Okamura}, {Ostriker}, {Owen}, {Pauls}, {Peoples}, {Peterson},
  {Petravick}, {Pier}, {Pope}, {Pordes}, {Prosapio}, {Rechenmacher}, {Quinn},
  {Richards}, {Richmond}, {Rivetta}, {Rockosi}, {Ruthmansdorfer}, {Sandford},
  {Schlegel}, {Schneider}, {Sekiguchi}, {Sergey}, {Shimasaku}, {Siegmund},
  {Smee}, {Smith}, {Snedden}, {Stone}, {Stoughton}, {Strauss}, {Stubbs},
  {SubbaRao}, {Szalay}, {Szapudi}, {Szokoly}, {Thakar}, {Tremonti}, {Tucker},
  {Uomoto}, {Vanden Berk}, {Vogeley}, {Waddell}, {Wang}, {Watanabe},
  {Weinberg}, {Yanny}, {Yasuda}, \& {SDSS Collaboration}}]{York2000}
{York}, D.~G., {Adelman}, J., {Anderson}, Jr., J.~E., {et~al.} 2000, \aj, 120,
  1579, \dodoi{10.1086/301513}

\bibitem[{{Zhu} {et~al.}(2018{\natexlab{a}}){Zhu}, {van de Ven},
  {M{\'e}ndez-Abreu}, \& {Obreja}}]{Zhu2018c}
{Zhu}, L., {van de Ven}, G., {M{\'e}ndez-Abreu}, J., \& {Obreja}, A.
  2018{\natexlab{a}}, \mnras, 479, 945, \dodoi{10.1093/mnras/sty1521}

\bibitem[{{Zhu} {et~al.}(2018{\natexlab{b}}){Zhu}, {van den Bosch}, {van de
  Ven}, {Lyubenova}, {Falc{\'o}n-Barroso}, {Meidt}, {Martig}, {Shen}, {Li},
  {Yildirim}, {Walcher}, \& {Sanchez}}]{Zhu2018a}
{Zhu}, L., {van den Bosch}, R., {van de Ven}, G., {et~al.} 2018{\natexlab{b}},
  \mnras, 473, 3000, \dodoi{10.1093/mnras/stx2409}

\bibitem[{{Zhu} {et~al.}(2018{\natexlab{c}}){Zhu}, {van de Ven}, {van den
  Bosch}, {Rix}, {Lyubenova}, {Falc{\'o}n-Barroso}, {Martig}, {Mao}, {Xu},
  {Jin}, {Obreja}, {Grand }, {Dutton}, {Macci{\`o}}, {G{\'o}mez}, {Walcher},
  {Garc{\'\i}a-Benito}, {Zibetti}, \& {S{\'a}nchez}}]{Zhu2018b}
{Zhu}, L., {van de Ven}, G., {van den Bosch}, R., {et~al.} 2018{\natexlab{c}},
  Nature Astronomy, 2, 233, \dodoi{10.1038/s41550-017-0348-1}

\bibitem[{{Zhuang} {et~al.}(2019){Zhuang}, {Leaman}, {van de Ven}, {Zibetti},
  {Gallazzi}, {Zhu}, {Falc{\'o}n-Barroso}, \& {Lyubenova}}]{Zhuang2019}
{Zhuang}, Y., {Leaman}, R., {van de Ven}, G., {et~al.} 2019, \mnras, 483, 1862,
  \dodoi{10.1093/mnras/sty2916}

\end{thebibliography}
\appendix
\section{Radial coverage test}\label{sec:radial_test}

The SAMI instrument has a fixed field of view ($15^{\prime\prime}$ diameter), meaning that each galaxy has a different maximum radial coverage. In particular, the most massive galaxies are larger than the SAMI field of view ($R_{\rm e} > 25^{\prime\prime}$) and therefore only their inner region is observed ($R_{max}$ $< 1 R_{\rm e}$).

\begin{figure}[ht!]
\centering
\includegraphics[scale=0.5]{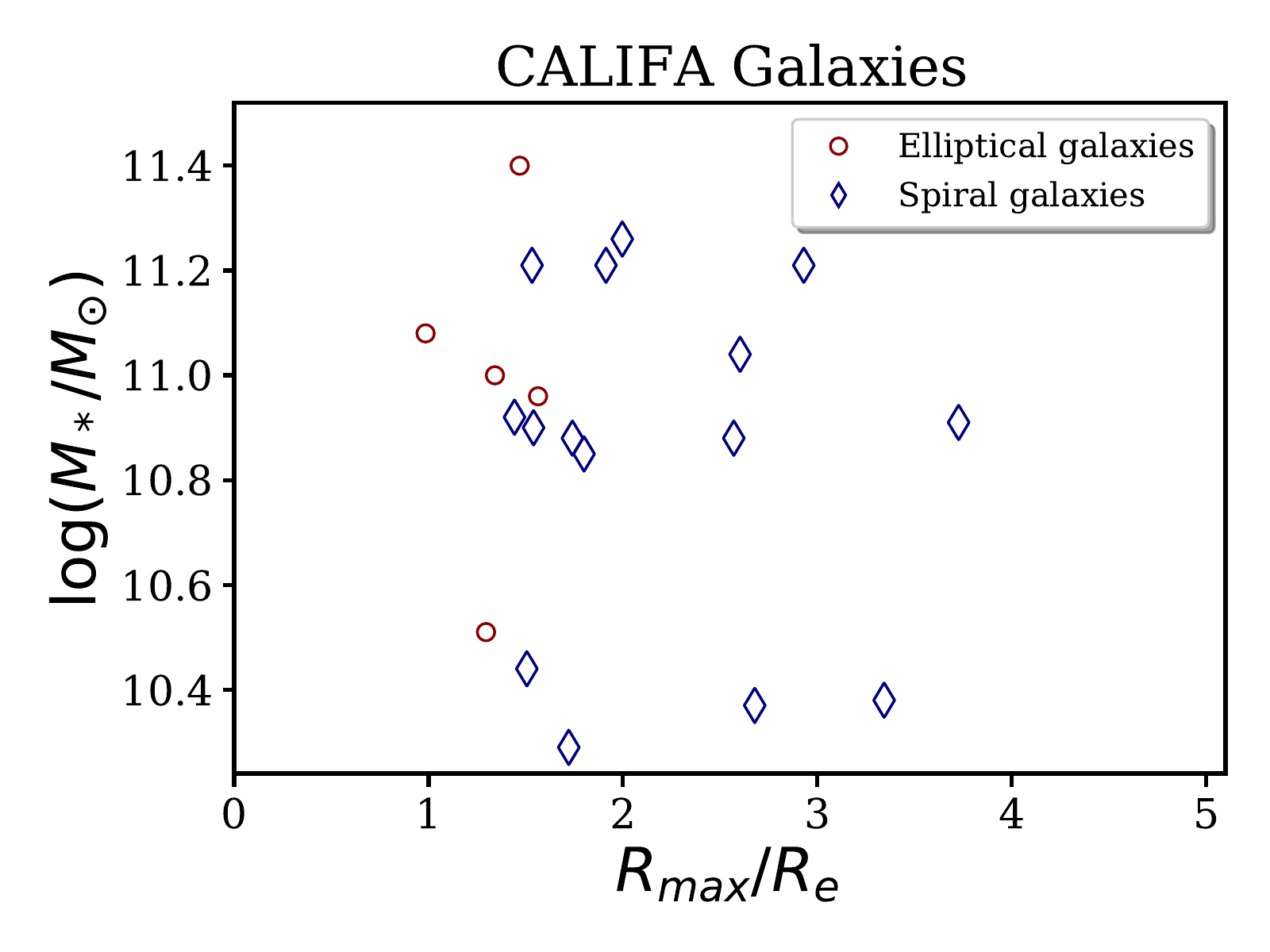}
\caption{Maximum radial extension (in units of $R_{\rm e}$) of the 20 CALIFA galaxies in our test sample. Elliptical galaxies are shown as dark red circles, spiral galaxies as  dark blue diamonds. Seven of the galaxies have kinematic maps which extend beyond 2 $R_{\rm e}$.}
\label{fig:califa_rmax}
\end{figure}

In order to test the reliability of results obtained from applying Schwarzschild models to galaxies with measurements that do not reach the same maximum radial extension and have a limited number of spatial bins, we selected a test sample consisting of 20 randomly selected CALIFA galaxies, covering different radial extents (Fig. \ref{fig:califa_rmax}). For each galaxy we have taken the CALIFA stellar kinematic maps \citep{Falcon-Barroso2017} and masked them at different radii, in order to have maps for each galaxy that extend up to $R_{max} = 0.5 R_{\rm e}$, $1 R_{\rm e}$, $1.5 R_{\rm e}$ and $2 R_{\rm e}$ (when possible), respectively. We then determined the best-fit model for each realization of the maps, in addition to fitting the whole galaxy (a set of up to 5 maps for each galaxy, depending on their radial coverage). We take the effective radius $R_{\rm e}$ from \cite{Falcon-Barroso2017}.

\begin{figure*}[ht!]
\centering
\includegraphics[scale=0.4, trim=2cm 0cm 0cm 0cm, clip=True]{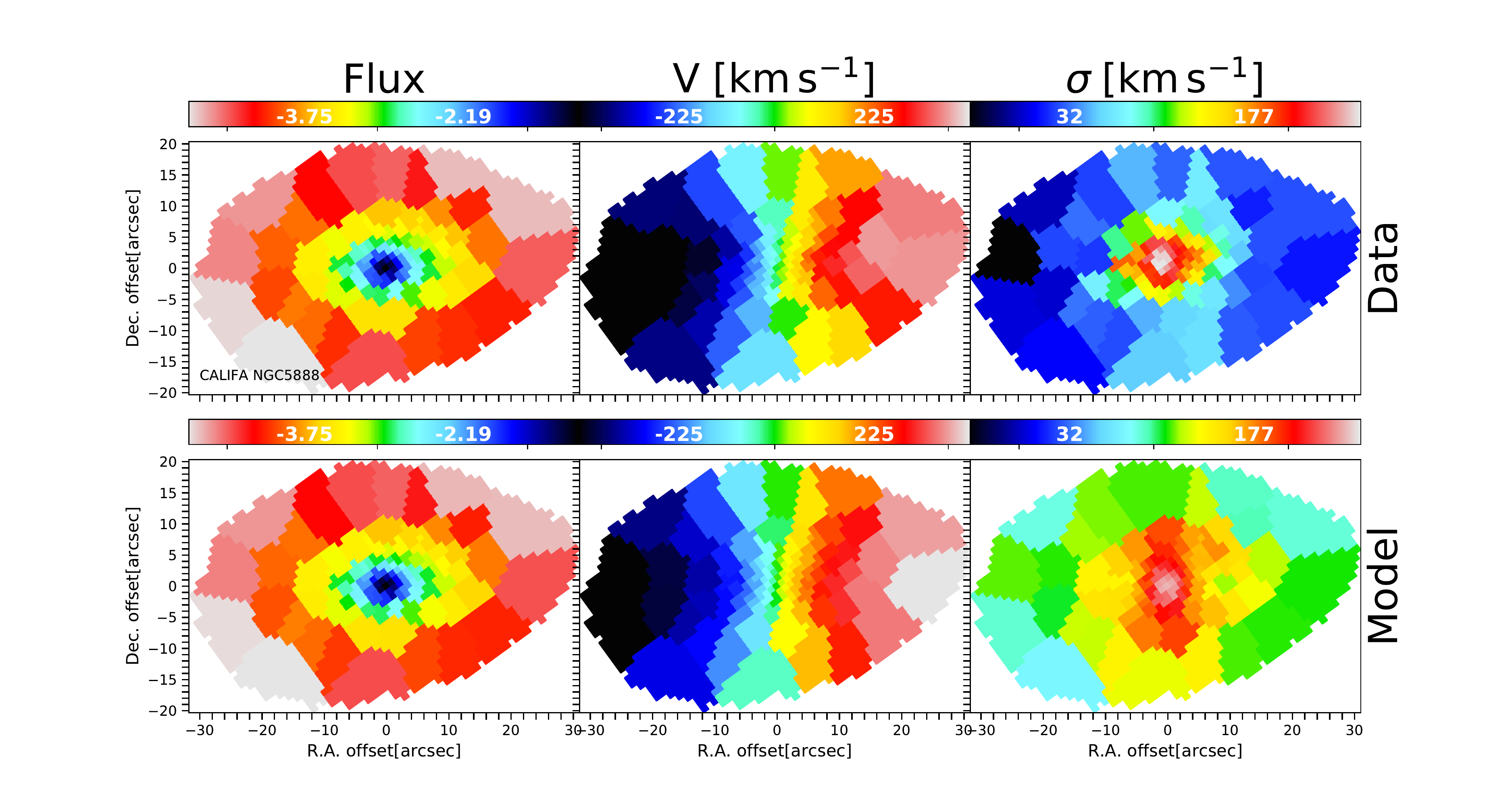}
\caption{Best-fit model for CALIFA test galaxy NGC5888 using 2-moments maps. Top: Observed luminosity, velocity and velocity dispersion Bottom: best-fit model luminosity, velocity and velocity dispersion. The model does not reproduce the velocity dispersion well.}
\label{fig:NGC5888_h2}
\end{figure*}

\begin{figure*}[ht!]
\centering

\includegraphics[scale=0.4, trim=2.5cm 0cm 0cm 0cm, clip=True]{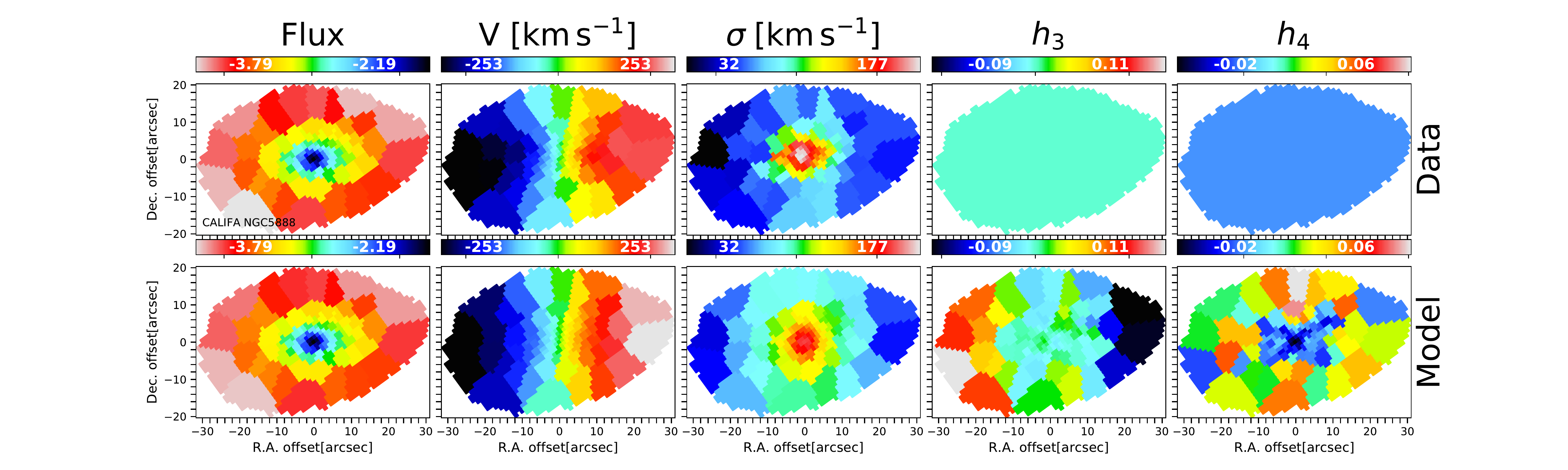}
\caption{Best-fit model for CALIFA test galaxy NGC5888 using 4-moments maps as per Fig. \ref{fig:NGC5888_h2}, including the higher-order stellar kinematic moments ($h_3$ and $h_4$, set to zero) in the fits. Even though the values of the observed $h_3$ and $h_4$ are set to zero, the model is better able to reproduce the velocity dispersion map compared to Fig. \ref{fig:NGC5888_h2}.}
\label{fig:NGC5888_h3h4}
\end{figure*}

For each of the 20 galaxies in our test sample we compare the retrieved orbital distributions, inclination angle of the galaxy, enclosed dark matter mass and enclosed total mass within 1$R_{\rm e}$, for the five different kinematic maps, to those obtained by \cite{Zhu2018c}. In most cases (16/20), the best-fit models reproduce the observed luminosity, velocity and velocity dispersion maps when all the parameters are unconstrained. However, the models reproduce the observations better (particularly the velocity dispersion) when we include the higher-order stellar kinematic moments ($h_3$ and $h_4$), even if the observed $h_3$ and $h_4$ are set to zero with large uncertainties set to 0.5. We show an example fit in Fig. \ref{fig:NGC5888_h2} and \ref{fig:NGC5888_h3h4}. The reduced $\chi^2$ decreases significantly from $\chi^2_{red} = 23.71 $ when the higher-order moments are not included in the fit, to $\chi^2_{red} = 4.52$ when $h_3$ and $h_4$ are free parameters.


In general, our retrieved best-fit values of orbital weights and enclosed mass are comparable to those found in \cite{Zhu2018c}. However, galaxies that are found to have low inclination angle ($\approx$ 40$^{\circ}$ - 50$^{\circ}$) in \cite{Zhu2018c} have a higher inclination angle in our best-fit model ($\approx$ 65$^{\circ}$). Moreover, due to the higher inclination angle, these galaxies show a lower fraction of cold orbits (required to reproduce the observed velocity dispersion). We note that those galaxies with a low observed inclination angle require stricter priors for the intrinsic shape parameters. 

\begin{figure}[ht!]
\centering
\includegraphics[scale=0.5]{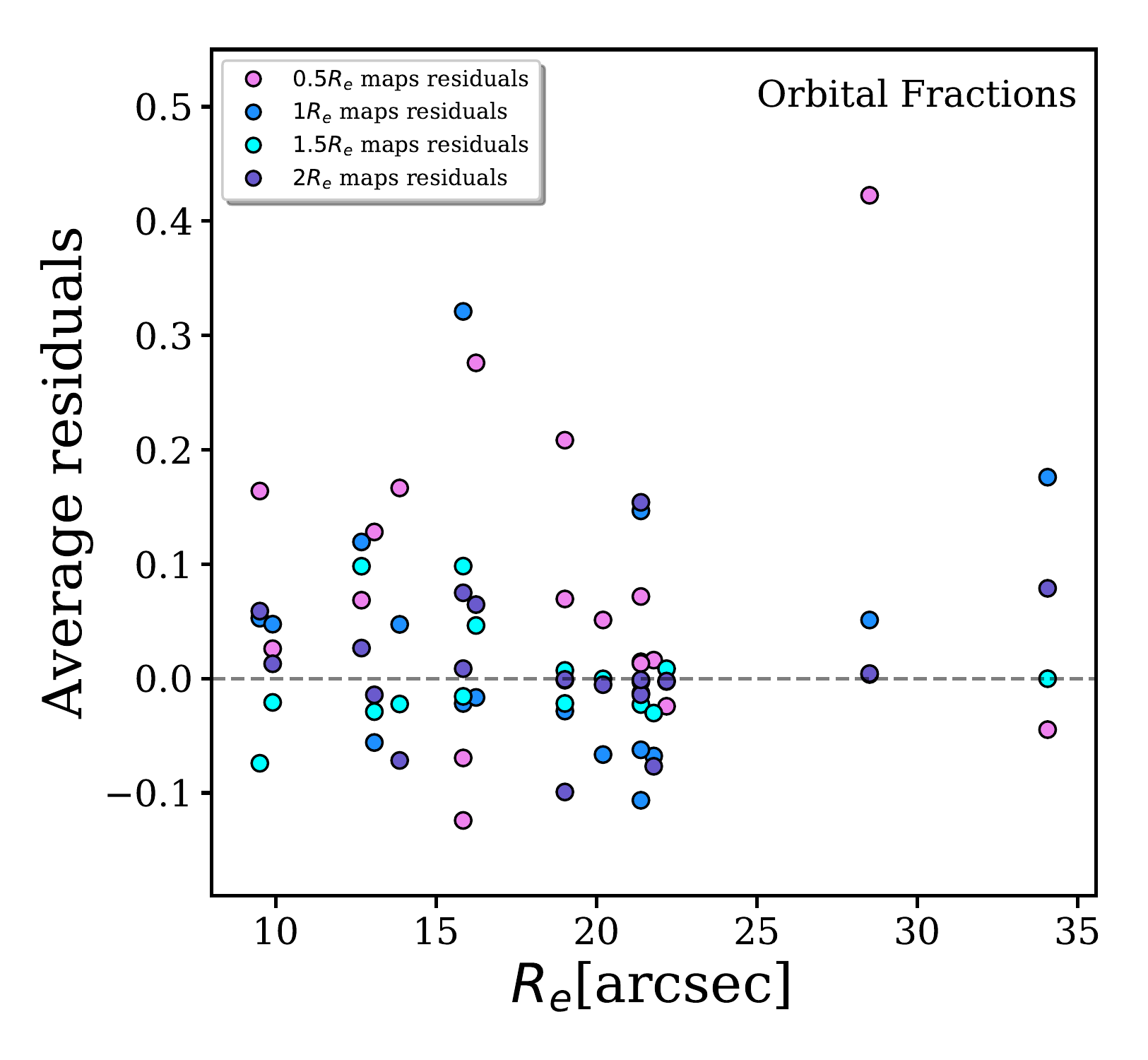}
\caption{Average residuals between the derived orbital fractions of each of the 4 maps from the values derived from the total maps for the galaxies in our test sample, as a function of $R_{\rm e}$. For each map, the residuals of the four orbital components are calculated following \ref{eq:res} and then averaged over the orbital components. Each point corresponds to the average value, color-coded by the value of $R_{max}$ of the map as shown in the bottom right corner. Comparing the derived values within 1$R_{\rm e}$ of the different maps for each galaxy, we find a general good agreement for all input $R_{max}$ maps, with the exception of those retrieved from the $R_{max} =$ 0.5$R_{\rm e}$ maps, which show a large scatter.}
\label{fig:ngc6411_weights}
\end{figure}

\begin{figure}[ht!]
\centering
\includegraphics[scale=0.5]{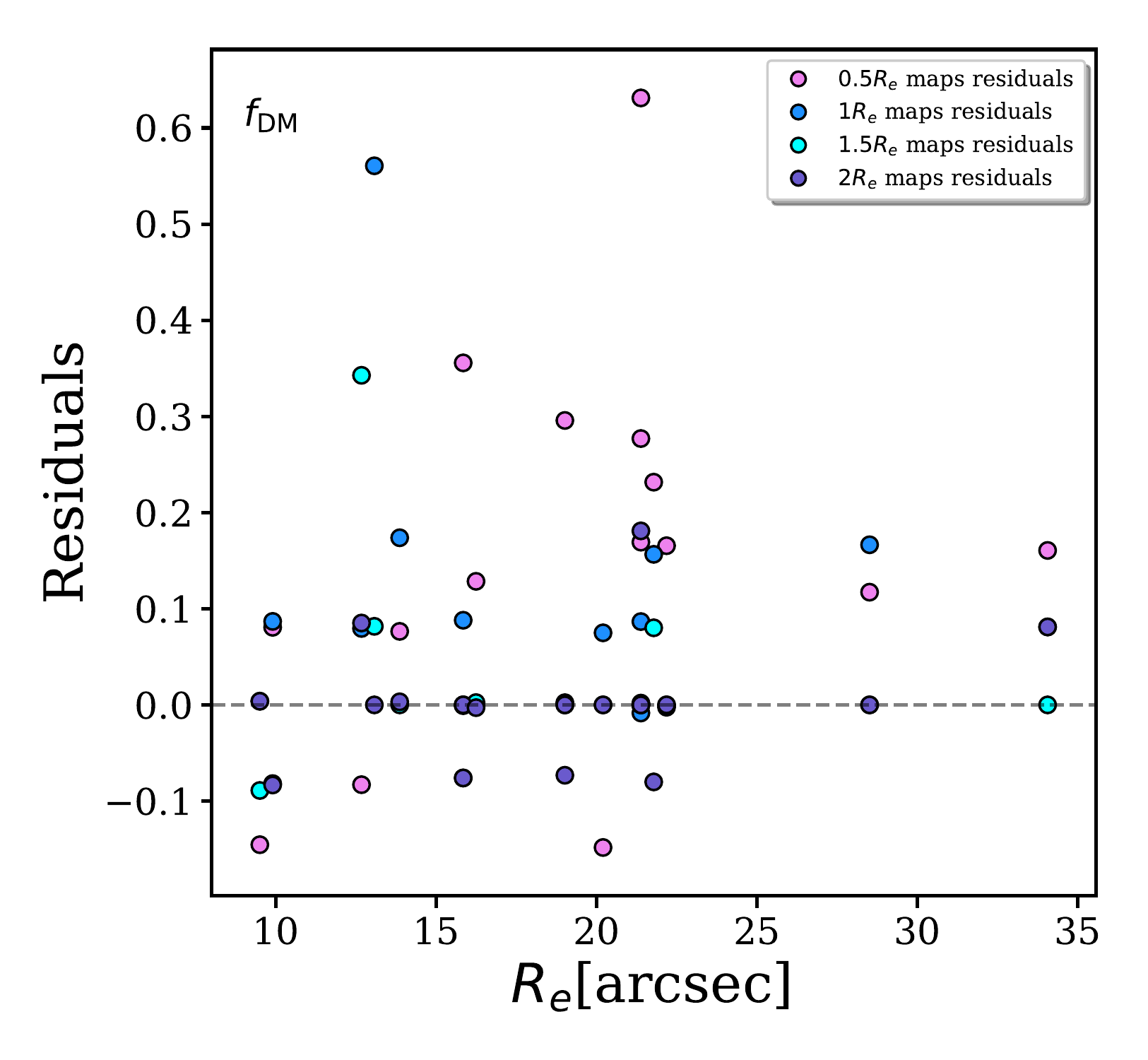}\includegraphics[scale=0.5]{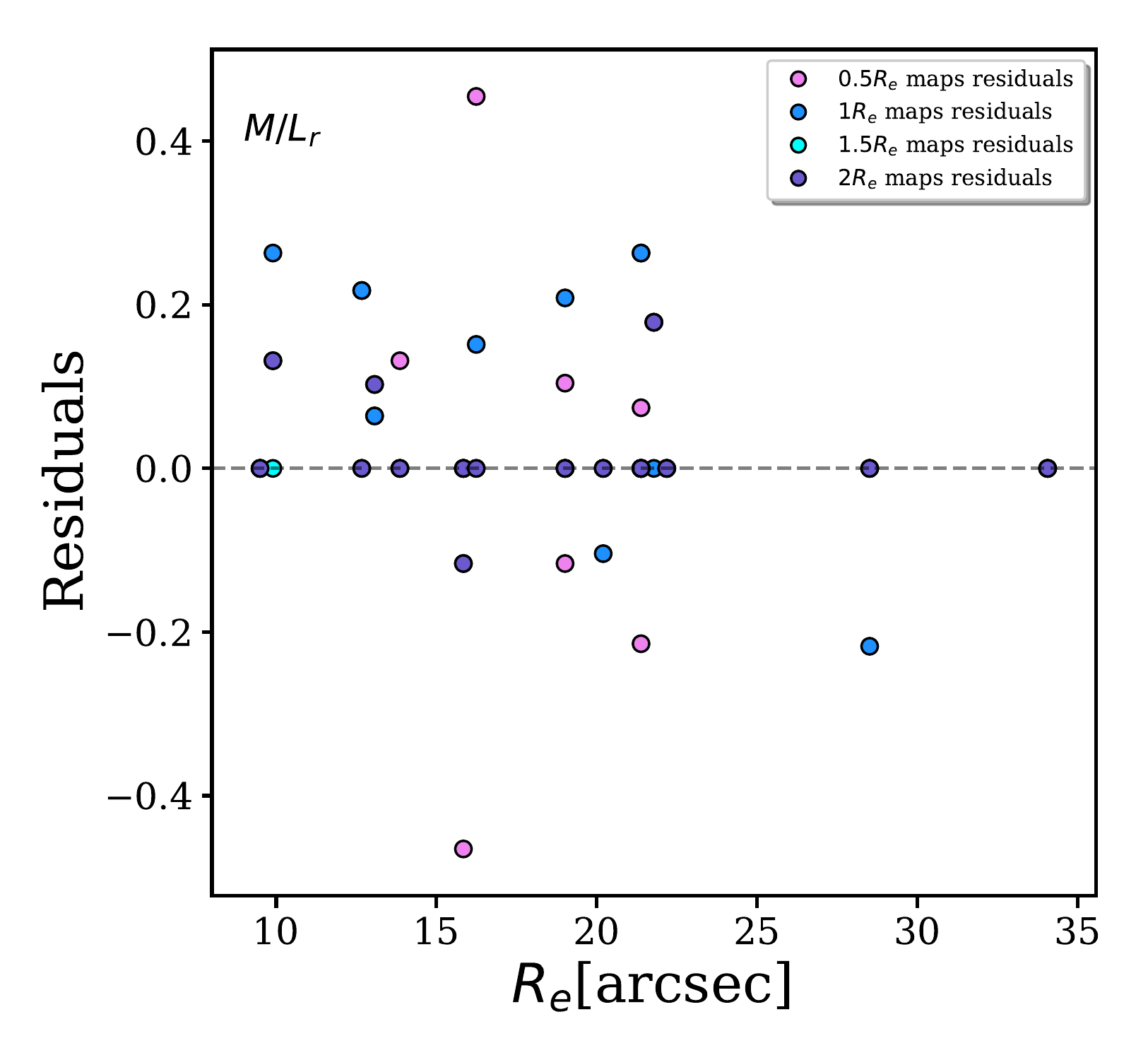}\\
\includegraphics[scale=0.5]{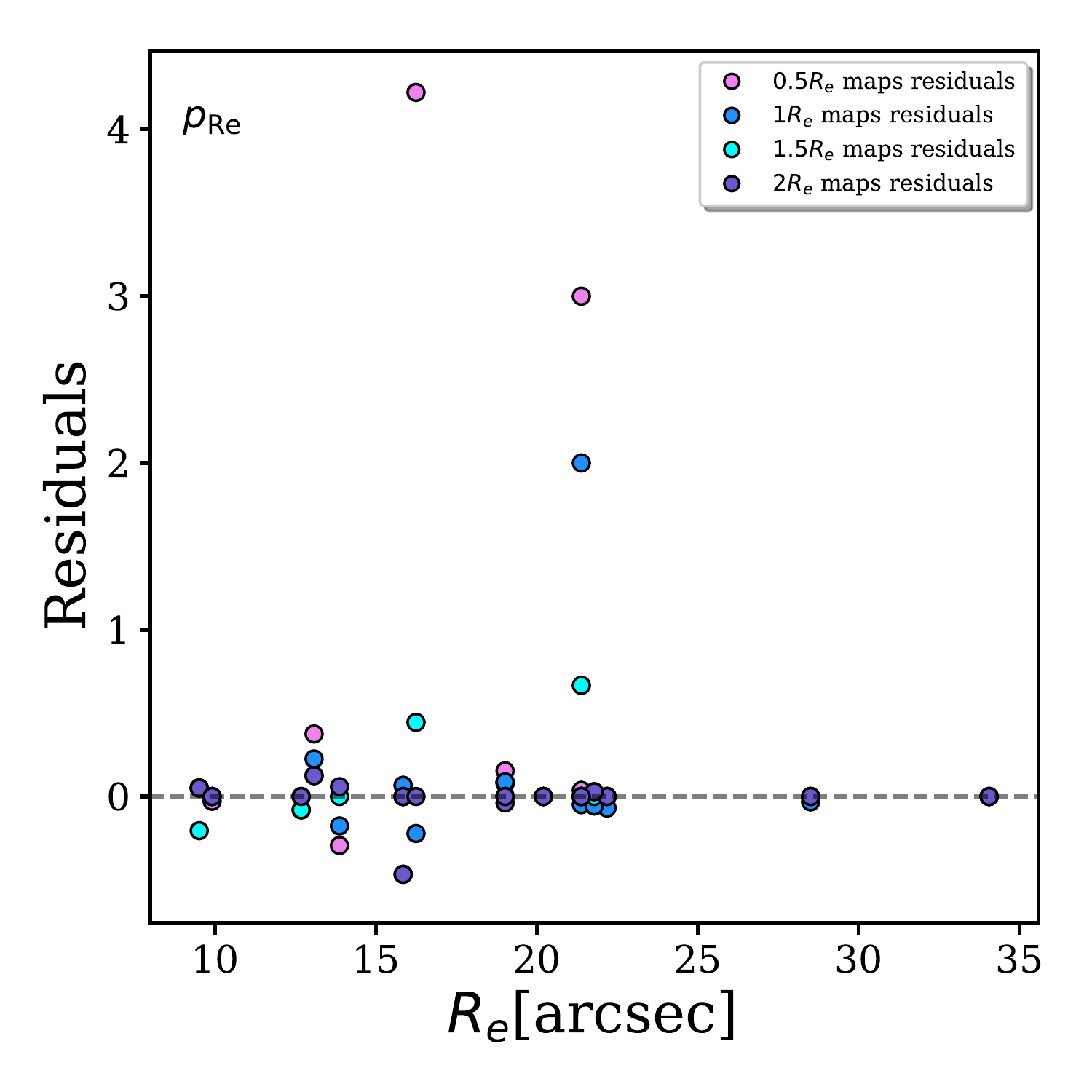}\includegraphics[scale=0.5]{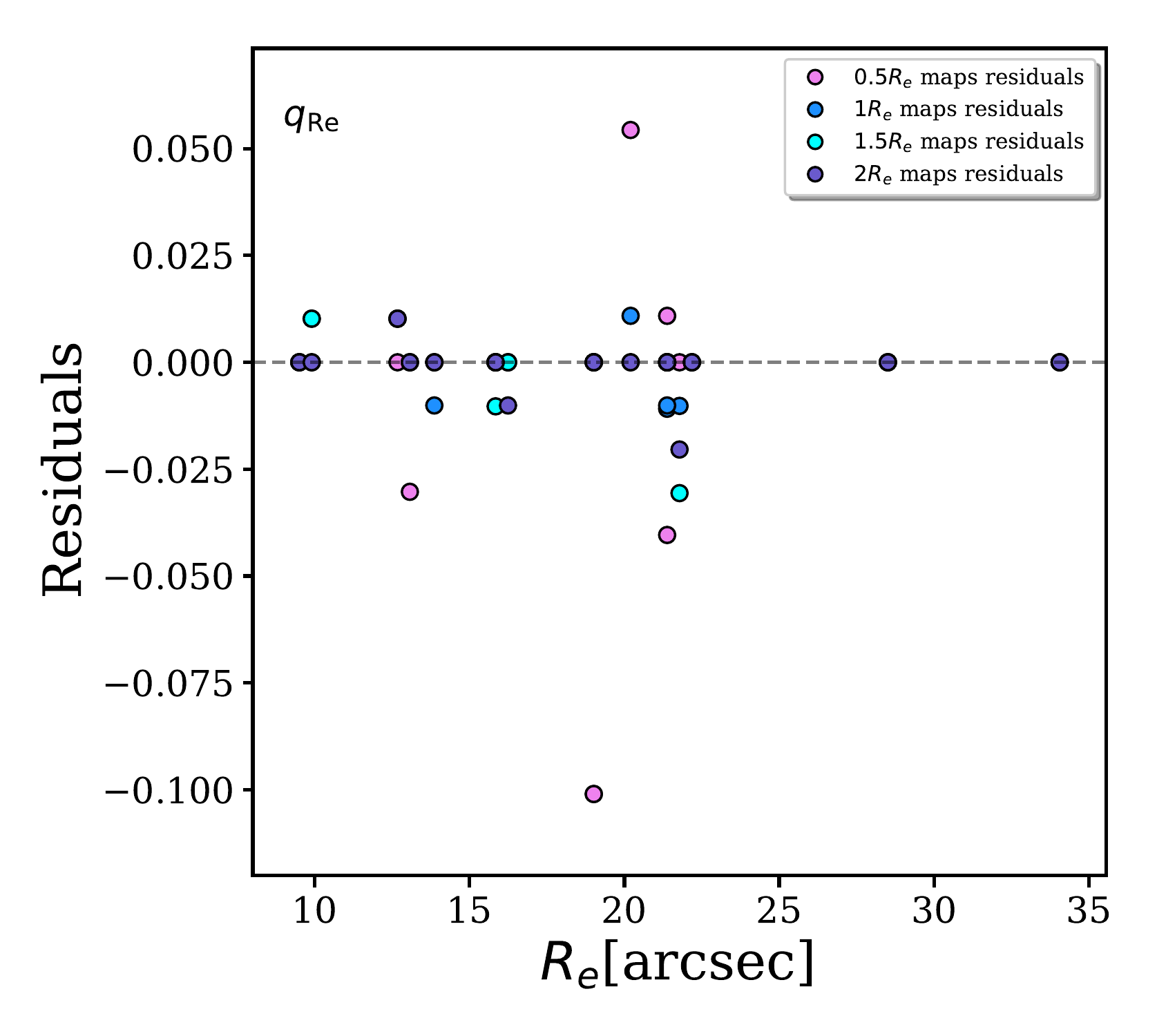}
\caption{Average residuals between the derived fraction of dark matter within 1$R_{\rm e}$ ($f_{\rm DM}$; top left), mass-to-light ratio in the $r$-band ($M_{\star}/L_r$; top right) and intrinsic axis ratios $p_{Re}$ (bottom left) and $q_{Re}$ (bottom right) of each of the 4 maps from the values derived from the total maps for the galaxies in our test sample, as a function of $R_{\rm e}$. For each map, the residuals of the four orbital components are calculated following \ref{eq:res}. Each point is color-coded by the value of $R_{max}$ of the map as shown in the bottom right corner. Comparing the derived values within 1$R_{\rm e}$ of the different maps for each galaxy, we find a general good agreement for all input $R_{max}$ maps, with the exception of those retrieved from the $R_{max} =$ 0.5$R_{\rm e}$ maps, which show a large scatter.}
\label{fig:ngc6411_weights_extra}
\end{figure}

Fig. \ref{fig:ngc6411_weights} shows the average residuals between the derived orbital fractions of each of the 4 maps from the values derived from the total maps for the galaxies in our test sample. For each map, the residual for each orbital component is given by:

\begin{equation}\label{eq:res}
    \delta = \frac{f_{{orb}_{TOT}} - f_{{orb}_{map}}}{f_{{orb}_{TOT}}} 
\end{equation}

where $f_{{orb}_{TOT}}$ is the orbital fraction for cold, warm, hot or counter-rotating (CR) - derived from the total map and $f_{{orb}_{map}}$ is the orbital fraction derived for one of the 4 kinematic maps - $R_{max} = 0.5 R_{\rm e}$, $1 R_{\rm e}$, $1.5 R_{\rm e}$ and $2 R_{\rm e}$.
Each point in Fig. \ref{fig:ngc6411_weights} shows the average of the 4 residuals (one for each orbital component), color-coded by the $R_{max}$ of the maps. 
We also show the residuals for the fraction of dark matter within 1$R_{\rm}$ ($f_{\rm DM}$), the mass-to-light ratio in the $r$-band ($M/L_r$) and the intrinsic axis ratios at 1$R_{\rm e}$ - $p_{\rm Re}$ and $q_{\rm Re}$ in Fig. \ref{fig:ngc6411_weights_extra}. The average residuals for each of the maps is shown in Table \ref{tab:ave_res}.

Comparing the derived values within 1$R_{\rm e}$ of the different maps for each galaxy, we find a general good agreement for all input $R_{max}$ maps, with the exception of those retrieved from the $R_{max} =$ 0.5$R_{\rm e}$ maps, which show a large scatter. We are therefore confident in the values estimated within 1$R_{\rm e}$ calculated using maps that extend to at least 1$R_{\rm e}$ for the analysis presented here.

\begin{table}[!ht]
	\begin{center}
		\begin{tabular}{ l l l l l l l } 
		     \hline
			 \multirow{2}{*}{\textbf{Radial Coverage}} & \multicolumn{4}{c}{\textbf{Residuals}} \\
			 
			  & Orbital Fractions & $f_{\rm DM}$ & $M/L_r$ & $p_{Re}$ & $q_{Re}$\\
			 \hline
			 0.5$R_{\rm e}$ & 0.080& 0.145 & 0.046 & 0.428 & -0.012 \\
			 1$R_{\rm e}$ &  0.030 & 0.086 & 0.047 & 0.101 & -0.001\\
			 1.5$R_{\rm e}$ & 0.001& 0.015 & 0.009 & 0.027 & -0.001\\ 
			 2$R_{\rm e}$ & 0.011  & 0.002 & 0.016 & -0.013 & -0.001\\  
			\hline
			
		\end{tabular}
	\end{center}
	\caption{Average residuals between the derived orbital fractions, fraction of dark matter within 1$R_{\rm e}$, mass-to-light ratio in the $r$-band ($M_{\star}/L_r$) and intrinsic axis ratios $p_{Re}$ and $q_{Re}$ of each of the 4 maps from the values derived from the total maps for the galaxies in our CALIFA test sample. Comparing the derived values within 1$R_{\rm e}$ of the different maps for each galaxy, we find a general good agreement for all input $R_{max}$ maps, with the exception of those retrieved from the $R_{max} =$ 0.5$R_{\rm e}$ maps, which have larger average residual.}
	\label{tab:ave_res}
\end{table}

\section{Example Galaxies 9403800123, 9011900793, 220465 and 9008500323}\label{sec:example_gals}

\begin{figure*}
\centering
\includegraphics[width=19cm, trim=2cm 2cm 2cm 2cm,clip=True]{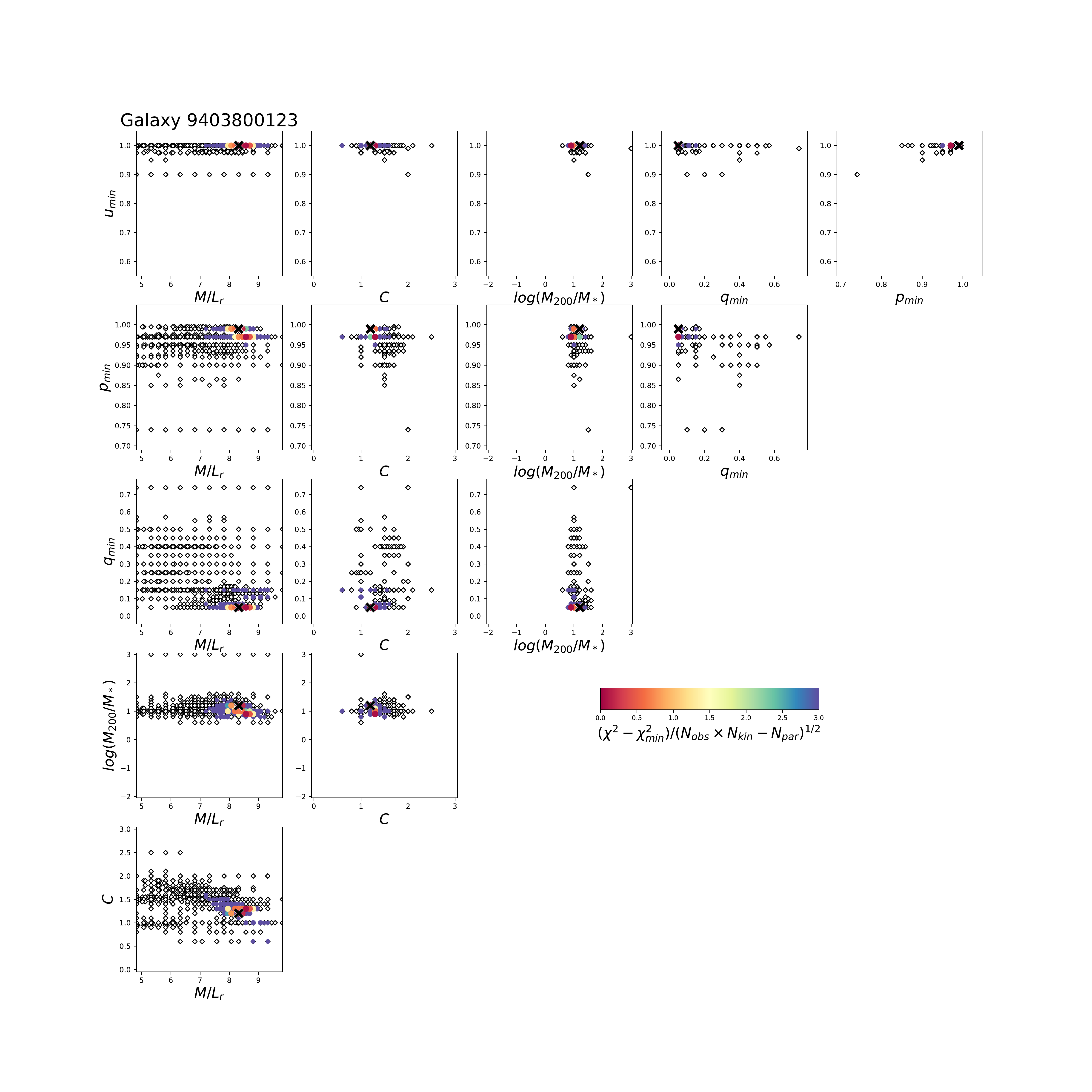}
\caption{Example galaxy 9403800123: model parameter grid. There are six free parameters: stellar mass-to-light ratio, $M_{\star}/L_r$ in solar units, the intrinsic shape of the flattest Gaussian component ($p_{min}$, $q_{min}$, $u_{min}$), the dark matter halo concentration, $\log c$, and dark matter fraction, $log M_{200}/M_{\star}$. The diamonds represent the parameters explored, with the best-fit model highlighted with a black cross. Models within the best-fit region are color-coded according to their $\chi^2$ values shown in the color bar. The best-fit values are well constrained.}
\label{fig:123_grid}
\end{figure*}

\begin{figure*}
\centering
\includegraphics[width=19cm, trim=2cm 2cm 2cm 2cm,clip=True]{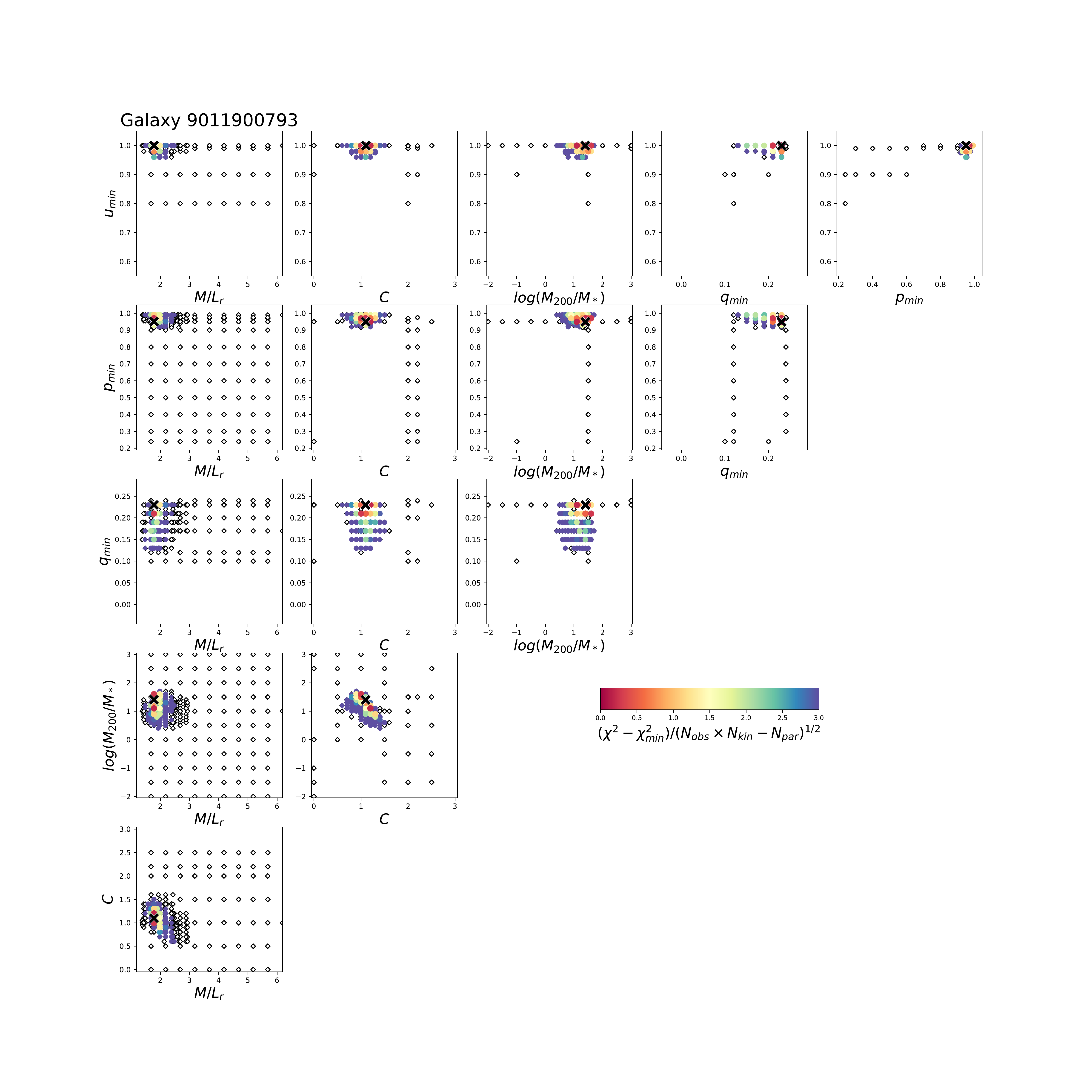}
\caption{Example galaxy 9011900793: model parameter grid. The diamonds represent the parameters explored, with the best-fit model highlighted with a black cross. Models within the best-fit region are color-coded according to their $\chi^2$ values shown in the color bar. }
\label{fig:793_grid}
\end{figure*}

\begin{figure*}
\centering
\includegraphics[width=19cm, trim=2cm 2cm 2cm 2cm,clip=True]{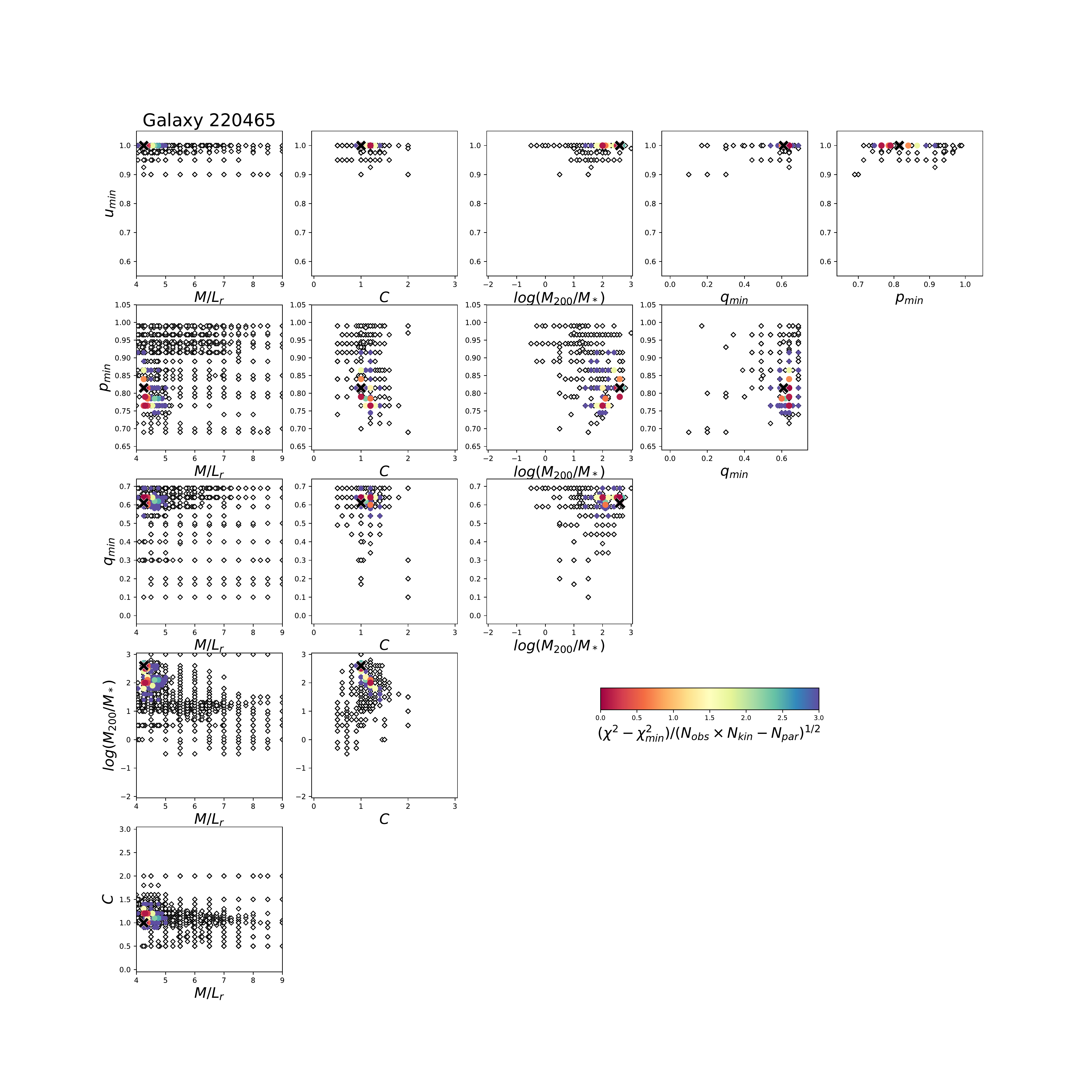}
\caption{Example galaxy 220465: model parameter grid. The diamonds represent the parameters explored, with the best-fit model highlighted with a black cross. Models within the best-fit region are color-coded according to their $\chi^2$ values shown in the color bar. }
\label{fig:465_grid}
\end{figure*}

\begin{figure*}
\centering
\includegraphics[width=19cm, trim=2cm 2cm 2cm 2cm,clip=True]{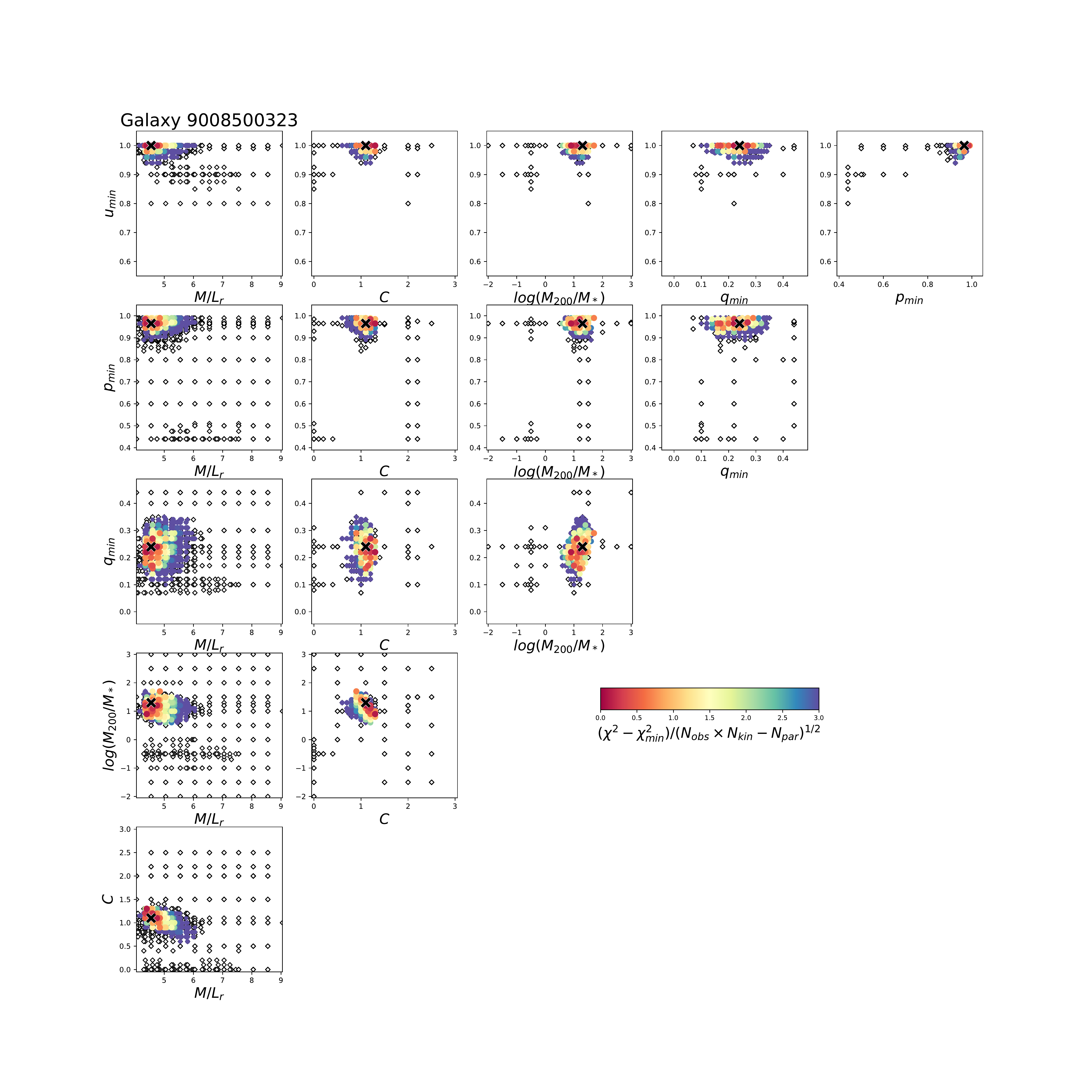}
\caption{Example galaxy 9008500323: model parameter grid. The diamonds represent the parameters explored, with the best-fit model highlighted with a black cross. Models within the best-fit region are color-coded according to their $\chi^2$ values shown in the color bar. }
\label{fig:323_grid}
\end{figure*}

\begin{figure}[ht!]
\centering
\includegraphics[scale=0.70]{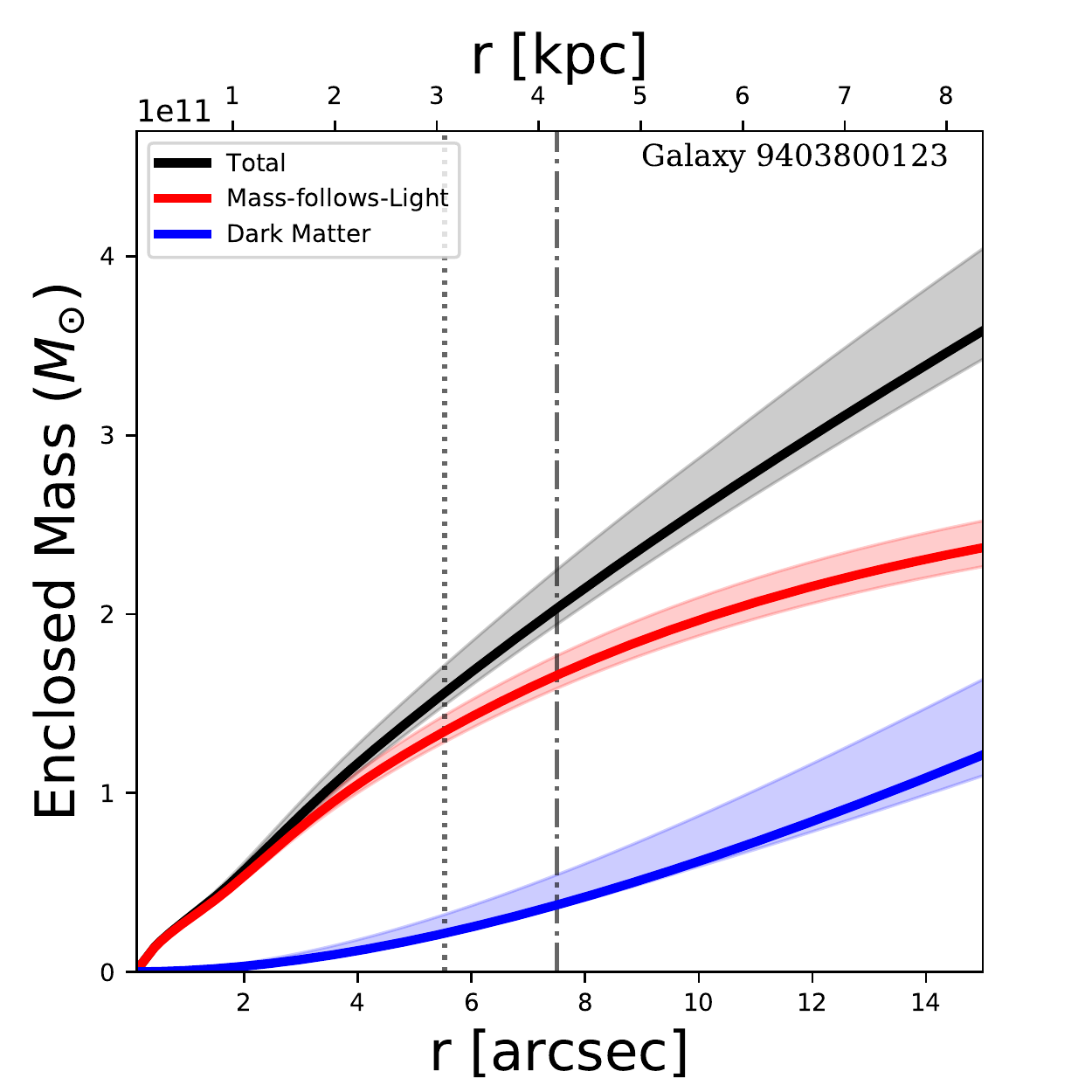}\includegraphics[scale=0.70]{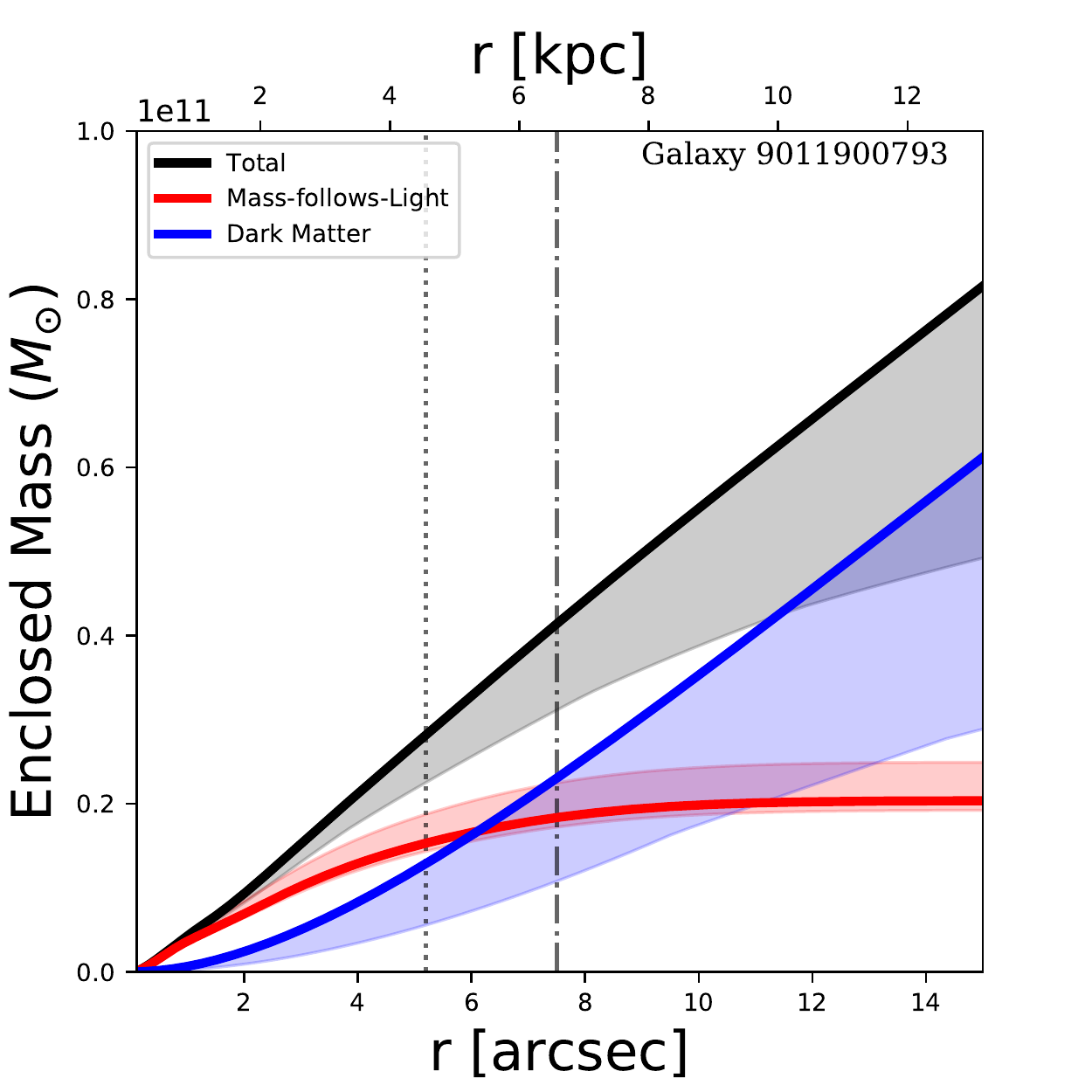}\\
\includegraphics[scale=0.70]{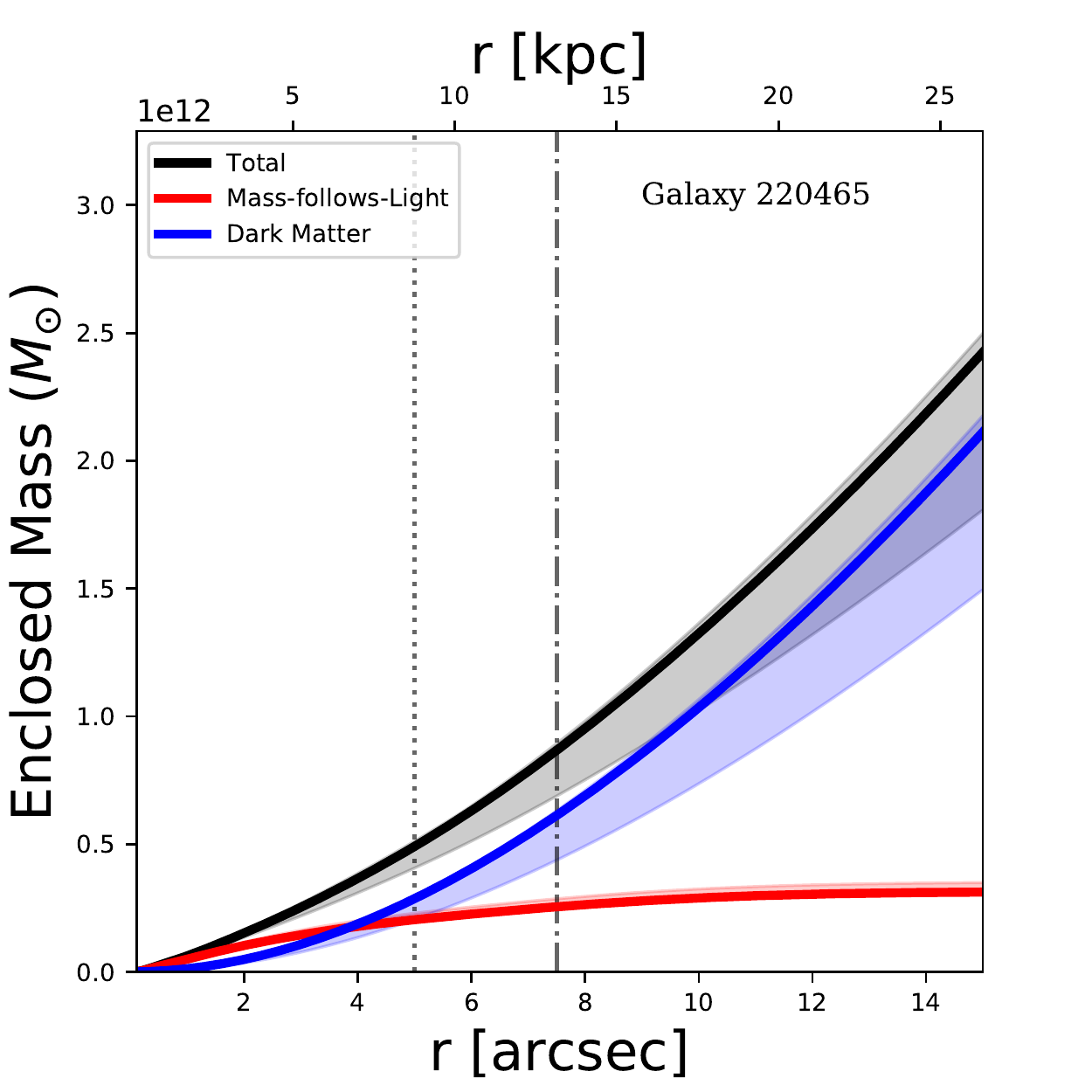}\includegraphics[scale=0.70]{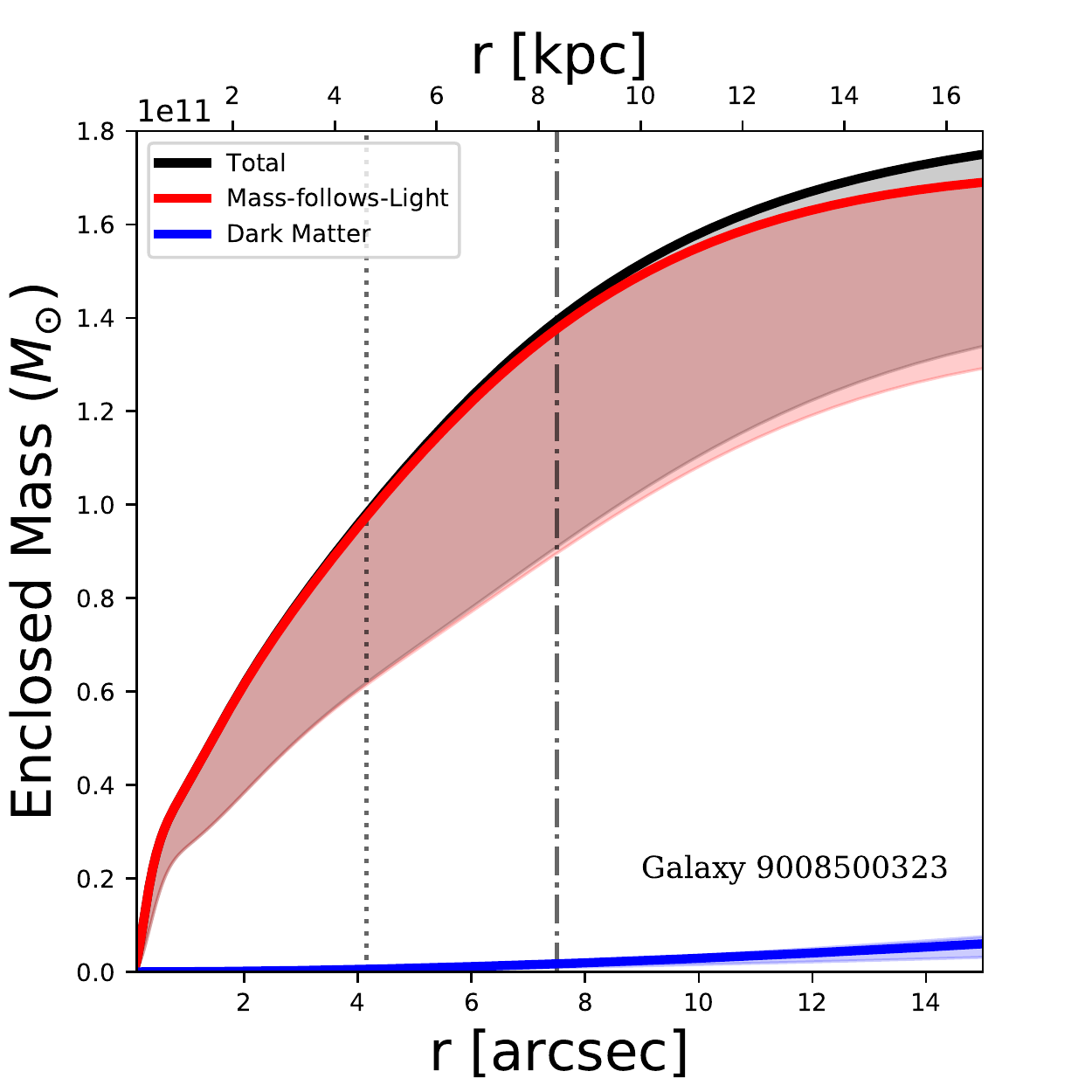}
\caption{Example galaxy 9403800123 (top left panel), 9011900793 (top right panel), 220465 (bottom left panel) and 9008500323 (bottom right panel): enclosed mass. Cumulative total mass (in black), stellar mass (in red) and dark matter mass (in blue) as a function of the radius of the galaxy. Solid lines are the cumulative profiles calculated from the best-fit, while the filled regions indicate the errors. Grey dotted and dash-dotted lines are located at 1$R_{\rm e}$ and at $R_{max}$, respectively. At larger radii the dark matter contribution becomes more important.}
\label{fig:mass}
\end{figure}

\begin{figure}[ht!]
\centering
\includegraphics[scale=0.6]{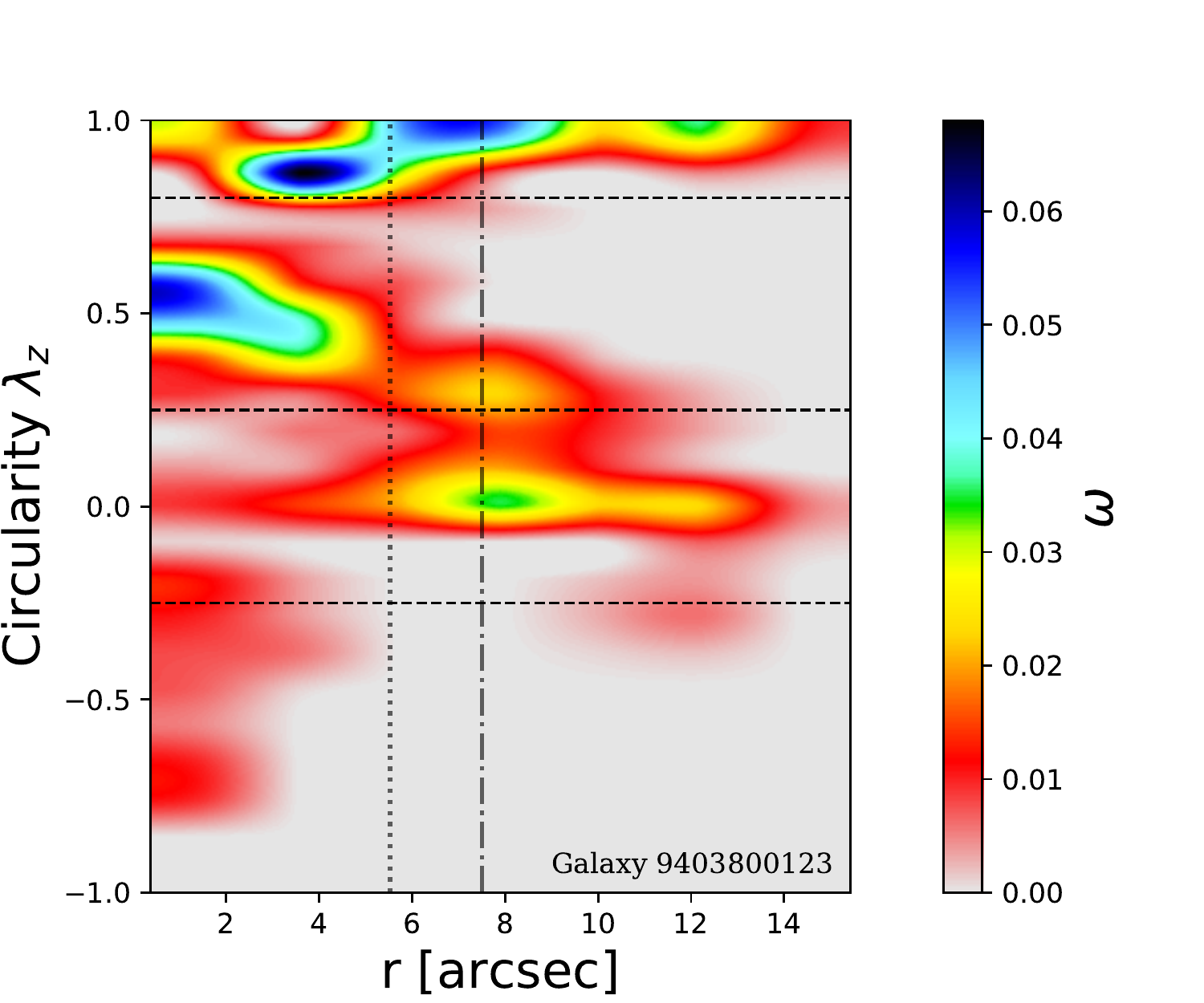}\includegraphics[scale=0.6]{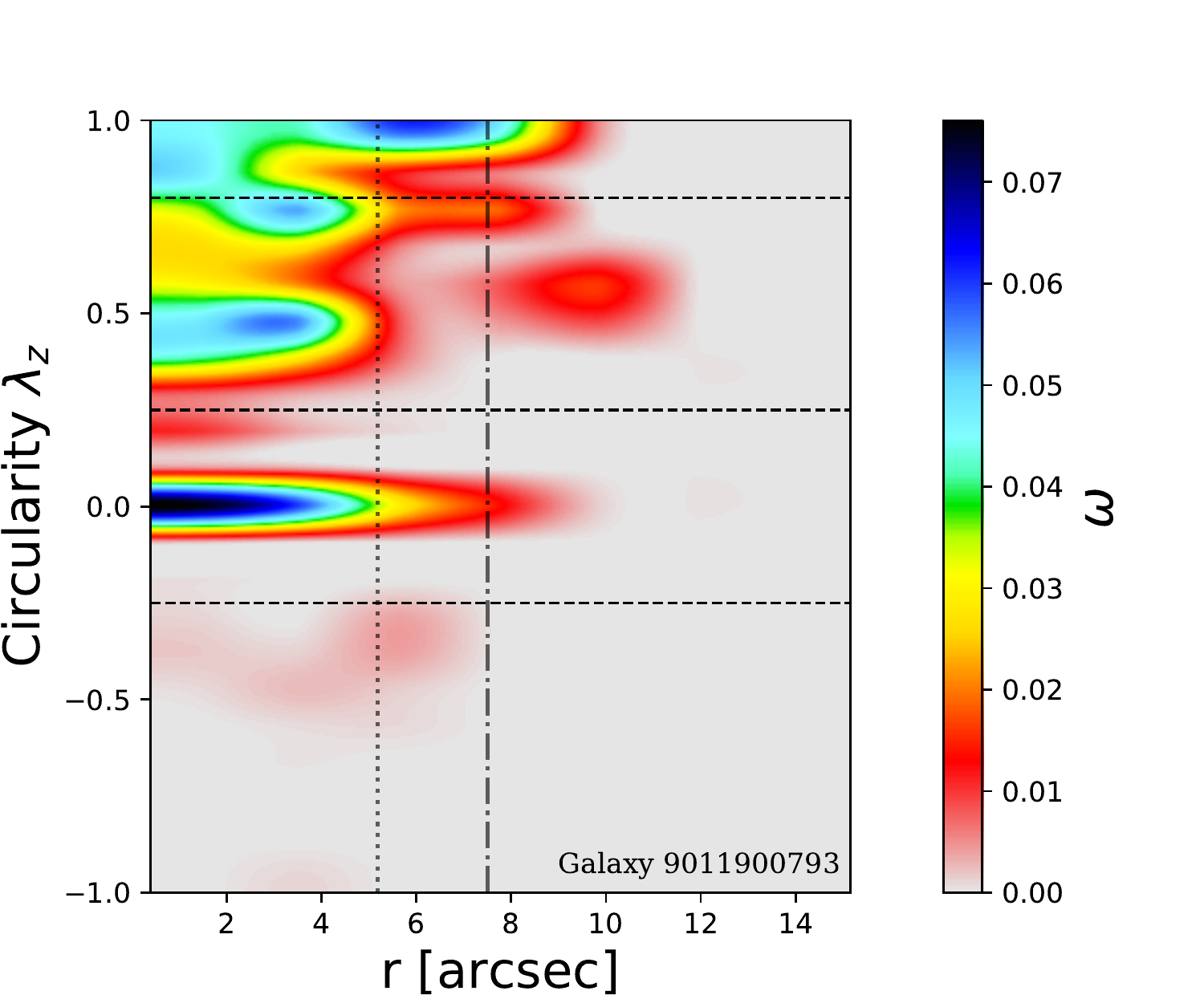}\\
\includegraphics[scale=0.6]{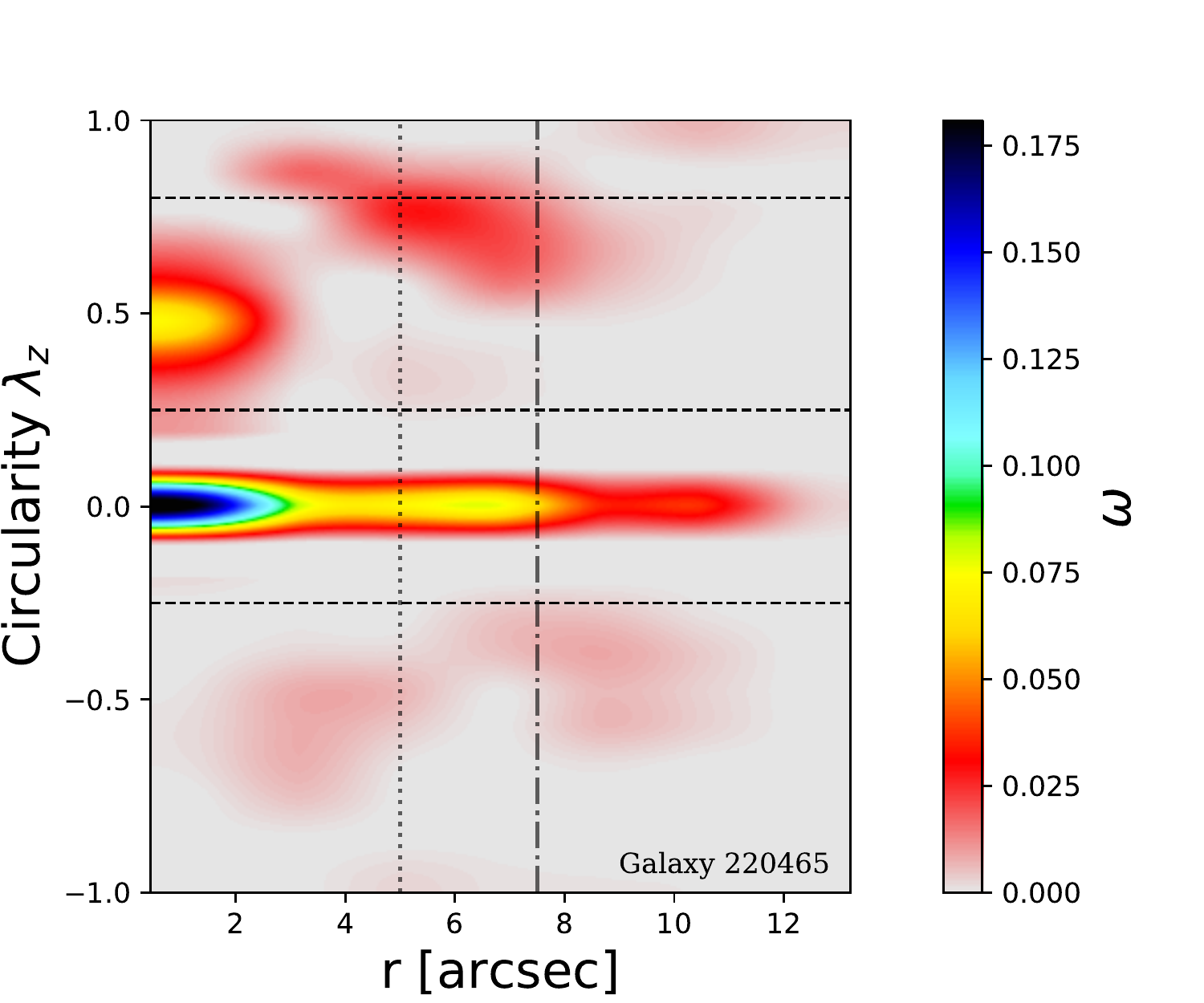}\includegraphics[scale=0.6]{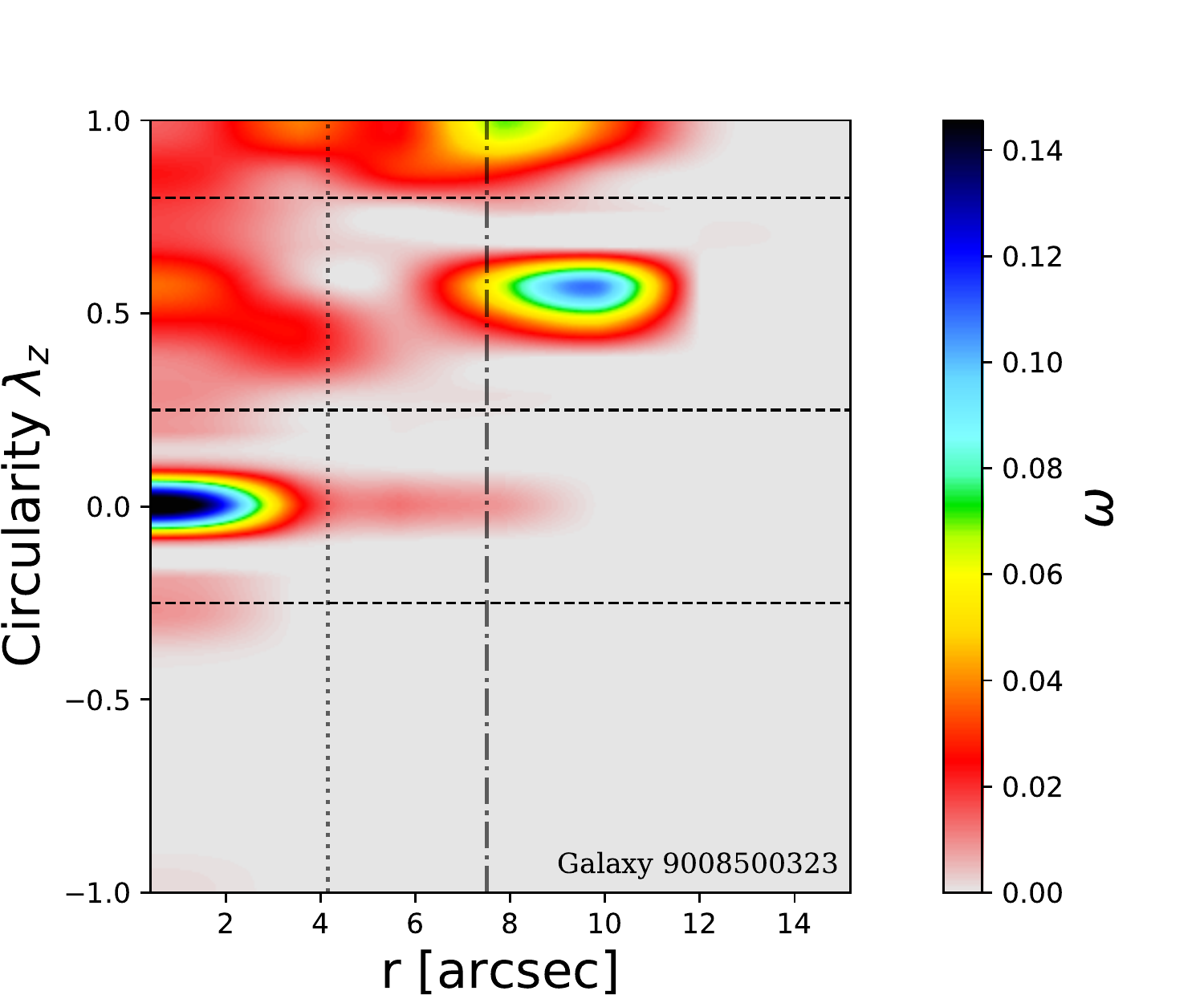}
\caption{Example galaxy 9403800123 (top left panel), 9011900793 (top right panel), 220465 (bottom left panel) and 9008500323 (bottom right panel): orbit circularity. The orbit distribution on the phase space of circularity $\lambda_z$ versus radius of the best-fit model. The color indicates the density of the orbits on the phase space, the three horizontal black dashed lines indicate $\lambda_z$ = 0.8, $\lambda_z$ = 0.25 and $\lambda_z$ = -0.25, dividing the orbits in four regions (cold, warm, hot and counter-rotating orbits). The vertical grey dotted and dash-dotted lines are located at 1$R_{\rm e}$ and at $R_{max}$, respectively. Galaxy 9403800123 is dominated by warm and cold orbits. Galaxies 9011900793, 220465, 9008500323 are dominated by hot orbits, but galaxy 9011900793 also has contributions from warm and cold orbits.}
\label{fig:orbits}
\end{figure}

\begin{figure}[ht!]
\centering
\includegraphics[scale=0.7]{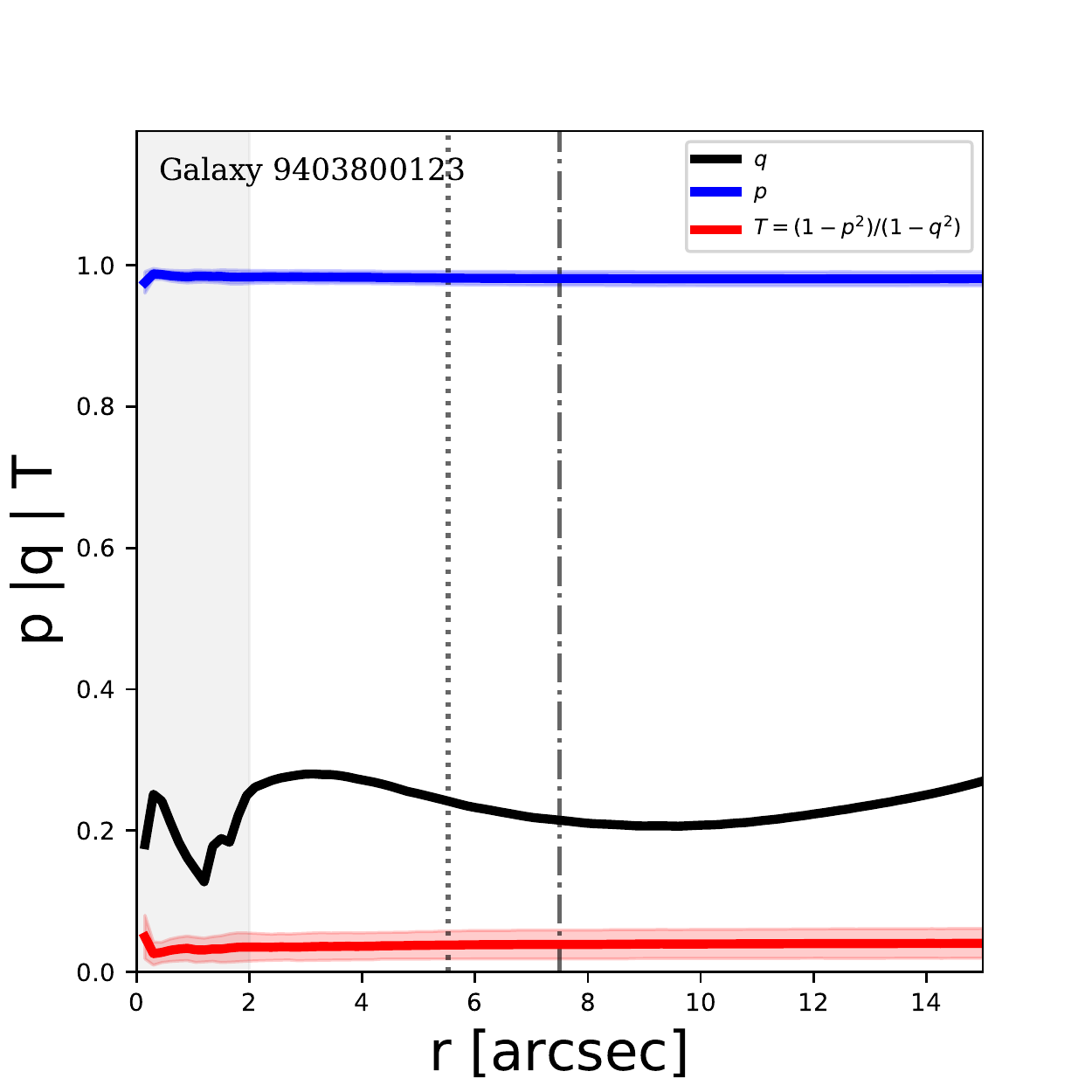}\includegraphics[scale=0.7]{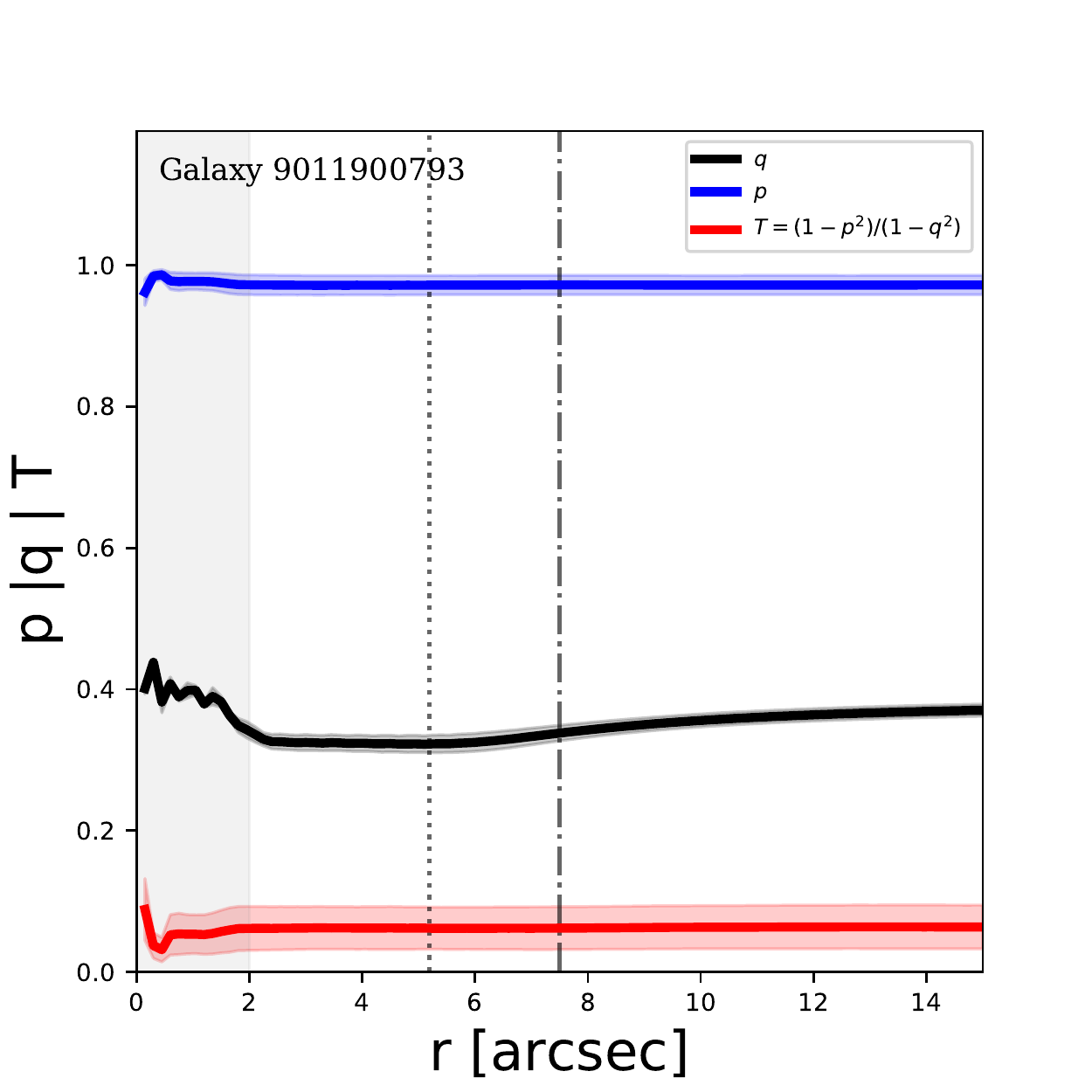}\\
\includegraphics[scale=0.7]{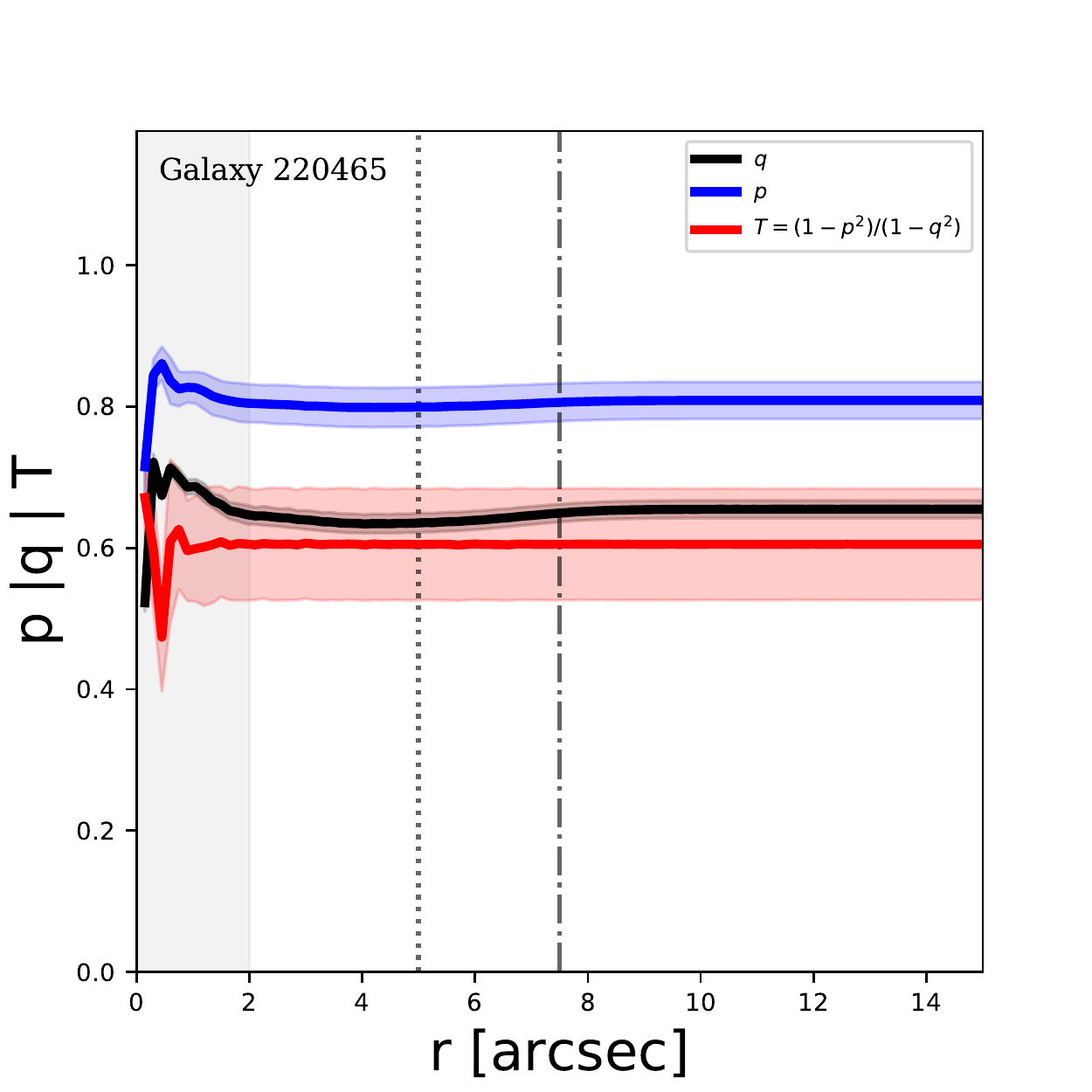}\includegraphics[scale=0.7]{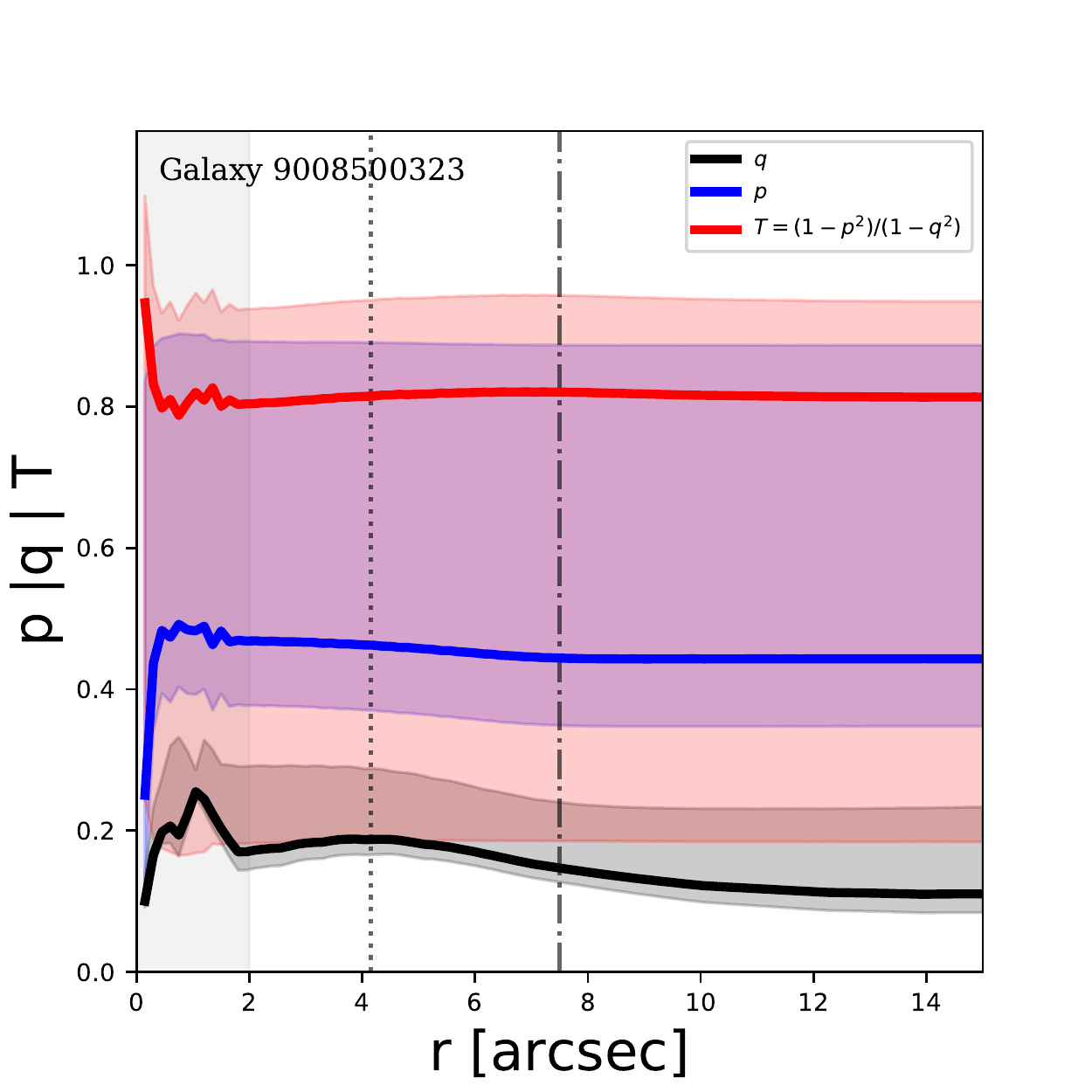}
\caption{Example galaxy 9403800123 (top left panel), 9011900793 (top right panel), 220465 (bottom left panel) and 9008500323 (bottom right panel): triaxiality. Variation of the axial ratios $p = b/a$, $q =
c/a$ and triaxial parameter $T = (1 - p^2)/(1 - q^2)$. The red, blue and black curves correspond to $p$, $q$ and $T$. The filled regions indicate the errors and the grey shaded region indicates the seeing limit ($r < 2^{\prime \prime}$). The vertical grey dotted and dash-dotted lines are located at 1$R_{\rm e}$ and at $R_{max}$, respectively. 9403800123 and 9011900793 are oblate in shape, while 220465 is triaxial and 9008500323 is close to prolate.}
\label{fig:pqt}
\end{figure}

\begin{figure}[ht!]
\centering
\includegraphics[scale=0.7]{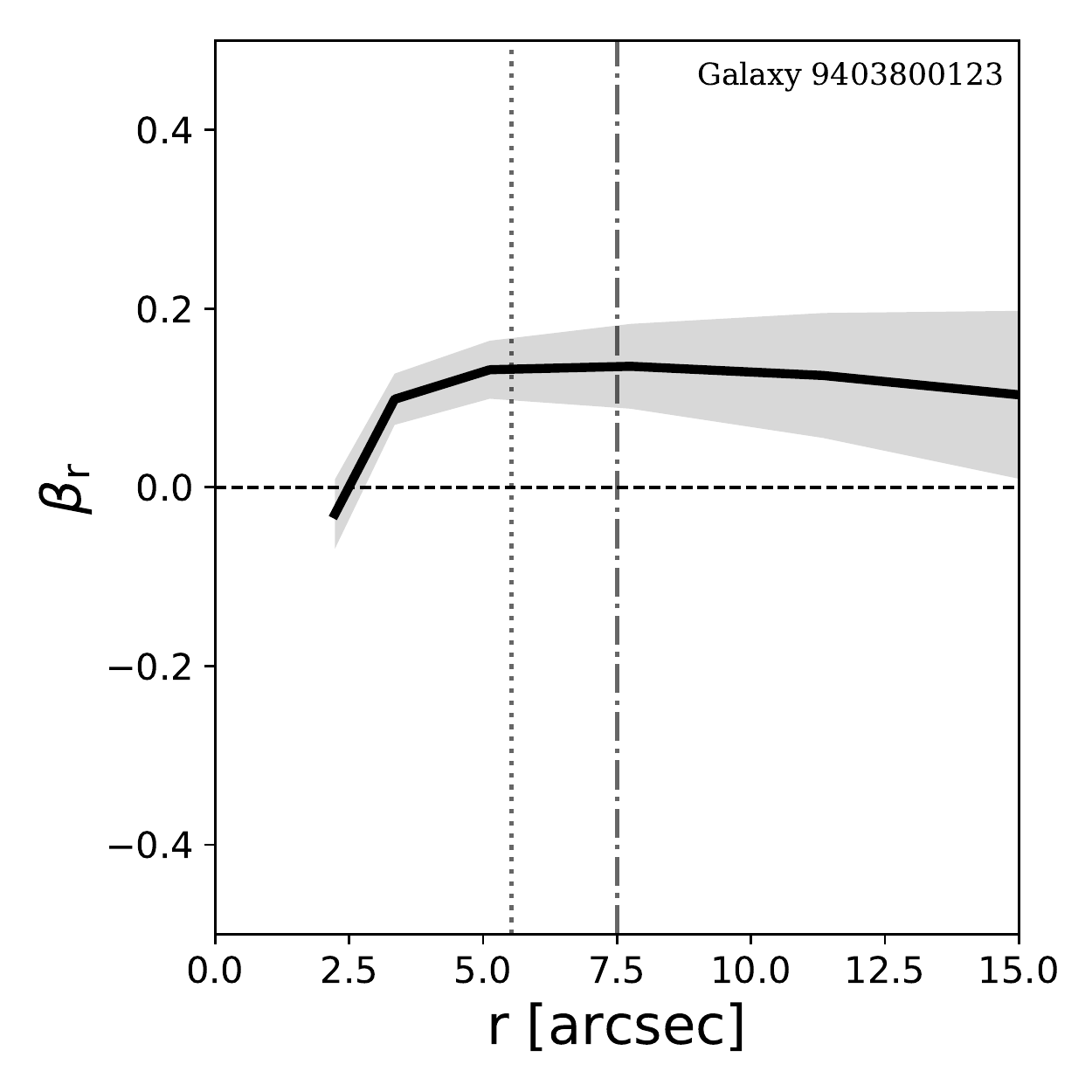}\includegraphics[scale=0.7]{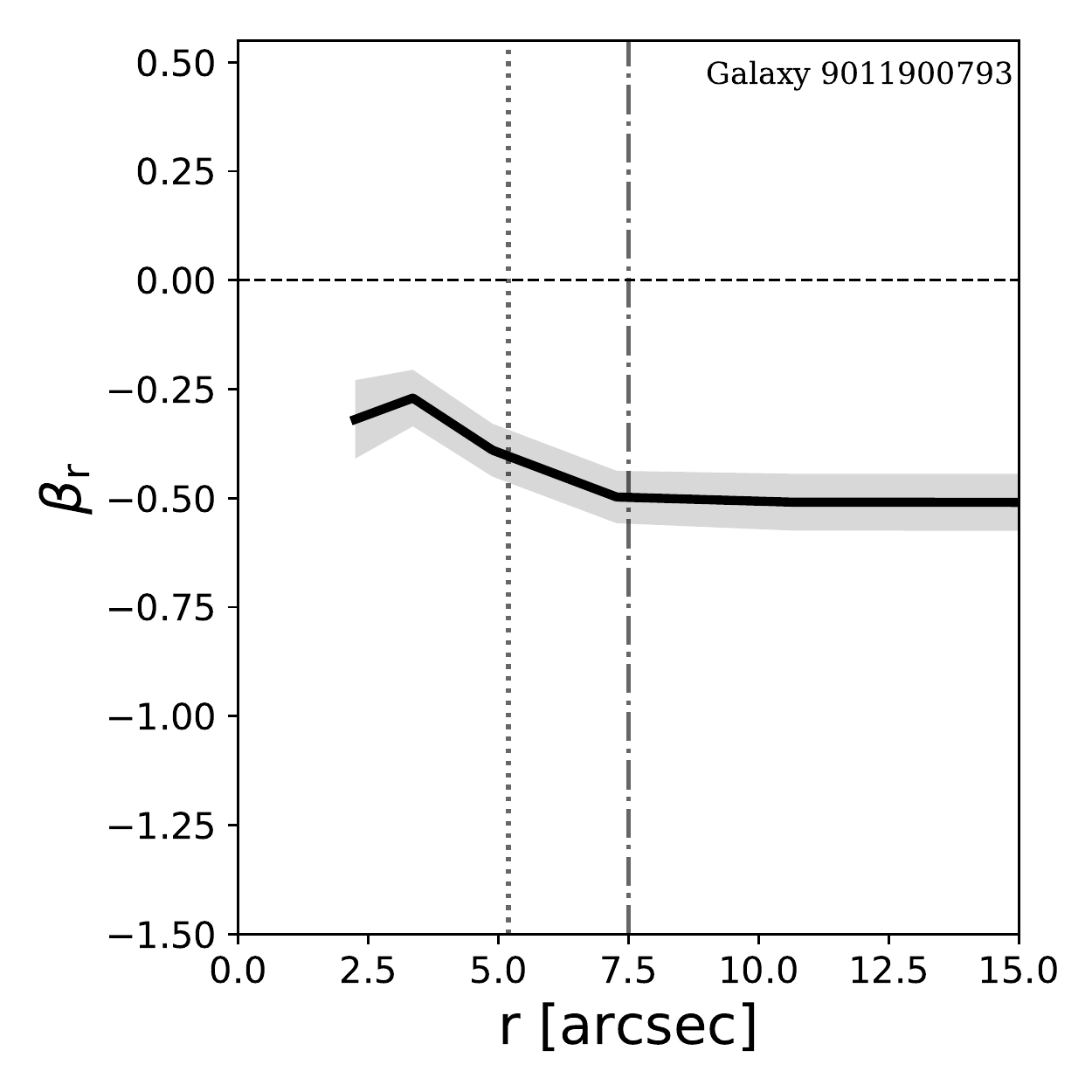}\\
\includegraphics[scale=0.7]{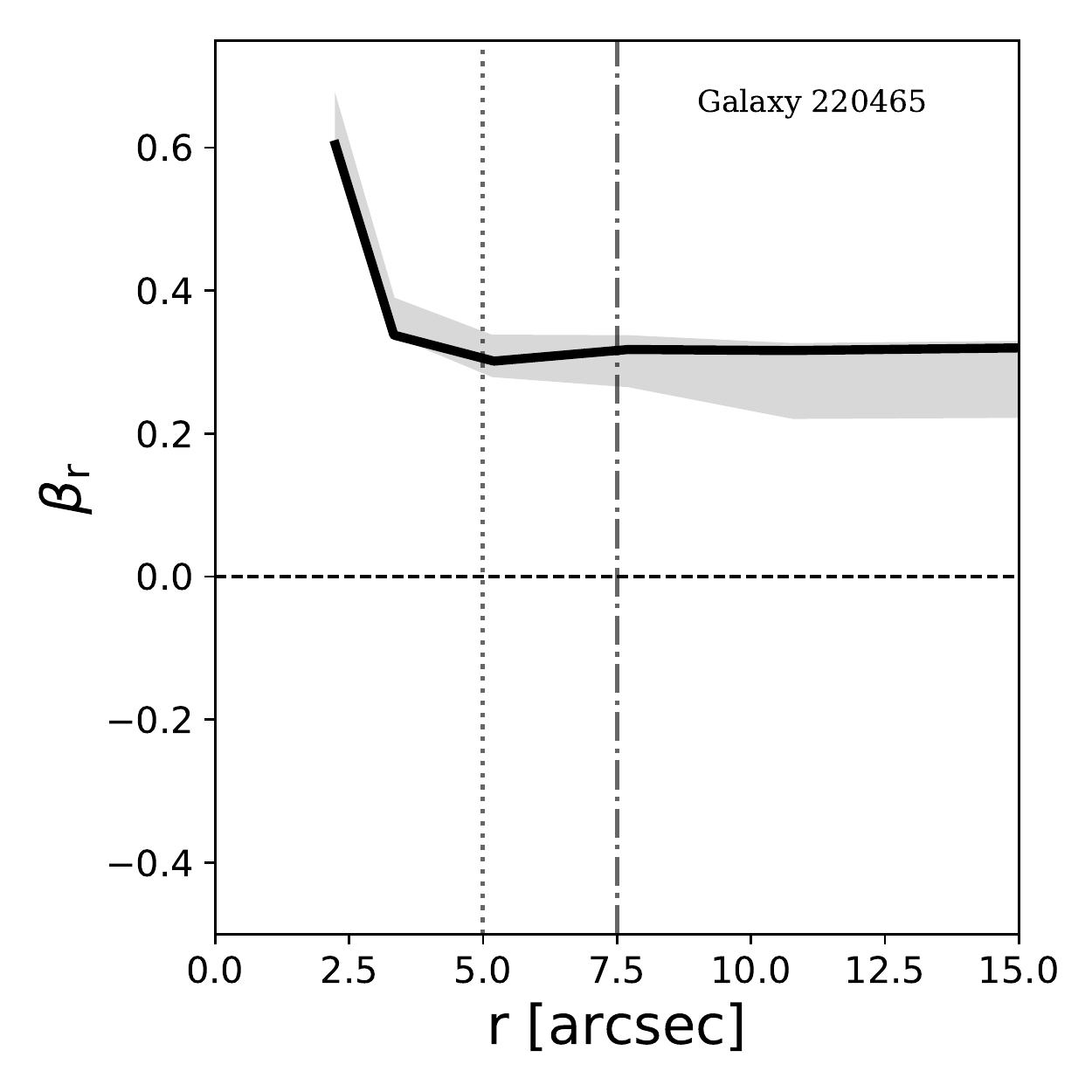}\includegraphics[scale=0.7]{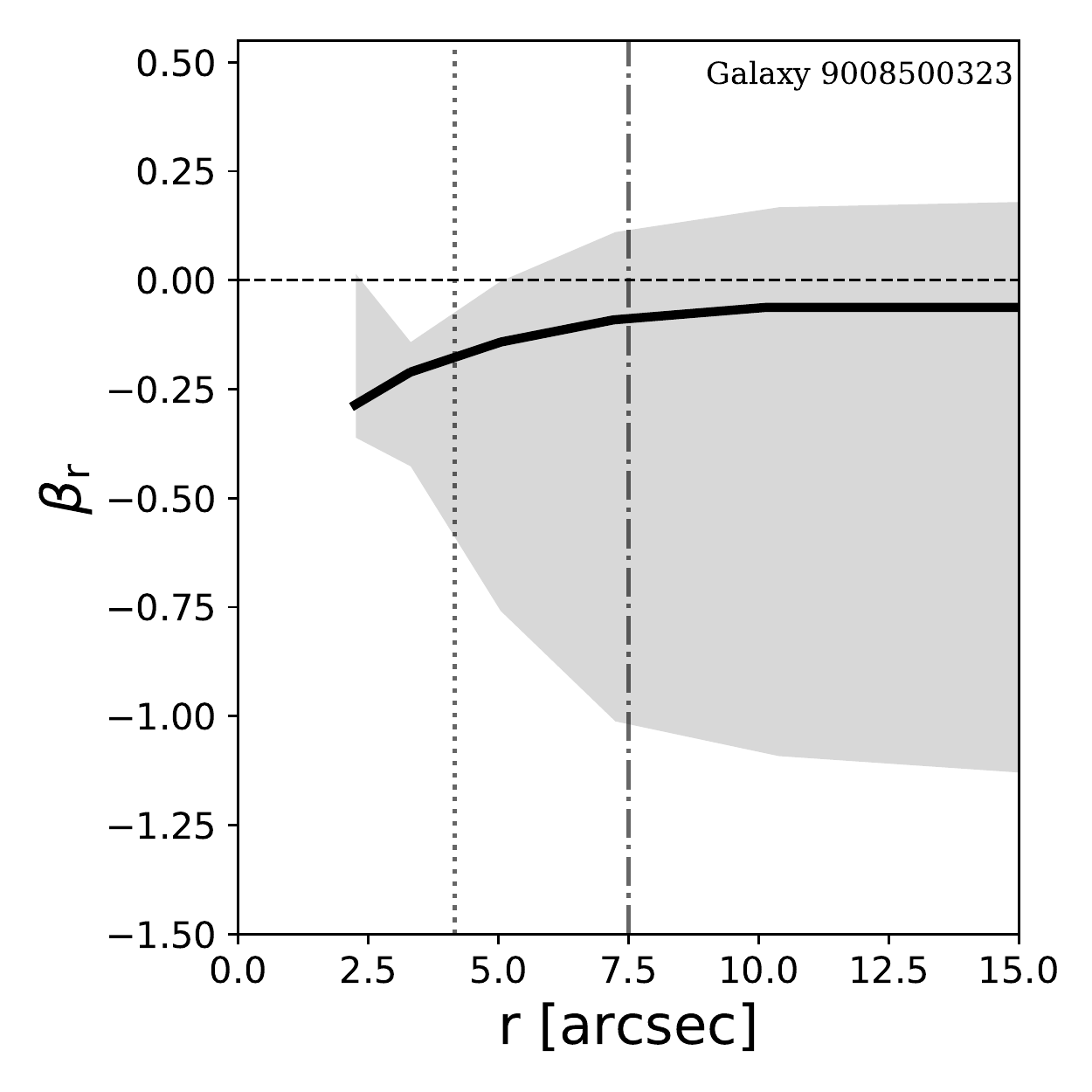}
\caption{Example galaxy 9403800123 (top left panel), 9011900793 (top right panel), 220465 (bottom left panel) and 9008500323 (bottom right panel): velocity anisotropy profile, $\beta_r$, as a function of the radius. The solid curves represent the velocity anisotropy profile obtained by the best-fit model. The filled region indicates the errors. The vertical grey dotted and dash-dotted lines are 1$R_{\rm e}$ and $R_{max}$, respectively. }
\label{fig:anisotropy}
\end{figure}

The parameter space for the complete model runs for example galaxies
9403800123, 9011900793, 220465 and 9008500323 (Fig. \ref{fig:123_bestfit},\ref{fig:793_bestfit},\ref{fig:465_bestfit} and \ref{fig:323_bestfit}) are shown in Fig. \ref{fig:123_grid}, Fig. \ref{fig:793_grid}, Fig. \ref{fig:465_grid} and in Fig. \ref{fig:323_grid}, respectively. The dots represent the parameters we have explored. Models within the best-fit region are color-coded according to their $\chi^2$ values. The largest red dot highlighted with a black cross indicates the best-fit model. Fig. \ref{fig:mass} to Fig. \ref{fig:anisotropy} show the obtained internal mass distribution, orbit circularity, triaxiality and tangential anisotropy for the four galaxies.

To test whether the parameter grid is well sampled in our iterative grid search, we run a super-sampled grid search for the 4 example galaxies. The best-fit parameters for both the default ($\sim$1250 models) and the “super-sampled” ($> 6000$) grid search are consistent with one another within the 1-$\sigma$ confidence level, converging on global minima.

We also tested whether including $h_5$ and $h_6$ make a significant difference to our best-fit model for the example galaxies. Fixing $h_5$ and $h_6$ to 0 and allowing the model to fit these higher moments does not significantly improve the fit. The variations in  $h_5$ and $h_6$ are quite small ($\sim 0.06$) and there are no significant changes in the kinematic fit, nor in the $\chi^2$ (derived from the fit to the measured moments) level (for example $\chi_{red}^2$ changed from 2.22 to 2.18 for example galaxy 9403800123) or morphology.

\section{Uncertainties on the model best-fit parameters}\label{app:errors}

\begin{figure}[ht!]
\centering
\includegraphics[width=5.5cm,trim=0.6cm 0cm 1.8cm 0cm,clip=True]{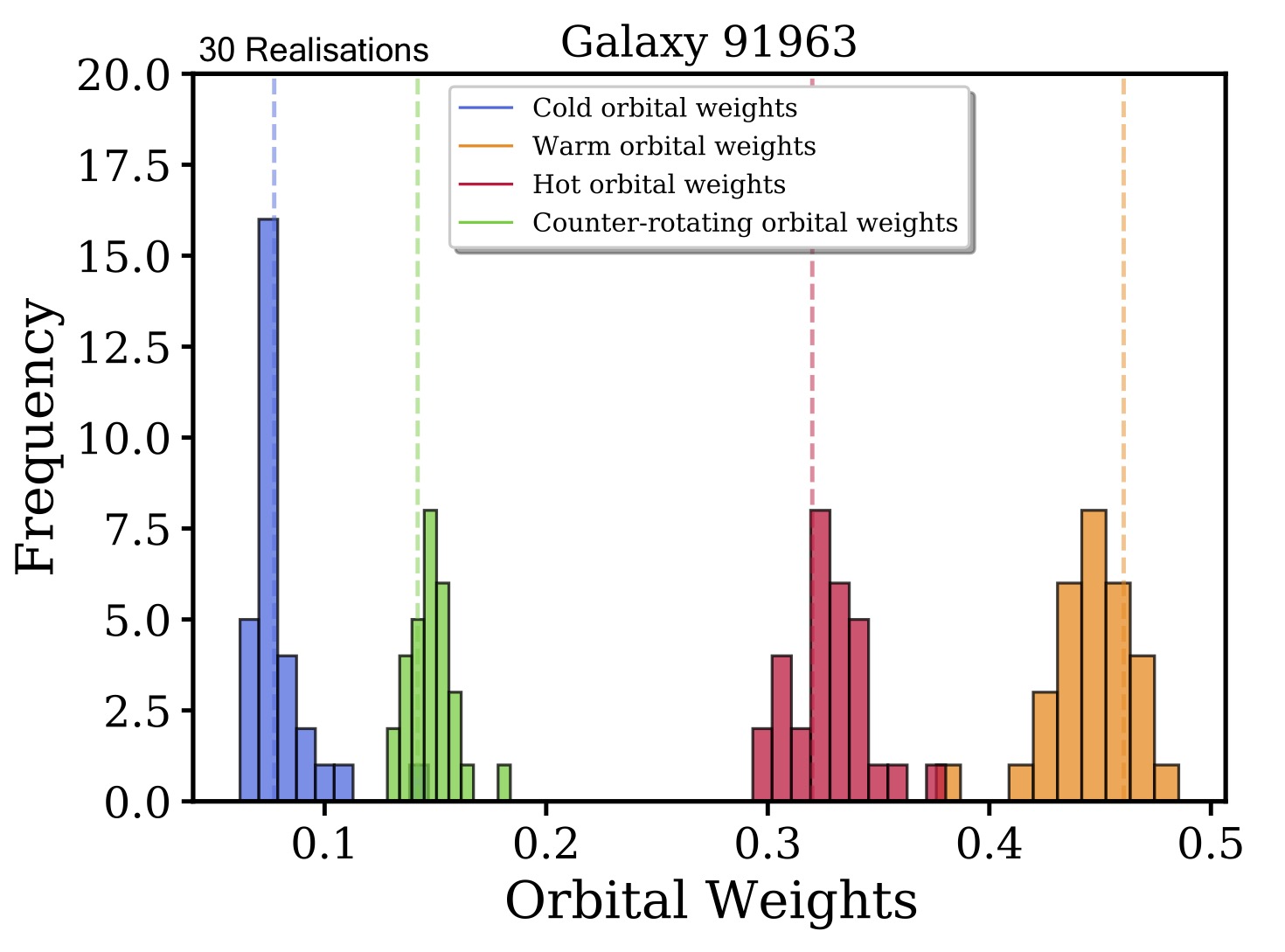}
\includegraphics[width=5.7cm,trim=0.6cm 0.7cm 1.5cm 0cm,clip=True]{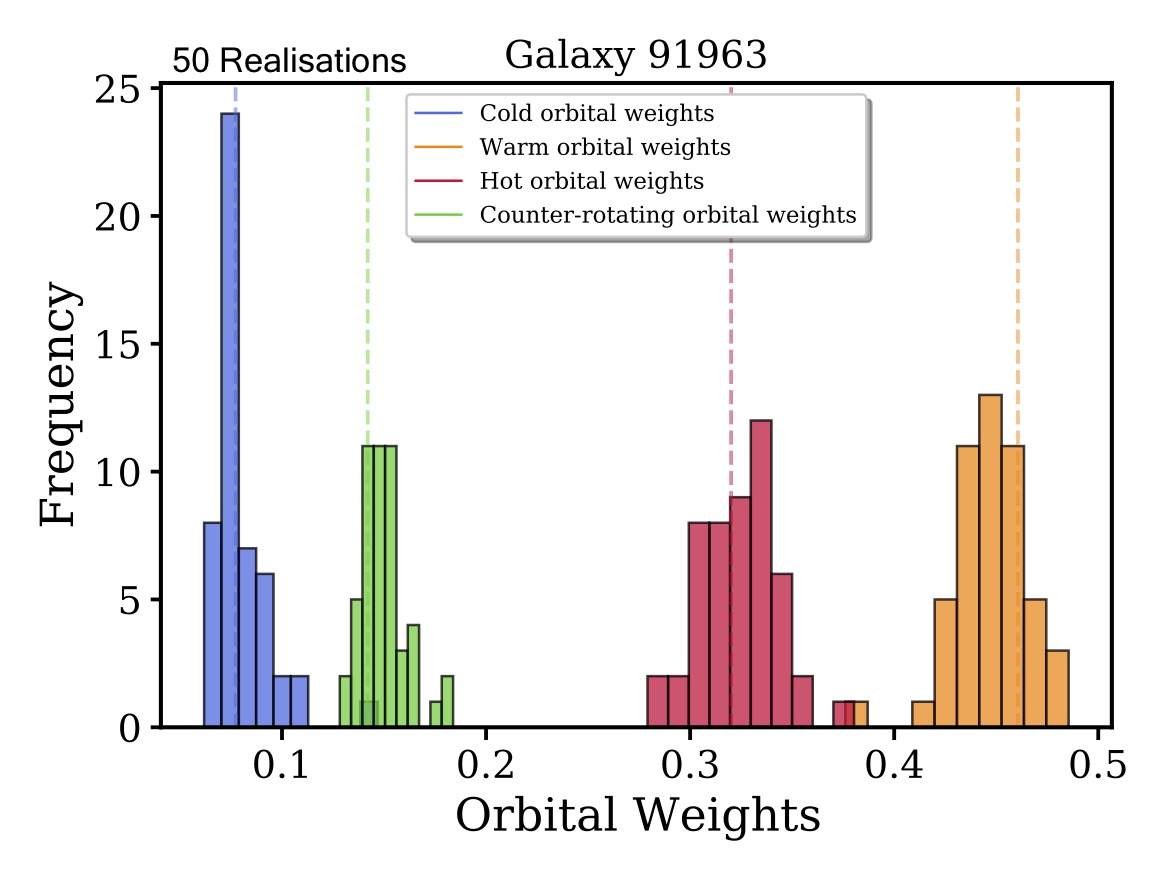}
\includegraphics[width=5.7cm,trim=0.6cm 0.7cm 1.5cm 0cm,clip=True]{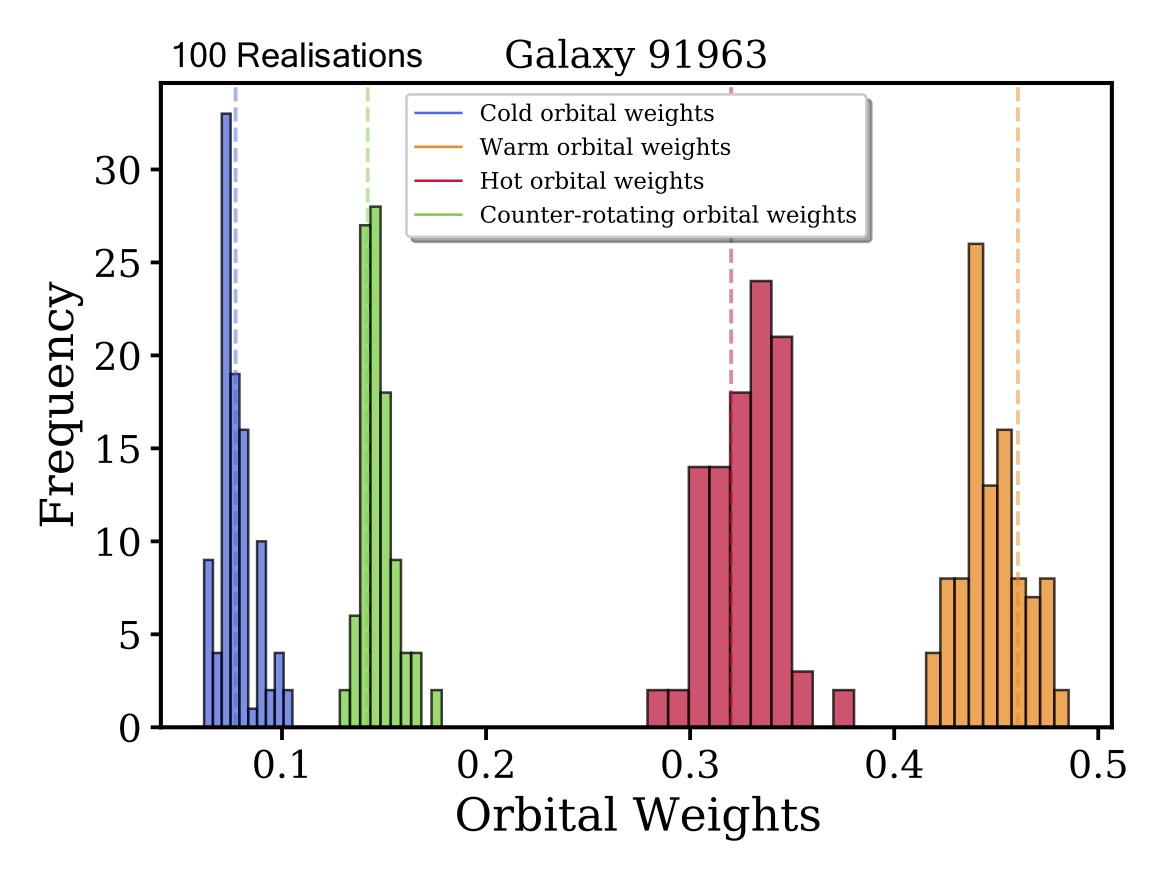}
\caption{Distribution of the orbital weights for the Monte Carlo realizations around the best-fit model values found for example galaxy 91963. Left-hand plot: 30 realizations; central plot: 50 realizations; right-hand plot: 100 realizations. The dashed lines represent the best-fit values. The unimodal distributions of the orbital components become more evident when increasing the number of realizations. We use 50 Monte Carlo realizations to derive the uncertainties for our galaxies to optimize the model run-time required.}
\label{fig:91963_mcrhist}
\end{figure}

In addition to the 1-$\sigma$ fluctuations from the best-fit model, we use Monte Carlo realizations to estimate the uncertainties on our best-fit values. This is particularly important to derive the uncertainties in the underlying model properties which are not accessible from the 1-$sigma$ confidence level directly. This approach factors in convergence issues and grid sampling, with no assumptions about how the model parameter uncertainties are distributed - only that the kinematic errors are Gaussian (a common assumption).
To this end, we select 16 SAMI galaxies ($\sim$ 10\% of the total sample), spanning different regions in the size - stellar mass plane. We apply Monte Carlo realizations, as described below, to each one of them, and we use the resulting variations from the best-fit parameters as the uncertainties for galaxies located in similar locations of the galaxy mass-size plane. For each galaxy, we take the kinematic values from the best-fit model and perturb them by adding noise, taken from a Gaussian distribution with standard deviation equal to the mean error of each observed kinematic moment ($V, \sigma, h_3, h_4$).
We keep the standard deviation as the uncertainty for each perturbed value. We tested repeating this process to have 30, 50 and 100 different realizations.
We then derive the best fit for each of the perturbed kinematic maps, using the same iterative grid search described in Sec. \ref{sec:bfmodel}. We compare the orbital weights retrieved from each realization and we find that there is in general good agreement, in particular when looking at the fitted inclination angle and the internal mass distributions values. The left-hand plot of Fig \ref{fig:91963_mcrhist} shows the average of the best-fit parameters derived for 30 Monte Carlo realizations of the best-fit model of example galaxy 91963. We find that the fraction of the orbits in passive galaxies follow a unimodal distribution. This becomes more evident when considering 50 or 100 realizations (right-hand panels of Fig. \ref{fig:91963_mcrhist}). 
We therefore decided to use 50 Monte Carlo realizations as a good compromise in deriving the uncertainties on the best-fit values for SAMI galaxies, since 100 realizations for even 16 galaxies are unfeasibly time-consuming. 

From the 50 best-fit models, we derive the standard deviation of the quantities of interest (e.g. fraction of cold orbits, etc.), and use this as the 1-$\sigma$ uncertainty around the value derived from the original best-fit model. With the Monte Carlo realizations we find typical uncertainties of 3-5\% for $p_{Re}$ and $q_{Re}$, $\sim$10-15\% for the fractions of orbits, $\sim$ 15\% for $\beta_r$ and $\sim$ 15\% for $\lambda_{Re,EO}$. The uncertainties for the fraction of dark matter are around 8\%. To the uncertainty of each parameter derived with this method we also add, in quadrature, the 1$\sigma$ confidence level from the parameter grid, which represents the model fluctuations. This method is applied to derive the uncertainties of all the quantities presented in this work.

\section{Velocity anisotropy parameter, $\beta_z$}\label{app:betaz}

We define the velocity anisotropy parameter, $\beta_z$, in cylindrical coordinates, following \cite{Cappellari2007}: 
\begin{equation}
\beta_z = 1 - \frac{\Pi_{zz}}{ \Pi_{RR}},
\end{equation}
with $\Pi_{kk}$ as defined in Equation \ref{eq:D_kk}. This parameter describes the global shape of the velocity dispersion tensor in the ($v_R, v_z$) plane. We calculate the value of $\beta_z$ within 1$R_{\rm e}$, excluding the inner regions ($r<2^{\prime \prime}$) since this is smaller than the FWHM of the PSF of our observations.

\begin{figure}[ht!]
\centering
\includegraphics[scale=0.60]{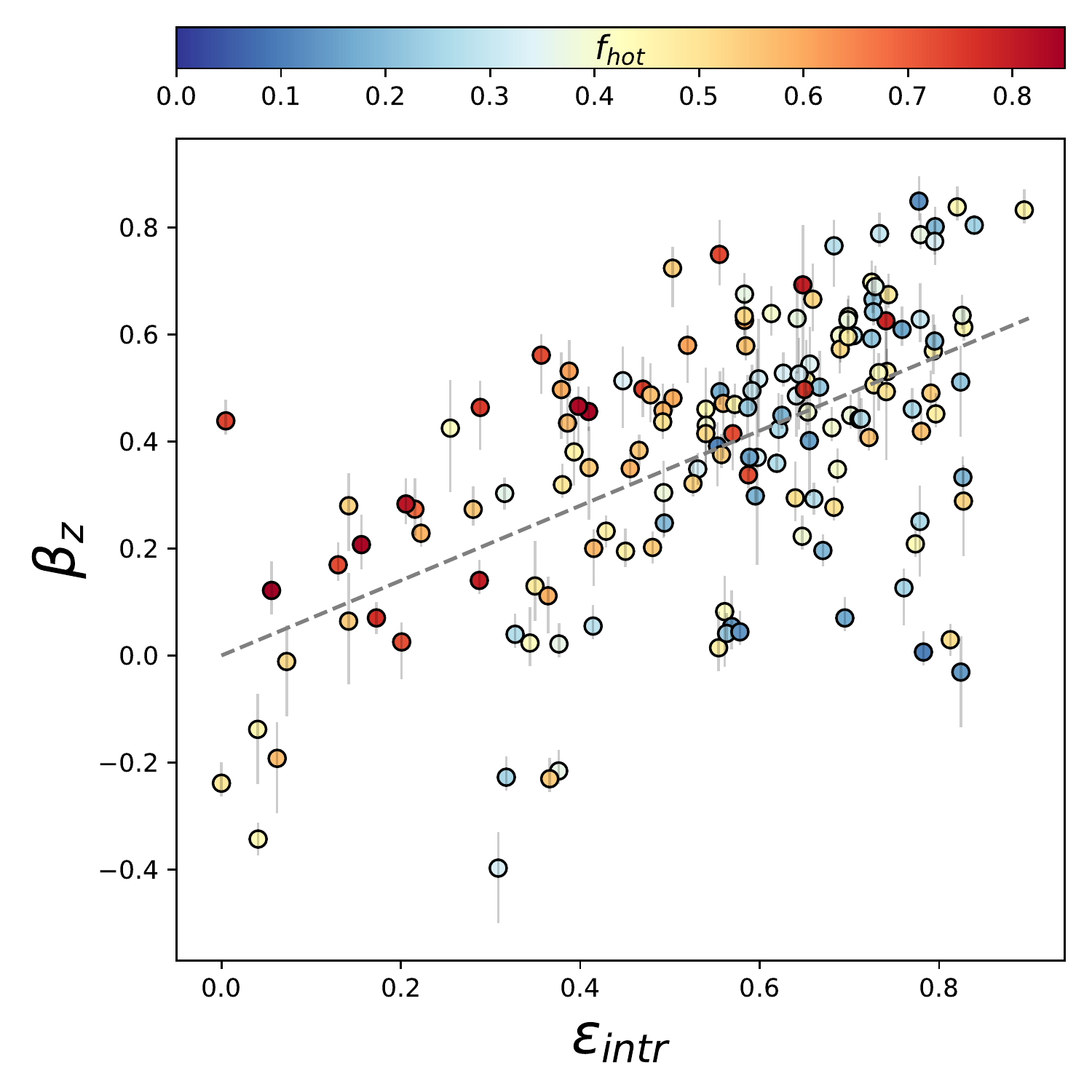}
\caption{Velocity dispersion anisotropy, $\beta_z$, within 1$R_{\rm e}$ as a function of intrinsic ellipticity ($\varepsilon = 1-q$), color-coded by their fraction of hot orbits. The grey line shows the relation $\beta_z = 0.7 \times \varepsilon_{intr}$ from \cite{Cappellari2007}. Galaxies with higher ellipticities have higher values of $\beta_z$. This means that flatter galaxies are more anisotropic than rounder galaxies. }
\label{fig:betaz}
\end{figure}

Fig. \ref{fig:betaz} shows the derived values of $\beta_z$ at 1$R_{\rm e}$, for each galaxy, as a function of intrinsic ellipticity ($\varepsilon = 1-q$). Galaxies with higher ellipticities have higher values of $\beta_z$. This means that flatter galaxies are more anisotropic than rounder galaxies. The grey line shows the relation $\beta_z = 0.7 \times \varepsilon_{intr}$ from \cite{Emsellem2007}. In general, we find higher values of $\beta_z$ compared to those seen in \cite{Cappellari2007} for the early-type galaxies in their sample from the SAURON survey. However, they applied axisymmetric Schwarzschild dynamical models to only 24 of their galaxies (a subsample that was consistent with axisymmetry), while the Schwarzschild dynamical models we use also include a set of box orbits that allow for triaxiality. Therefore, the scatter that we see in our relation, is likely to be due to the contribution from hot orbits. This is better shown by color-coding the galaxies in the $\beta_z - \varepsilon$ plane by their fraction of hot orbits. As seen in Fig. \ref{fig:betaz}, we have contributions $>20\%$ from hot orbits in all of the galaxies in our sample. The negative $\beta_z$ values that we find can be explained with the velocity ellipsoids not being cylindrically aligned, as mentioned in Sec. \ref{sec:beta}.

\section{Ratio of ordered to random motion}\label{app:vsigma}

\begin{figure}[ht!]
\centering
\includegraphics[scale=0.60]{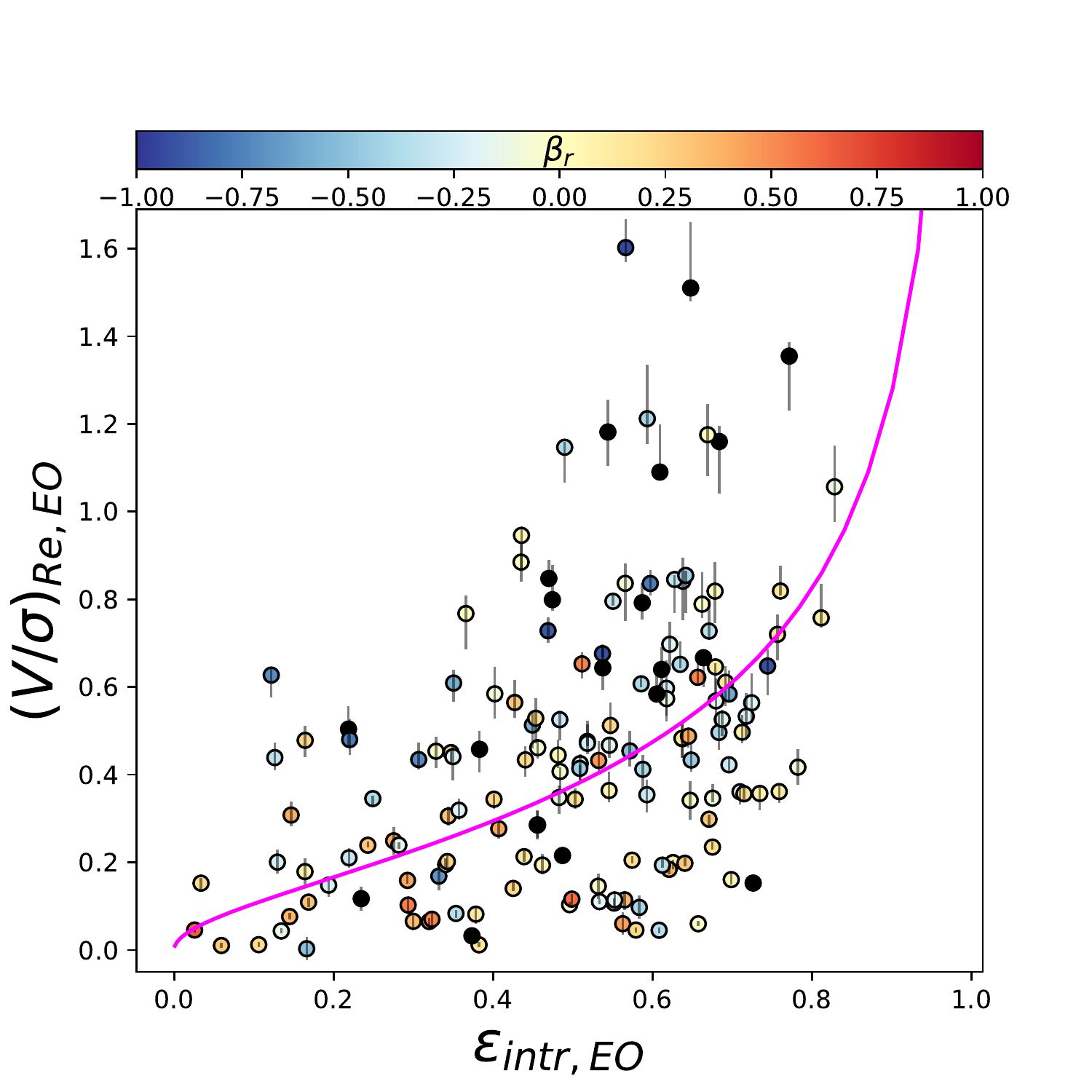}
\caption{$V/\sigma$ as a function of the ellipticity $\varepsilon_{intr,EO}$ derived from MGE fits to the edge-on projected maps, calculated at 1$R_{\rm e}$. The magenta line corresponds to the relation $\beta_z = 0.7 \varepsilon$ for edge-on galaxies \citep{Emsellem2007}. Data points are color-coded by the velocity anisotropy $\beta_r$. As expected, $V/\sigma$ increases with increasing intrinsic ellipticity.}
\label{fig:vsigma_eps}
\end{figure}

For completeness, we also measure the ratio of ordered to random motion
$V/\sigma$, also measured within 1$R_{\rm e}$, using the definition from \cite{Cappellari2007}:
\begin{equation}
    \left(V/\sigma \right)^2 = \frac{\sum_{i=0}^{N_{spx}} F_i V_i^2}{\sum_{i=0}^{N_{spx}} F_i \sigma_i^2}.
\end{equation}
Results obtained using $V/\sigma$ are similar to those obtained for $\lambda_{Re,EO}$ (see Fig. \ref{fig:lambda_r_beta} in Sec. \ref{sec:lambda_r}) and are shown in Fig. \ref{fig:vsigma_eps}.
\end{document}